\documentclass[12pt]{article}

\usepackage{amsmath,amsfonts,amssymb,latexsym}
\usepackage{mathtools}
\newcommand{\sfac}{\mathsf{a}} 

\usepackage{graphicx}
\usepackage{caption}
\usepackage{subcaption}

\usepackage[T1]{fontenc}
\usepackage[utf8]{inputenc}

\usepackage{microtype}

\usepackage[a4paper,left=2.2cm,right=2.2cm,top=2.6cm,bottom=2.6cm]{geometry}

\usepackage{setspace}
\PassOptionsToPackage{dvipsnames}{xcolor}
\usepackage{xcolor}
\usepackage{silence}
\usepackage{tikz}
\usepackage{pgfplots}
\pgfplotsset{compat=1.18}
\usepackage{tikz-3dplot}
\usetikzlibrary{arrows.meta,calc,decorations.markings}

\usepackage{amsthm}

\newtheorem{theorem}{Theorem}[section]

\newtheorem{lemma}{Lemma}[section]

\newtheorem{remark}{Remark}[section]

\numberwithin{equation}{section}

\usepackage{comment}

\def\be{\begin{equation}}
\def\ee{\end{equation}}
\def\bq{\begin{eqnarray}}
\def\eq{\end{eqnarray}}
\def\beq{\begin{eqnarray}}
\def\eeq{\end{eqnarray}}
\newcommand{\spec}{\operatorname{spec}}

\usepackage[colorlinks=true,
linkcolor=MidnightBlue,
citecolor=MidnightBlue,
urlcolor=MidnightBlue]{hyperref}
\begin{document}


\title{\textsc{Persistence and Transition Varieties in Scalar Field Cosmology}}

\author{\Large{\textsc{Spiros Cotsakis$^{1,2}$\thanks{\texttt{skot@aegean.gr}}}}\\
$^{1}$Clare Hall, University of Cambridge, \\
Herschel Road, Cambridge CB3 9AL, United Kingdom\\
$^{2}$Institute of Gravitation and Cosmology,  RUDN University\\
ul. Miklukho-Maklaya 6, Moscow 117198, Russia\\
}

\date{July 2026}

\maketitle

\begin{abstract}\noindent
We develop a unified bifurcation-theoretic description of Friedmann--Robertson--Walker cosmologies with a scalar field,
a barotropic fluid of index \(\gamma\), and spatial curvature. For the strict exponential potential
\(V(\phi)=V_0e^{\lambda\phi}\), the slope \(a=\sqrt{3/2}\,\lambda\) acts as a distinguished parameter, and the local
phase portrait is organised by a small transition set in the \((\gamma,a)\) plane: the kinetic endpoint threshold
\(|a|=3\), the curvature-induced scalar threshold \(a^2=3\), the scalar--fluid exchange curve \(a^2=\tfrac92\gamma\),
and the degeneracy axes \(\gamma=\tfrac23\) and \(\gamma=2\). 
For the quadratic (massive) potential \(V(\phi)=\tfrac12 m^2\phi^2\), the effective slope is dynamical. We introduce a
compact bounded-slope closure \(\zeta=\arctan\lambda\) yielding a closed autonomous four-dimensional system in
\((X,Y,\Omega_k,\zeta)\) without time-rescaling. This formulation exposes invariant gates and robust equilibrium continua
(lines and, on \(\gamma=\tfrac23\), a connecting segment) together with vertical \(\gamma\)-thresholds controlling
loss and recovery of normal hyperbolicity. Near the corresponding organising loci we compute translated jets, perform centre(-like) reductions, and extract canonical normal forms governing persistence and transitions.
We then assemble the local normal forms into an explicit stratification of the exponential parameter plane and a
canonical pull-back stratification for the massive extensions, and formulate the corresponding physical path maps into
the relevant unfolding charts. This makes explicit how the fluid and curvature modes supply the deformation directions
needed to realise the versal effects within the FRW class itself and leads to a direct regime-level interpretation.
In the strict exponential class, slow roll and ultra slow roll appear as persistent attracting balances and
nonhyperbolic bottleneck passages, respectively, while on the massive side the quadratic invariant slices recover the
oscillatory sector identified previously in the mode-interaction setting, including its periodic-orbit / invariant-torus
mechanism. The resulting framework organises not only robust local effects such as critical slowing, curvature
leakage, and tracker exchange, but also the admissible sequences of such episodes along massive trajectories.
\end{abstract}
\newpage
\tableofcontents
\newpage

\section{Introduction}
\label{sec:intro}

A familiar way to formulate stability in the Einstein--matter system is through the symmetry ladder of solutions.
For example, in cosmology one moves from homogeneous and isotropic Friedmann--Robertson--Walker (FRW) models
to anisotropic and inhomogeneous classes and ultimately to the fully generic Cauchy problem for the Einstein equations
as a constrained hyperbolic system~\cite{ycb09}. This viewpoint is indispensable, but it addresses only one notion of generality: generic perturbations of a given spacetime within increasingly large classes. A complementary notion arises already within a fixed symmetry class, where the reduced Einstein–matter equations form a parameter-dependent dynamical system and hence a bifurcation problem.
From this perspective, the relevant question is not only well-posedness, regularity  and stability under generic Cauchy perturbations (the Cauchy problem), but also structural stability and its versal unfoldings at each step of the symmetry ladder—what one may call the Thom problem for general relativity, in the spirit of Ren\'e Thom~\cite{thom72,arnold85,kuz23,montaldi21}.  The resulting bifurcation geometry smoothly deforms the dynamics and `organises' robust transitions and persistence domains (i.e.\ parameter space regions on which the topology of the phase portraits changes, or the qualitative phase portrait is unchanged, respectively) that need not, in general,  be visible from symmetry breaking perturbation analyses alone.
In this paper we develop this viewpoint for FRW scalar-field cosmologies with exponential and quadratic potentials, including fluid and curvature, and show how extra state space  directions and a bounded slope mode generate organising centres, transition varieties, and unfolding (deformation) histories within a unified framework.

Scalar-field cosmologies provide a flexible dynamical framework for modelling accelerated expansion,
scaling regimes, and multi-stage cosmic histories (cf.~\cite{coley1}-\cite{ChervonBook} and refs. therein). Their qualitative behaviour is controlled not only by the potential but also by which additional components are present (fluids, curvature)~\cite{6}-\cite{u3}, and by
where the system sits relative to organising thresholds at which hyperbolicity is lost~\cite{MI-I}. Near such
thresholds, standard attractor intuition breaks down and robust transitional phenomena emerge
(critical slowing, exchange of attractors, slow curvature leakage, and drift along equilibrium
manifolds). The goal of this paper is to isolate these organising thresholds, compute the associated
translated jets and reduced normal forms, and assemble them into a persistence/transition geometry
that can be read directly.

In the context of scalar-field cosmology, bifurcation geometry enters through singularity theory  and works at three levels (absent in hyperbolic treatments due to the implicit function theorem engagement): firstly,  by dealing with degeneracies at the level of equilibrium solutions (i.e., non-hyperbolic sites - points and continua - usually parameter dependent `branches') and their corresponding parameter space loci,  via \emph{singular sets and discriminants} (i.e., bifurcation sets) before centre manifold and normal form reductions; secondly, after reduction in the context of \emph{organising-centre  geometry} of the state-(auxiliary-)parameter space (i.e.,  thresholds,  gates,   episodes, and reduced jets); and last, at the \emph{versal deformation level} where the unfolding parameter space rules the dynamics completely and introduces transition varieties and stratifications through bifurcation diagrams which deploy all possible stable perturbations of the system.

A central theme emerging already in the first level of degenerate sites and loci is the distinction between \emph{parameter-driven} and \emph{state-variable} organisation. In the strict exponential class \(V(\phi)=V_0 e^{\lambda\phi}\), the slope \(a=\sqrt{\tfrac32}\lambda\) is a genuine distinguished (i.e., external) parameter, and organising loci appear as curves
and axes in the \((\gamma,a)\) plane ($\gamma$ denotes the fluid parameter). In the harmonic (quadratic, or `massive') class \(V(\phi)=\tfrac12 m^2\phi^2\), by contrast, the effective slope evolves, so the organising picture becomes that of an augmented state space with slow drift in a slope direction. A single orbit can therefore experience several regime changes without any distinguished parameter variation: its \emph{frozen-slope projection}
\((\gamma,a_{\rm eff}(N))\) (with \(N=\ln \texttt{a}\) the e-fold time (\(\texttt{a}\) being the scale factor)) traces a curve in \((\gamma,a)\) and can
cross the pulled-back exponential `stratification' (i.e., in the distinguished rather than unfolding parameter space), triggering the same qualitative transitions identified in the parameter-driven setting.

A key technical step in the harmonic extension is a compact closure of the slope dynamics.
For \(V(\phi)=\tfrac12 m^2\phi^2\) the usual slope variable \(\lambda:=-V_{,\phi}/V\) is unbounded as
\(\phi\to0\), obscuring global phase--space structure. In particular, the slope direction becomes non-compact and the naive \(\lambda\)-closure introduces an apparent singular boundary at \(\phi=0\), so that trajectories can run off to \(|\lambda|\to\infty\) and the global invariant-set/equilibrium organisation is difficult to represent in a single compact state space. This also complicates direct comparison with the strict exponential class, where the slope is a fixed parameter and the organising loci are naturally described in the \((\gamma,a)\) plane. We introduce instead the bounded slope mode
\(\zeta=\arctan\lambda\in(-\pi/2,\pi/2)\), which regularises the slope direction without rescaling time and yields a closed autonomous four--dimensional system for the state and slope variables \((X,Y,\Omega_k,\zeta)\) in e-fold time.  This makes the additional slope variable explicit and identifies the invariant submanifolds on which the slope evolution degenerates (\(\zeta'=0\)), allowing the organising centre analysis of the harmonic class to proceed on the same footing as for the strict exponential case.

Further, to connect and compare with the harmonic case, we present a consolidated dynamical and singularity-theoretic treatment~\cite{ar93}-\cite{wig03} of exponential scalar-field cosmologies with fluid and curvature, with the organising geometry made explicit through translated jets,
centre reductions and normal forms, from which the versal deformation geometry is more or less directly obtainable. A key organising episode for the scalar-only case is the coupled event at \(|a|=3\) (\(a=\sqrt{\tfrac32}\lambda\)), where the kinetic endpoints exhibit a hilltop-type linear degeneracy while the scalar equilibrium simultaneously reaches its existence boundary on the constraint set, producing a
multi-site organising centre. The additional state-space directions associated with the fluid and curvature
fractions then enter the reduced jets as independent deformation directions, supplying the local unfolding
coefficients that underlie transition varieties and persistence domains. This framework clarifies which
features persist under component extensions and supports a systematic route to extracting targeted physical
predictions from the same organising geometry.

\paragraph{Reader's routes.}
Readers primarily interested in the main organising picture and conclusions can read
Sections~\ref{sec:intro}--\ref{sec:mainresults} and then jump to the Discussion
(Section~\ref{sec:discussion}), together with the short glossary in
Appendix~\ref{app:glossary}. Readers who would like a quick entry into the organising
terminology before proceeding to the detailed reductions may also consult Appendix~\ref{app:toy},
which illustrates the main notions on a few simple low-dimensional toy systems. As a minimum,
we suggest the following two entry points: Theorem~\ref{thm:Ri-summary} lists the main results,
and Table~\ref{tab:results-dashboard} points to the detailed derivations in later sections.
Sections~\ref{sec:exp-Z2-gate}--\ref{sec:primer} contain the detailed equilibrium classifications, translated jets,
centre-manifold reductions, normal-form derivations, as well as the versal constructions that
underpin the summary in Sections~\ref{sec:intro}--\ref{sec:mainresults}; they may be read
selectively as needed.

\paragraph{Guide to the paper.}
We begin with a concise summary of the harmonic and exponential landscapes and of the relevant
linearised stability, organising geometry and notation, and unfolding inputs in
Section~\ref{sec:backgroundandnotation}. Section~\ref{sec:mainresults} states the main results
of the present work. We then introduce the basic exponential system and the choice of variables in
Section~\ref{sec:exp-Z2-gate}, and perform the shifted-equilibrium and organising-centre tests.
The minimal scalar--fluid and scalar--fluid--curvature extensions are developed in
Sections~\ref{sec:fluid-org} and \ref{sec:curvature}, respectively. The corresponding massive
(quadratic) fluid and curvature extensions are treated in Section~\ref{sec:massive-extensions}.
We assemble these pieces into a unified unfolding and bifurcation-diagram narrative in
Section~\ref{sec:unfoldings}, and then develop the associated transition varieties and persistence
framework in Section~\ref{sec:strata}. Section~\ref{sec:primer} provides an expository
``primer'' that translates the stratified normal-form picture into a short dictionary of robust
dynamical effects (critical slowing, curvature leakage, tracker exchange, etc.) and a practical
workflow for reading off dominant regimes from the charts. We conclude with a discussion in
Section~\ref{sec:discussion}. Key terminology and correspondences are collected in
Appendix~\ref{app:glossary}, while Appendix~\ref{app:toy} gives a brief pedagogical guide to
the main organising notions through simple toy examples.

\section{Background, variables, and organising notions}
\label{sec:backgroundandnotation}
In this Section, after a brief discussion of  background notation and a mention of linear stability results, we provide a general  description of the state variables (phase space) and parameter spaces (distinguished and unfolding). In particular, we motivate the use of singularity and bifurcation theory in scalar field cosmology, and then present a `high-level' (no specific details) introduction to the main problem addressed in this paper. This  facilitates our description of the main results  in the next Section.

\subsection{Background and notation}
\label{subsec:BackNot}
We consider spatially homogeneous and isotropic FRW cosmologies with a minimally coupled scalar field
\(\phi\) with potential \(V(\phi)\), optionally coupled to a one-fluid component with barotropic index
\(\gamma\in(0,2]\)\footnote{We adopt the standard one-fluid range \(\gamma\in(0,2]\) throughout; most local computations extend formally to other \(\gamma\)-ranges, but their physical interpretation may change (e.g.\ effective/phantom components).}, pressure $p$, energy density $\rho$, and allowing nonzero spatial curvature \(k\). Here $\gamma$ is the standard barotropic index of the additional perfect-fluid component,
so that $w_m=\gamma-1$. Thus $\gamma=1$ corresponds to dust-like matter,
$\gamma=4/3$ to radiation, and $\gamma=2$ to a stiff fluid. The value
$\gamma=2/3$ corresponds to $w_m=-1/3$, namely the threshold at which the fluid
redshifts in the same way as the spatial-curvature term; this is why it reappears below as
a distinguished organising value. More generally, in the present framework $\gamma$ may
also be viewed as an effective barotropic parameter for whichever additional homogeneous
component is retained in the FRW background.

The
FRW line element (signature $(-,+,+,+)$) is
\be
ds^{2}=-dt^{2}+\texttt{a}(t)^{2}\left(\frac{dr^{2}}{1-k r^{2}}+r^{2}d\Omega_{2}^{2}\right),
\ee
where $\texttt{a}(t)$ denotes the scale factor of the universe as a function of the proper time,  $d\Omega_{2}^{2}=d\vartheta^{2}+\sin^{2}\vartheta\, d\varphi^{2}$, \(k=0,\pm 1\),  and the energy density and pressure assume their usual forms $\rho_\phi,p_\phi$ for the scalar field,  
while we use the notation $\rho_m,p_m$ for those of the fluid content. 

The dynamical systems formulation of FRW scalar-field cosmology is classical: using expansion--normalised variables and a
monotone time variable (often \(N=\ln \texttt{a}\)), the Einstein--matter system reduces to an autonomous flow on a constrained state
space. The qualitative behaviour is then described in terms of equilibria (kinetic, potential-dominated, fluid-dominated, curvature-dominated, and  scalings), their invariant manifolds, and global monotonicity properties; comprehensive reviews and many model-specific studies exist
(see e.g.\ \cite{coley1}-\cite{u3} and refs. therein).
Using e-fold time \(N=\ln \texttt{a}\) and the standard Hubble-normalised variables 
\begin{equation}\label{eq:stand-variables}
X:=\frac{\dot\phi}{\sqrt6\,H},\qquad
Y:=\frac{\sqrt{V}}{\sqrt3\,H},\qquad
\Omega_m:=\frac{\rho_m}{3H^2},\qquad
\Omega_k:=-\frac{k}{a^2H^2},
\end{equation}
the dynamics closes (after specifying \(V\)) as an autonomous system subject to the Friedmann constraint
\begin{equation}\label{eq:friedmann-constraint-bg}
X^2+Y^2+\Omega_m+\Omega_k=1.
\end{equation}
Throughout this paper, ``curvature'' refers exclusively to the spatial curvature of the
homogeneous FRW background, encoded by the discrete parameter $k$ and by its
Hubble-normalised variable $\Omega_k=-k/(a^2H^2)$. We do not consider metric
perturbations or perturbative curvature variables (such as Bardeen potentials or
comoving-curvature perturbations), so ``curvature'' is always used here in the background
FRW sense.

The autonomous dynamical systems obtained in terms of the Hubble-normalised variables \eqref{eq:stand-variables} subject to the constraint \eqref{eq:friedmann-constraint-bg} are  formulated in~Refs. \cite{coley1}-\cite{u3}, and will form the starting basis of all our work in the present paper\footnote{Hereafter, we shall not explicitly cite these papers when we use the variables \eqref{eq:stand-variables} and the autonomous dynamical systems derived  therein. The present formulation contains important extensions and/or modifications of the dynamical systems formulations of Refs.~\cite{coley1}-\cite{u3} which will be clearly visible to the reader in the following.}. We note that  the admissible state space will be a compact semialgebraic set of the form \eqref{eq:friedmann-constraint-bg} with boundary components corresponding to vanishing potential, fluid vacuum, or flat/curved subcases. We shall be dealing with polynomial maps and so our treatment generally preserves the semialgebraic character of the constraint (cf. the Tarski-Seidenberg theorem \cite{ar83}, Sect. 37).

\subsection{Bifurcation theory and singularity theory: the missing bridge}
\label{subsec:BifnSingBridge}
All dynamical systems in this work have the autonomous form $\dot{x}=f(x,\lambda)$, with $f:\mathbb{R}^n\times\mathbb{R}^k\to\mathbb{R}^n$, where $x$ denote the \emph{state variables} and $\lambda$ the \texttt{distinguished parameters}\footnote{Here and throughout this paper, we use text typewriter to emphasise selected technical terms at their first appearance; these terms together with other standard terminology  are collected in Appendix A.} present in the \texttt{map germ} $f$ defining the bifurcation problem (sometimes we write $f_\lambda(x)$ to denote the parameter-dependent vector field). It is important to note that dynamics takes place in open sets of the form $U\times V\subset \mathbb{R}^n\times\mathbb{R}^k$, where $U, V$ denote the \emph{phase space} and \emph{base space} of the problem respectively. Whereas in hyperbolic treatments $D_xf(x,\lambda)$ has (maximal) rank $n$ (and so $\textrm{det}D_xf(x,\lambda)$ is nonzero everywhere),  we  presently do not make the assumption that locally our defining systems are topologically equivalent to their linear parts, so we  allow the spectrum to cross the imaginary axis. As we briefly mentioned in the Introduction, bifurcation and singularity theory in scalar field cosmology create three new distinct levels of `organising structure':
\begin{itemize}
\item \textbf{Organising sites and loci:}
At the most basic level, to every equilibrium \texttt{site} (non-hyperbolic point $(x_0,\lambda_0)$ in the simplest case) in the \texttt{landscape} $\mathcal{Z}$, we associate  the \emph{singular set} of the bifurcation problem (defined by) $f$
\be\label{eq:singset}
\Sigma_{f}=\{\,(x,\lambda)\in\mathcal{Z}\mid \textrm{det}D_xf(x,\lambda)=0 \},
\ee
and using the projection $\pi_f:\mathcal{Z}\to\mathbb{R}^k$, with $\pi_f(x,\lambda)=\lambda$,    the \emph{bifurcation set} (or the \emph{discriminant}) of the bifurcation problem $f$, is the subset of the base space  $\Delta_f=\pi(\Sigma_{f})$, i.e., 
\be\label{eq:bifnset}
\Delta_f=\{\,\lambda\in\mathbb{R}^k\mid \exists\, x:\,\, (x,\lambda)\in \Sigma_{f}\}.
\ee
We imagine that as the parameter $\lambda$ changes, the system  finds itself in states determined by bifurcation values of $\lambda$, i.e., those in a bifurcation set. Physically speaking, during such organising \texttt{thresholds} the dynamics changes dramatically and one expects further \texttt{episodes}, \texttt{loci} and other \texttt{organising sets} to take action. Such effects are ideally revealed and studied by recasting the original formulation of the system in terms of \texttt{centre manifold} and \texttt{normal form reductions}. Such tools are designed precisely  to study organising \texttt{degeneracies} that are closely related to singular sets and bifurcations (for broader applications of centre manifold theory to mathematical modelling problems arising from partial differential equations, including spatio-temporal pattern formation, lubrication flows, beam and shell theories, homogenisation, and spatial discretisations, see Roberts~\cite{roberts}).

\item \textbf{Organising centres:}
Once the system is recast in a normal form, the true nature of its \texttt{organising geometry} becomes  apparent. The central issue here is described by the structure of the system's \texttt{organising centre} (terminology due to R. Thom), guiding and delimiting the various organising degeneracies. For example, if we consider the system $\dot{x}=f(x,\lambda)$  in the simplest case of dimension one, the various degeneracies are classified in the so-called $A_k$-types, an equilibrium is of type $A_k$ iff the first non-vanishing derivative is of the order $(k+1^{\textrm{th}})$, so that a singular point of type $A_1$ is nondegenerate ($f'(x)=0, f''(x)\neq 0$), but various nontrivial degeneracies arise at degenerate equilibria of types $A_{\geq 2}$. 

For the systems considered in this paper, the general organising geometry contains far more interesting degeneracies  due to a variety of factors: firstly, because the translated jets require higher-order terms for finite determinacy; secondly, the presence of \texttt{organising loci} in the distinguished parameter spaces which further complicates the dynamics;  thirdly,  the existence of \texttt{gates} where hyperbolicity, or \texttt{normal hyperbolicity},  is lost or regained; and lastly, because of \texttt{organising episodes} due to collision/annihilation of parameter-dependent sites. In fact, a central issue affecting the overall dynamics of bifurcation geometry is dependence on distinguished parameters, which in the present problem are given by the pair $(\gamma,a)$. This issue is handled in two stages: first, by constructing a certain set of topological perturbations of the normal form called the \texttt{versal unfolding} (or deformation), and secondly, by enumerating the possible \texttt{bifurcation diagrams} which can occur. This is discussed next.

\item \textbf{Unfolding geometry and dynamics:} Once we have determined the \texttt{codimension}, the \texttt{unfolding space} (parameters), and constructed the versal unfolding and diagrams, the most important remaining issue is to examine their behaviour under further perturbations  (`imperfect bifurcations' in other terminology, cf. \cite{GolubitskySchaefferI}), in an effort to  determine their \texttt{persistence}. Possible sources of non-persistence in the versal dynamics are faithfully tracked by \texttt{transition varieties}, which we shall later  analyse in detail for different models of scalar field cosmology, while their complements provide the requisite \texttt{stratifications}.

\end{itemize}

Hence, local bifurcation theory provides the mechanism behind the organising loci identified in later sections:
near a non-hyperbolic equilibrium (or equilibrium manifold) one reduces the flow to a centre manifold and,
after near-identity simplification, obtains a low-dimensional normal form whose leading terms govern
persistence and transition. Singularity theory adds an efficient bookkeeping layer: families of reduced
germs are organised by their transition sets in unfolding space (fold, cusp, swallowtail, and related finite-codimension hierarchies), and
the complement decomposes into strata on which diagram types persist.

What is often missing in the cosmology literature is a systematic extraction of the relevant reduced germs
\emph{directly} from the FRW mixtures via translated jets, together with a clear identification of which
physical directions supply the unfolding coefficients after reduction. In particular, when the extra
directions are state variables (fluid and curvature fractions) rather than external knobs, the
``unfolding parameters'' must be understood in the local reduced sense: as coefficients induced in the
reduced normal form by motion in the extra state space directions. In the next subsection we provide an introduction to the problem of this missing information.


\subsection{Organising scalar field cosmology}
In the present paper, terms such as ``versal unfolding'' are used in the standard local sense
attached to the reduced germs obtained after centre-manifold and normal-form reduction.
By contrast, expressions such as ``organising centre'', ``organising locus'', and the associated
stratified geometry refer to the larger geometric rôle these reduced local models play inside
the cosmological dynamical setting. Thus we do not claim a global singularity-theoretic
classification of the full FRW systems in the strict Arnol'd~\cite{ar93} or Golubitsky--Schaeffer~\cite{GolubitskySchaefferI} sense;
rather, we use those tools locally and then assemble the resulting reduced pictures into a
broader organising framework adapted to scalar-field cosmology.

\subsubsection{Exponential versus quadratic potentials}
In the strict exponential class \(V(\phi)=V_0 e^{\lambda\phi}\) the slope parameter
\(a=\sqrt{\tfrac32}\lambda\) is constant, so the organising geometry  is naturally represented in the
external parameter plane \((\gamma,a)\) (equivalently \((\gamma,\lambda)\)).
In the harmonic class \(V(\phi)=\tfrac12 m^2\phi^2\), the standard slope variable
\(\lambda:=-V_{,\phi}/V\) evolves and becomes unbounded as \(\phi\to0\).
To obtain a compact autonomous closure we introduce the bounded slope mode
\be\label{bsm}
\zeta:=\arctan\lambda \in(-\tfrac{\pi}{2},\tfrac{\pi}{2}),
\ee
so that the massive extensions close as a four-dimensional system in \((X,Y,\Omega_k,\zeta)\)
(with \(\Omega_m\) eliminated by \eqref{eq:friedmann-constraint-bg}).

For \(V(\phi)=V_0 e^{\lambda\phi}\), the slope is constant and the system closes in low dimension. The equilibrium taxonomy
in the scalar-only, scalar--fluid and scalar--curvature mixtures is well understood: kinetic points on the compactified
boundary, scalar-dominated points, scaling points (when they exist), and curvature-dominated structures. Many works describe
existence and stability conditions for these equilibria and the associated late-time attractors as functions of \((\gamma,\lambda)\)
(or \((\gamma,a)\)); see e.g.\ \cite{coley1}-\cite{u3} and refs. therein.

However, what is typically \emph{not} made explicit is the organising geometry that governs transitions between diagram types.
In particular:
(i) the kinetic endpoint thresholds at which the linearisation degenerates are usually treated as isolated bifurcation
curves rather than as coupled multi-site episodes;
(ii) the manner in which additional full-component directions (fluid and curvature) enter local reductions as unfolding
coordinates is often stated heuristically rather than derived from translated jets; and
(iii) degeneracies involving equilibrium manifolds (e.g.\ the \(\gamma=\tfrac23\) phenomenon) require normal hyperbolicity
logic rather than isolated equilibrium bifurcation logic, and are rarely incorporated into a unified persistence picture.

For non-exponential potentials the slope variable \(\lambda(\phi)=-V_{,\phi}/V\) evolves, and one must either:
(a) enlarge the state space by adding a slope variable (and sometimes auxiliary variables), or
(b) use alternative compactifications (e.g.\ compactifying a Hubble/mass ratio) to control the flow near problematic regions.
For the quadratic potential \(V(\phi)=\tfrac12 m^2\phi^2\), the slope is unbounded as \(\phi\to 0\) and the naive \(\lambda\)
closure is singular. Global regular formulations exist (notably via compactification in other directions) and provide a
useful qualitative picture of massive dynamics (\cite{coley1}-\cite{u3} and refs. therein).

From the viewpoint of persistence and organising geometry, two obstacles remain.
First, it is not straightforward to place the massive class on the same footing as the strict exponential class, because
the slope is no longer a parameter: transitions in the massive flow correspond to \emph{internal} drift through slope regimes.
Second, once full components (fluid and curvature) are included, the massive equilibrium structure is no longer organised only by
isolated sites: robust equilibrium continua and normal hyperbolicity thresholds become central, and these have not been
integrated into a single unfolding/stratification narrative.

\subsubsection{Unfoldings, transition sets, and strata}

The systems considered in this paper also possess robust invariant hypersurfaces such as
\(\Omega_k=0\) (flat models) and \(\Omega_m=0\) (fluid vacuum), together with equilibrium sites
and equilibrium continua (lines/segments) that persist under fluid/curvature fraction additions.
When equilibria are isolated, organising loci occur where a transverse eigenvalue crosses zero,
enlarging the centre subspace and triggering a change in the local phase portrait.
When equilibria form a manifold (e.g.\ a line or segment), the relevant organising mechanism is
loss (and recovery) of \emph{normal hyperbolicity}: the tangential direction is neutral for geometric
reasons, and qualitative change occurs when a \emph{normal} eigenvalue passes through zero, cf. \cite{wig88}. 

Near a chosen organising site, we translate to a local coordinate \(\mathbf u\) and write the system as
\be
\dot{\mathbf u}=J\,\mathbf u + Q(\mathbf u)+O(\|\mathbf u\|^3),
\ee
where $J$ is the linear Jordan part and $Q$ is a quadratic homogeneous polynomial. We then perform a centre manifold reduction, and simplify by near-identity changes to obtain a low-dimensional normal form. In this local normal form setting, an \emph{unfolding} is a family of reduced vector fields depending
on parameters \(\mu\); the associated \emph{transition set} is the subset of \(\mu\)-space where the
reduced diagram is not structurally stable (multiple equilibria, changes in multiplicity, etc.).
The complement of the \texttt{transition set} decomposes into connected components (the \emph{strata}), on which
the reduced dynamics is persistent.

In the strict exponential class, the unfolding parameters include genuine external quantities (notably
the slope `offset' \(a-3\)) together with contributions induced by excursions in the additional state-space
directions (fluid and curvature fractions) after reduction.
In the massive class, the same organising stratification is encountered dynamically: the \texttt{frozen-slope projection}
\((\gamma,a_{\rm eff}(N))\) traces a curve in \((\gamma,a)\), and the orbit experiences regime changes when
this curve crosses the pull-back of an organising locus.

\section{Main results of this paper}
\label{sec:mainresults}

This paper develops a unified dynamical systems and singularity theory description of FRW cosmologies with a scalar field,
a barotropic fluid, and spatial curvature. The central output is an explicit \emph{organising geometry}:
a small collection of parameter loci and invariant state loci where hyperbolicity (or normal hyperbolicity) is lost,
together with the reduced normal forms that control the corresponding transitions. These local results are then assembled
into an explicit stratification of parameter space for the strict exponential mixtures and a canonical \texttt{pull-back
stratification} for the harmonic extensions, so that qualitative regimes and regime changes can be read off from the same
charts. In this way, the dynamics of all possible perturbations to nearby systems is fully described.


More concretely, this paper supplies the missing bridge by:
(i) computing translated jet systems and centre reductions at the organising sites in the strict exponential mixtures and
identifying the small transition set in the \((\gamma,a)\) plane;
(ii) introducing a compact bounded-slope massive closure \(\zeta=\arctan\lambda\) that yields a regular autonomous formulation
and allows for a detailed study of invariant gates, robust equilibrium continua, and vertical normal hyperbolicity thresholds; and
(iii) assembling both classes into a single persistence/stratification narrative in which the exponential stratification
serves as a reference stratification and the massive extensions are interpreted as dynamical passages through its pull back.

\paragraph{Meaning of the pull-back relation.}
The pull-back language is meant in a literal geometric sense and does \emph{not} claim any global equivalence between the
exponential and massive systems.

We now work in the context of singular sets and discriminants as defined in \eqref{eq:singset} and \eqref{eq:bifnset}, and consider the reduced phase space
\(U=\{(X,Y,\Omega_k)\}\). We write \(U_{\textrm{exp}}\) and \(U_{\textrm{har}}\) for the corresponding exponential and harmonic phase spaces, respectively. The exponential base space is
\(V_{\textrm{exp}}=\{(\gamma,a)\}\), with \(a=\sqrt{3/2}\,\lambda\), while the frozen-slope base space for the harmonic closure is
\[
V_{\textrm{frozen}}
=
\{(\gamma,\zeta)\mid \zeta'=0,\ \zeta\in(-\pi/2,\pi/2)\}.
\]
Define the \texttt{frozen-slope map}
\be\label{fro}
\pi: U_{\textrm{har}}\times V_{\textrm{frozen}}\longrightarrow V_{\textrm{exp}},
\qquad
(X,Y,\Omega_k,\zeta;\gamma)\longmapsto
(\gamma,a_{\rm eff}(\zeta)).
\ee
Here \(a_{\rm eff}\) is the instantaneous effective slope induced by \(\zeta\); explicitly,
\(a_{\rm eff}(\zeta)=\sqrt{3/2}\tan\zeta\), in agreement with \eqref{eq:aeff-zeta}. Thus a frozen value of the harmonic slope variable determines an instantaneous exponential parameter value. The organising loci
\(\Gamma_i\subset V_{\textrm{exp}}\), \(i=1,\dots,5\), listed in Theorem~\ref{thm:Ri-summary}, item~\textbf{(R1)}, then lift through \eqref{fro} to hypersurfaces \(\pi^{-1}(\Gamma_i)\) in the harmonic state space. Crossing such a lifted hypersurface along a harmonic orbit is precisely the mechanism by which the \emph{instantaneous} local phase portrait in the \((X,Y,\Omega_k)\) harmonic directions changes in the same way as in the parameter-driven exponential setting.

\begin{table}[t]
\centering
\caption{Organising loci, leading reduced dynamics, and the associated robust effects.
Equation references point to the detailed derivations in Sections~\ref{sec:exp-Z2-gate}--\ref{sec:primer}.}
\label{tab:results-dashboard}
\begin{tabular}{p{0.18\textwidth} p{0.25\textwidth} p{0.33\textwidth} p{0.17\textwidth}}
\hline
\textbf{Locus} & \textbf{Site / mechanism} & \textbf{Leading reduced form (schematic)} & \textbf{Where proved} \\
\hline
$|a|=3$ & kinetic endpoints $K_\pm$; hilltop-type degeneracy &
$\dot U=(3-a)U - \tfrac{3}{2}\,U^3+\cdots$
(critical slowing; directional exits) &
\eqref{eq:red-Kp-sfc-unfolded}, \eqref{eq:red-Km-sfc-unfolded}; \S\ref{sec:unfoldings} \\[0.8ex]

$a^2=3$ & curvature direction at $S$ becomes centre &
$\dot W = -2W^2+\cdots$ (algebraic curvature leakage) &
\eqref{eq:red-S-a2eq3-sfc} \\[0.8ex]

$a^2=\tfrac92\gamma$ & $S$--$F$ exchange (transcritical) &
transcritical exchange of sink/saddle between $S$ and $F$ &
\eqref{eq:eigs-S-sfc}, \eqref{eq:trdet-F2-sfc} \\[0.8ex]

$\gamma=\tfrac23$ & equilibrium manifold; loss/recovery of normal hyperbolicity &
normal eigenvalue crosses $0$ along $\mathcal L_{MC}$ &
\eqref{eq:line-MC-g23} and related reductions \\[0.8ex]

\(\gamma=2\)
&
stiff-fluid boundary; forced neutral coincidences at several sites
&
higher-codimension centre dynamics; boundary degeneracy
&
\(\S \ref{subsubsec:special-gamma-jets}\)--\(\S \ref{subsubsec:nf-sfc}\)
\\[0.8ex]

massive closure & $\zeta=\arctan\lambda$; equilibrium continua and gates &
pull-back stratification via $(\gamma,a_{\rm eff}(\zeta))$; gate hypersurfaces &
\S\ref{sec:massive-extensions}, \S\ref{sec:strata} \\
\hline
\end{tabular}
\end{table}

For quick orientation, Theorem~\ref{thm:Ri-summary} lists the main results in a form that can be read independently of the proofs,
and Table~\ref{tab:results-dashboard} gives a compact dashboard with pointers to the detailed derivations in Sections~\ref{sec:exp-Z2-gate}--\ref{sec:primer}.
\begin{theorem}[Summary of main results]\label{thm:Ri-summary}
Consider FRW cosmologies with a minimally coupled scalar field with potential \(V(\phi)\), optionally a one-fluid component
(with barotropic index \(\gamma\in(0,2]\)), and spatial curvature, formulated in the Hubble-normalised variables
\((X,Y,\Omega_m,\Omega_k)\) with e-fold time \(N=\ln \texttt{a}\).
Then the following hold (items R1-3 correspond to the exponential class, R4-6 to the harmonic class).

\begin{enumerate}
\item[\textbf{(R1)}] \textbf{Structure of equilibria: Strict exponential organising set.}
For \(V(\phi)=V_0 e^{\lambda\phi}\) with \(a=\sqrt{\tfrac32}\lambda\), the local phase portraits of the scalar--fluid--curvature
(SFC) mixtures are organised in the \((\gamma,a)\) plane by the loci \(\Gamma_i\ i=1,\dots , 5\),
\be\label{loci}
|a|=3,\qquad a^2=3,\qquad a^2=\tfrac92\gamma,\qquad \gamma=\tfrac23,\qquad \gamma=2,
\ee
corresponding respectively to kinetic endpoint (hilltop) degeneracy, curvature-induced loss of hyperbolicity at the scalar site,
scalar--fluid stability exchange, loss of isolation along the \(M\)--\(C\) equilibrium manifold, and stiff-fluid coincidences. (cf. Sections \ref{sec:exp-Z2-gate}, \ref{sec:fluid-org}, \ref{subsec:scalar-curv} for loci \(\Gamma_i\ i=1,2\), and Section \ref{subsec:scalar-fluid-curv} for loci \(\Gamma_i\ i=3, 4, 5\).)

\item[\textbf{(R2)}] \textbf{Centre reduction/normal forms: Coupled hilltop episode and explicit reduced coefficients.}
The kinetic threshold \(|a|=3\) constitutes a coupled multi-site organising episode: the kinetic endpoints exhibit a hilltop-type
linear degeneracy while the scalar equilibrium simultaneously reaches its existence boundary on the constraint set.
Translated jets and explicit centre reductions determine the leading reduced normal forms and identify the reduced coefficients
that act as unfolding coordinates once extra state-space directions (fluid and curvature fractions) are included. (cf. Sections \ref{sec:exp-Z2-gate}, \ref{sec:fluid-org}, \ref{subsec:scalar-curv} for loci \(\Gamma_i\ i=1,2\), and Section \ref{subsec:scalar-fluid-curv} for loci \(\Gamma_i\ i=3, 4, 5\).)

\item[\textbf{(R3)}] \textbf{Versal deformation: Transition varieties and stratification (exponential case).}
The organising loci induce a stratification of the exponential parameter plane into regions on which the local diagram types
persist. Near \(|a|=3\), the macroscopic \((\gamma,a)\) picture admits a refinement via codimension-three transition-variety charts
in the reduced unfolding coordinates. (Sections \ref{sec:unfoldings}, \ref{subsec:strata-transition}, \ref{subsec:strata-exponential}.)

\item[\textbf{(R4)}] \textbf{Bounded slope formulation: Bounded-slope closure for the harmonic class.}
For \(V(\phi)=\tfrac12 m^2\phi^2\), introducing the bounded slope mode \(\zeta=\arctan\lambda\in(-\tfrac\pi2,\tfrac\pi2)\) yields a
regular autonomous closure (without time-rescaling) for the massive extensions, making the slope direction global and revealing
robust invariant hypersurfaces associated with degeneracies of the slope evolution (\(\zeta'=0\)). (Section \ref{subsec:massive-system}.)

\item[\textbf{(R5)}] \textbf{Structure of equilibria: Equilibrium continua and vertical normal hyperbolicity thresholds.}
In the massive mixtures, finite-slope equilibria organise into robust equilibrium continua (including the \(M\) and \(C\) lines and
a connecting segment at \(\gamma=\tfrac23\)), and the dominant organising mechanisms are vertical \(\gamma\)-thresholds controlling
loss/recovery of normal hyperbolicity transverse to these continua. \ref{subsec:massive-gates-eq}.)

\item[\textbf{(R6)}] \textbf{Normal forms/versal effects: Pull-back stratification and dynamical regime changes.}
A frozen-slope projection induces a canonical pull-back of the exponential stratification to the massive state space.
Consequently, a single massive orbit can undergo multiple qualitative regime changes without external parameter variation, when the
drifting effective slope crosses pulled-back organising loci. The associated universal effects are read off from the reduced normal
forms and stratified charts. (Sections \ref{subsec:massive-jets-nf}, \ref{subsec:massive-subcases-compare},  \ref{sec:unfoldings}, \ref{subsec:strata-transition}, \ref{subsec:strata-massive}.)
\end{enumerate}
\end{theorem}

\subsection{Organising set for the strict exponential mixtures}
\label{subsec:mainresults-exponential}

For the strict exponential potential \(V(\phi)=V_0 e^{\lambda\phi}\) we use \(a=\sqrt{\tfrac32}\lambda\) and treat
\((\gamma,a)\) (equivalently \((\gamma,\lambda)\)) as external parameters.  Sections~\ref{sec:exp-Z2-gate}--\ref{sec:curvature} compute the principal equilibria,
their translated quadratic jets, and the relevant centre reductions in the scalar-only, scalar--fluid, scalar--curvature,
and full SFC classes.  The resulting local dynamics is organised by the following five loci (see Fig.~\ref{fig:bifn-map-gamma-lambda}):

\begin{itemize}
\item \textbf{Kinetic endpoint threshold (hilltop component):} \(|a|=3\) (equivalently \(|\lambda|=\sqrt6\)).
At \(K_\pm\) the linearisation becomes fully (or partially) degenerate and the reduced kinetic dynamics acquires a
centre direction; the leading reduced behaviour is cubic, producing critical slowing down and long kinetic episodes.
This threshold participates in a coupled multi-site organising episode because it coincides with the existence boundary
of other equilibria.

\item \textbf{Curvature-induced scalar threshold:} \(a^2=3\) (equivalently \(\lambda^2=2\)).
At the scalar site \(S\) the curvature direction becomes centre; the curvature variable admits a fold-type reduction with
quadratic leading term, yielding universal algebraic curvature decay/growth near the threshold.

\item \textbf{Scalar--fluid exchange curve:} \(a^2=\tfrac92\gamma\) (equivalently \(\lambda^2=3\gamma\)).
The scalar site \(S\) and the scaling site \(F\) exchange stability across this curve (transcritical mechanism), producing
the robust ``tracker versus scalar-domination'' dichotomy.

\item \textbf{Fluid--curvature degeneracy axis:} \(\gamma=\tfrac23\).
The equilibria at \((X,Y)=(0,0)\) cease to be isolated; the fluid and curvature equilibria lie on an equilibrium manifold,
and the relevant organising notion is loss/recovery of \emph{normal hyperbolicity} rather than an isolated equilibrium
bifurcation.

\item \textbf{Stiff-fluid axis:} \(\gamma=2\).
Additional neutral directions occur at multiple sites (including kinetic and fluid/curvature sectors), producing forced
higher-codimension centre dynamics; we record this boundary case for completeness.
\end{itemize}
A further organising contribution is developed in Section~\ref{sec:unfoldings}: the kinetic threshold \(|a|=3\) is treated
as a coupled multi-site episode rather than an isolated bifurcation curve. Using translated jets and explicit centre
reductions (including worked reductions in the full SFC jet), we identify the reduced coefficients that play the role of
unfolding coordinates in the local normal forms. This supplies an explicit codimension-three transition variety skeleton
used later to refine the macroscopic \((\gamma,a)\) picture near the kinetic gate.

\begin{figure}[t]
\centering
\begin{tikzpicture}[scale=2.2]
  \draw[->] (0,0) -- (2.2,0) node[below] {$\gamma$};
  \draw[->] (0,-2.6) -- (0,2.6) node[left] {$\lambda$};

  \draw[densely dashed] (2/3, -2.4) -- (2/3, 2.4) node[above right] {$\gamma=\tfrac23$};
  \draw[densely dashed] (2, -2.4) -- (2, 2.4) node[above left] {$\gamma=2$};

  \draw[densely dashed] (0, {sqrt(2)}) -- (2.1, {sqrt(2)}) node[right] {$\lambda^2=2$};
  \draw[densely dashed] (0, {-sqrt(2)}) -- (2.1, {-sqrt(2)});
  \draw[densely dashed] (0, {sqrt(6)}) -- (2.1, {sqrt(6)}) node[right] {$|\lambda|=\sqrt6$};
  \draw[densely dashed] (0, {-sqrt(6)}) -- (2.1, {-sqrt(6)});

  \draw[thick] plot[domain=0:2,samples=100] (\x, {sqrt(3*\x)});
  \draw[thick] plot[domain=0:2,samples=100] (\x, {-sqrt(3*\x)});
  \node at (1.55,1.8) {$\lambda^2=3\gamma$};
\end{tikzpicture}
\caption{Principal organising loci in the \((\gamma,\lambda)\) plane for the strict exponential SFC class.}
\label{fig:bifn-map-gamma-lambda}
\end{figure}

As an example, Figure~\ref{fig:quartic-unfolding-schematic} provides a schematic bifurcation/phase-line atlas for a quartic-leading reduced germ appearing in the scalar–fluid class, illustrating how strata correspond to persistent phase-line types.

\begin{figure}[t]
\centering
\begin{tikzpicture}[x=1.2cm,y=1.2cm,>=stealth]

\begin{scope}
\draw[->] (-3.2,0) -- (1.4,0) node[below] {$\mu_1$};
\draw[->] (0,-1.6) -- (0,2.4) node[left] {$\mu_3$};

\draw[thick] plot[domain=-3:0,samples=120] (\x,{(\x*\x)/4}); 
\node[above right] at (-2.0,1.2) {$\mu_3=\mu_1^2/4$};

\draw[thick] (-3.2,0) -- (1.2,0); 

\node at (-2.1,1.9) {\small $0$ eq};
\node at (-2.1,-0.5) {\small $2$ eq};
\node at (-2,0.6) {\small $4$ eq};

\fill (0,0) circle (1.2pt);
\node[above right] at (0,0) {\small $(0,0)$};

\node[below] at (-1.0,-1.45) {\small (a) $\mathbb{Z}_2$ quartic unfolding: regions by \# equilibria};
\end{scope}

\begin{scope}[shift={(6.2,1.4)}]

\def\phaseline#1#2{
  \draw (-1.8,#1) -- (1.8,#1);
  \node[below] at (0,#1) {\scriptsize $#2$};
}

\phaseline{0.6}{\mu_3>\mu_1^2/4}
\draw[->] (-1.6,0.6) -- (-0.6,0.6);
\draw[->] (-0.4,0.6) -- (0.6,0.6);
\draw[->] (0.8,0.6) -- (1.6,0.6);

\phaseline{-0.2}{\mu_3<0}
\fill (-1.0,-0.2) circle (1.2pt);
\fill (1.0,-0.2) circle (1.2pt);
\node[above] at (-1.0,-0.2) {\scriptsize $-u_0$};
\node[above] at (1.0,-0.2) {\scriptsize $u_0$};
\draw[->] (-1.6,-0.2) -- (-1.15,-0.2);
\draw[->] (-0.85,-0.2) -- (0.85,-0.2);  
\draw[->] (0.85,-0.2) -- (-0.85,-0.2);
\draw[->] (1.15,-0.2) -- (1.6,-0.2);

\phaseline{-1.0}{\mu_1<0,\ 0<\mu_3<\mu_1^2/4}
\fill (-1.3,-1.0) circle (1.2pt);
\fill (-0.6,-1.0) circle (1.2pt);
\fill (0.6,-1.0) circle (1.2pt);
\fill (1.3,-1.0) circle (1.2pt);
\node[above] at (-1.3,-1.0) {\scriptsize $-u_2$};
\node[above] at (-0.6,-1.0) {\scriptsize $-u_1$};
\node[above] at (0.6,-1.0) {\scriptsize $u_1$};
\node[above] at (1.3,-1.0) {\scriptsize $u_2$};

\draw[->] (-1.6,-1.0) -- (-1.35,-1.0);
\draw[->] (-1.15,-1.0) -- (-1.25,-1.0);
\draw[->] (-0.45,-1.0) -- (0.45,-1.0);
\draw[->] (0.75,-1.0) -- (0.65,-1.0);
\draw[->] (1.35,-1.0) -- (1.6,-1.0);

\node[below] at (0,-2.8) {\small (b) Phase line sketches for $\dot u=u^4+\mu_1u^2+\mu_3$};
\end{scope}

\end{tikzpicture}
\caption{Schematic organisation of the $\mathbb{Z}_2$-equivariant quartic unfolding
$\dot u=u^4+\mu_1u^2+\mu_3$ by organising curves in $(\mu_1,\mu_3)$ and the corresponding one-dimensional phase-line types.
(Here “eq” denotes equilibria of the reduced normal form.)}
\label{fig:quartic-unfolding-schematic}
\end{figure}

\subsection{Harmonic extensions: bounded-slope closure and equilibrium continua}
\label{subsec:mainresults-massive}

For the quadratic potential \(V(\phi)=\tfrac12 m^2\phi^2\), the effective slope \(\lambda(\phi)=-V_{,\phi}/V\) is not a
parameter but a state-dependent quantity.  A key technical contribution of this paper (cf. Section \ref{sec:massive-extensions}) is a compact, bounded-slope closure:
we introduce \(\zeta=\arctan\lambda\in(-\tfrac\pi2,\tfrac\pi2)\), which yields a closed autonomous \(4\)D system for
\((X,Y,\Omega_k,\zeta)\) with regular slope evolution and without time-rescaling.
This closure reveals organising mechanisms that differ qualitatively from the strict exponential case:

\begin{itemize}
\item The finite-slope equilibria form robust equilibrium continua (lines \(M\) and \(C\), and a connecting equilibrium segment
on the special axis \(\gamma=\tfrac23\)), rather than isolated parameter-controlled sites.

\item The dominant degeneracies are \emph{vertical} \(\gamma\)-thresholds controlling loss/recovery of normal hyperbolicity
transverse to these continua (notably \(\gamma\in\{0,\tfrac23,2\}\) at the fluid line \(M\), and \(\gamma=\tfrac23\) at the
curvature line \(C\)).

\item The slope dynamics induces invariant \emph{gates} (e.g.\ \(Y=0\) and \(\zeta=0\)) that control when the internal drift in
the effective slope stalls or reverses direction, providing a natural mechanism for multi-stage histories in massive models.
\end{itemize}

This also clarifies the relation of the present harmonic results to those of Ref.~\cite{MI-I}.
The bounded-slope quadratic SF--fluid--curvature system contains the lower-dimensional
quadratic subcases as invariant slices (for example \(\Omega_k=0\), \(\Omega_m=0\), and
\(\Omega_k=\Omega_m=0\)), so the oscillatory sector identified in \cite{MI-I}, including its
periodic-orbit / invariant-torus mechanism, is recovered on the corresponding invariant
quadratic sectors. What is new here is that these inherited quadratic dynamics are placed
inside a larger organising geometry with fluid, curvature, invariant gates, and robust
equilibrium continua. Thus the present paper extends the Ref.~\cite{MI-I} picture rather than replacing
it: the local oscillatory mechanism survives on the quadratic slices, while the full bounded-slope
system explains how such behaviour fits into a broader pull-back stratification of massive
scalar-field cosmology.

\subsection{From organising loci to strata and versal physical effects}
\label{subsec:mainresults-strata-primer}

Section~\ref{sec:strata} converts the organising loci into an explicit stratification of the strict exponential
\((\gamma,a)\) plane: the complement of the organising set decomposes into connected components (strata) on which the local
diagram types persist. For the massive extensions, the same organising picture is encountered dynamically: the frozen-slope
projection \((X,Y,\Omega_k,\zeta)\mapsto(\gamma,a_{\rm eff}(\zeta))\) induces a canonical pull-back stratification of the
massive state space. A single massive orbit can therefore cross pull-back stratum boundaries as \(a_{\rm eff}(N)\) drifts,
triggering the same qualitative regime changes identified in the parameter-driven setting.

Section~\ref{sec:primer} then translates the resulting normal forms and strata into a short dictionary of robust dynamical
effects (critical slowing near the kinetic threshold, algebraic curvature leakage near \(a^2=3\), tracker-versus-domination
exchange across \(a^2=\tfrac92\gamma\), and normal-hyperbolicity transitions at \(\gamma=\tfrac23\)), together with a
workflow for reading dominant regimes directly from the stratified charts.

\begin{remark}[How to read the bifurcation geometry.]
Unfolding-parameter space is organised by a \emph{transition set} (where local diagram types change).
Its complement decomposes into \emph{strata} on which the qualitative phase portrait persists.
Near an organising locus we replace the full system by an appropriate reduced normal form (centre reduction),
while for massive extensions we use the frozen-slope projection to interpret orbits as dynamical passages through the
pull-back stratification of the exponential organising set.
Key terminology and correspondences are collected in App.~\ref{app:glossary}.
\end{remark}

We also show in Section~\ref{sec:primer} how the same stratified picture organises the physically familiar
regimes of slow roll, ultra slow roll, and oscillatory behaviour: in the strict exponential class,
SR and USR appear as persistent attracting balances and nonhyperbolic bottleneck passages,
respectively, while on the massive side the quadratic invariant slices recover the oscillatory
sector identified in \cite{MI-I}, including its periodic-orbit / invariant-torus mechanism.

Physically, three loci may be kept in mind from the start. The line $\Gamma_1:|a|=3$
(equivalently $\lambda^2=6$) is the kinetic-endpoint threshold: it is the point at which the
scalar equilibrium reaches the kinetic boundary of the compactified state space, and hence
organises stiff, kinetic-like episodes and the associated critical slowing near $K_\pm$. The
line $\Gamma_2:a^2=3$ (equivalently $\lambda^2=2$) is the scalar acceleration threshold:
for the scalar-dominated exponential branch it is precisely the boundary between accelerated
and decelerated scalar expansion, and in the curvature extension it is simultaneously the
point where the curvature direction becomes marginal. By contrast, $\Gamma_3:a^2=\tfrac92\gamma$
is the scalar--fluid exchange/tracker threshold. The later sections make these readings
precise through the corresponding reduced jets and normal forms.

\paragraph{Technical development.}
Sections~\ref{sec:exp-Z2-gate}--\ref{sec:primer} provide the detailed equilibrium analyses, translated jets, centre reductions and normal form
derivations that support the summary statements in Sections~\ref{sec:intro}--\ref{sec:mainresults}.
Readers may consult these sections selectively: each subsection begins by stating which item(s) in
Theorem~\ref{thm:Ri-summary} and Table~\ref{tab:results-dashboard} it establishes.

\section{Exponential potential: equilibria and the hilltop multi-site gate}
\label{sec:exp-Z2-gate}

The main results of this section are summarised in Theorem~\ref{thm:sec4}.

\begin{theorem}[Organising locus and multi-site hilltop gate: scalar-only exponential class]
\label{thm:sec4}
Consider the scalar-only exponential system \eqref{eq:f1-basic}--\eqref{eq:f2-basic}
on the constraint $X^2+Y^2=1$, with slope parameter $a=\sqrt{3/2}\,\lambda$.
Then the organising structure is:

\begin{itemize}
\item \textbf{Equilibrium events and episode: multi-site gate ($2$D / codim-$3$ within the $\mathbb Z_2$ amplitude class):}
at $|a|=3$ (equivalently $|\lambda|=\sqrt6$) the scalar-dominated equilibrium $S$
collides with the kinetic boundary and coincides with $K_\pm$ (steady-state to steady-state
boundary collision), cf.\ Remark \ref{remark4-2}.

\item \textbf{Centre \& normal form reductions: Hilltop amplitude germ (planar, $\mathbb Z_2$-equivariant):}
at the same parameter value $|a|=3$, the translated systems at $K_\pm$ and at $S$
share a common quadratic-order jet and simultaneously have vanishing linear part.
The common organising centre is the triangular $\mathbb Z_2$-equivariant amplitude
(hilltop) germ \eqref{OS:single-sf}, realised at multiple equilibrium sites.

\item \textbf{Cubic-order control / versal deformation:}
the missing $U^2$ term in the endpoint jet forces cubic-order retention and the
resulting nondegenerate versal deformation has codimension $3$ (with unfolding
directions supplied by additional physical components in later sections).
\end{itemize}
\end{theorem}
The additional fluid and curvature components introduced in Sects.~\ref{sec:fluid-org}--\ref{sec:curvature} supply the physically distinguished \emph{versal unfolding} directions of this gate.


\subsection*{Structure of this section}
We begin with the Einstein--scalar equations for a flat FRW universe with an
exponential potential and introduce a preliminary dimensionless scaling
$(x,y,z,\tau)$.  In this representation the origin is a highly degenerate
equilibrium of the \emph{uncompactified scalar-only} dynamical system (the
linearisation vanishes), so the leading qualitative information is carried by the
lowest nonzero jet.  We therefore interpret the right--hand side $g=(g_1,g_2,g_3)$
as a \emph{bifurcation problem} in the sense of Golubitsky--Schaeffer
and use the resulting codimension estimate to motivate a systematic reduction of
the effective unfolding space.
This leads to compactified expansion--normalised variables $(X,Y)$ and a
two--dimensional autonomous system in $N=\ln a$, for which a distinguished
normalisation yields the simplest possible scalar--only ``basic system''.
The equilibria and their physical interpretation are standard, but our emphasis is
on the variable-choice principle that reduces the codimension of the underlying
bifurcation germ to a range compatible with mode interaction analysis.
We end by identifying a ``synchronised threshold'' at $|\lambda|=\sqrt6$ at which the
scalar-dominated site meets the kinetic boundary and the kinetic endpoints acquire a
hilltop-type degeneracy, thereby motivating the multi-site gate viewpoint developed
further once fluid and curvature are promoted to full components.

\subsection{Einstein--scalar equations and exponential potential}
\label{subsec:einstein-scalar-exp}

We consider a flat FRW universe sourced by a canonical scalar field $\phi$
with energy density and pressure
\begin{equation}
\rho_\phi=\frac12\dot\phi^{2}+V(\phi),\qquad
p_\phi=\frac12\dot\phi^{2}-V(\phi),
\label{eq:rho-p}
\end{equation}
and exponential potential
\begin{equation}
V(\phi)=V_{0}e^{\lambda\phi},
\label{eq:expV}
\end{equation}
where $V_{0}>0$ and $\lambda$ is constant.
The Friedmann constraint reads
\begin{equation}
3H^{2}=\frac12\dot\phi^{2}+V(\phi),
\label{eq:friedmann}
\end{equation}
while the Klein--Gordon equation and the acceleration equation are
\begin{equation}
\ddot\phi+3H\dot\phi+V_{,\phi}=0,
\label{eq:KG}
\end{equation}
\begin{equation}
\dot H+H^{2}=\frac{\ddot \sfac}{\sfac}
=-\frac16(\rho+3p)=\frac13\Bigl(V(\phi)-\dot\phi^{2}\Bigr).
\label{eq:accel}
\end{equation}
For \eqref{eq:expV} we have $V_{,\phi}=\lambda V$.

\subsection{A preliminary scaling and the codimension issue}
\label{subsec:prelim-scaling}

We introduce the dimensionless variables
\begin{equation}
x=\alpha\sqrt{V(\phi)},\qquad
y=\beta\dot\phi,\qquad
z=\gamma H,\qquad
\tau=\delta t,
\label{eq:prelimvars}
\end{equation}
with coefficients to be determined.\footnote{The scaling constant $\gamma$ in
\eqref{eq:prelimvars} is unrelated to the barotropic index $\gamma$ introduced later.}
Denoting differentiation with respect to
$\tau$ by a prime, Eqs.~\eqref{eq:KG}--\eqref{eq:accel} yield the 3D system
\begin{equation}
x'=\frac{\lambda}{2\beta\delta}\,xy,
\label{eq:xprime-prelim}
\end{equation}
\begin{equation}
y'=-\frac{\lambda\beta}{\alpha^{2}\delta}x^{2}
-\frac{3}{\gamma\delta}\,yz,
\label{eq:yprime-prelim}
\end{equation}
\begin{equation}
z'=\frac{\gamma}{3\alpha^{2}\delta}x^{2}
-\frac{\gamma}{3\beta^{2}\delta}y^{2}
-\frac{1}{\gamma\delta}\,z^{2}.
\label{eq:zprime-prelim}
\end{equation}
Setting $g=(g_{1},g_{2},g_{3})$ for the right--hand side of
\eqref{eq:xprime-prelim}--\eqref{eq:zprime-prelim}, we regard $g$ as a \emph{bifurcation problem} in the sense of Golubitsky--Schaeffer~\cite{GolubitskySchaefferI}.
The equilibrium at the origin\footnote{Concretely, one
fixes a reference slope $\lambda_*$ and writes $\ell=\lambda-\lambda_*$.  The germ is
taken at $(x,y,z,\ell)=(0,0,0,0)$.}
\begin{equation}
O(x,y,z,\ell)=(0,0,0,0)
\label{eq:O-prelim}
\end{equation}
satisfies
\begin{equation}
g|_{O}=0,\qquad \det Dg|_{O}=0.
\label{eq:g-deg}
\end{equation}

\paragraph{The bifurcation-problem viewpoint.}
We classify nearby realisations up to \emph{strong equivalence} (smooth changes of
state variables, together with admissible reparametrisations of the unfolding
parameters).  The codimension $\operatorname{codim} g$ measures the minimal number of
independent perturbation directions required to obtain a versal deformation of the
germ.  For this class of degenerate equilibria one has
\begin{equation}
\operatorname{codim} g \ge 8,
\label{eq:codim8}
\end{equation}
see Ref.~\cite{GolubitskySchaefferI}.
Estimate \eqref{eq:codim8} provides a quantitative obstruction to using
\eqref{eq:prelimvars} as a starting point for mode interactions: the resulting
unfolding space is far too large to be physically interpretable.  Our objective is
therefore to seek a representation with significantly lower effective codimension
-- ideally at most three -- in harmony with the variable--choice strategy adopted
in \cite{MI-I}.

\subsection{Compactified variables and evolution in \texorpdfstring{$\ln a$}{\ln \texttt{a}}}
\label{subsec:compactified}

We introduce compactified variables
\begin{equation}
X=\frac{x}{z},\qquad
Y=\frac{y}{z},
\label{eq:XYdef}
\end{equation}
and from now on denote differentiation with respect to $N=\ln \sfac$
by a dot, so that for any quantity $u$,
\begin{equation}
\dot u=\frac{u'}{z}.
\label{eq:dotrule}
\end{equation}
Using \eqref{eq:XYdef} and \eqref{eq:dotrule} together with
\eqref{eq:xprime-prelim}--\eqref{eq:zprime-prelim}, we obtain the reduced planar
system
\begin{equation}
\dot X=
\frac{\lambda}{2\beta\delta}XY
-
X\left(
\frac{\gamma}{3\alpha^{2}\delta}X^{2}
-
\frac{\gamma}{3\beta^{2}\delta}Y^{2}
-
\frac{1}{\gamma\delta}
\right),
\label{eq:Xdots-general}
\end{equation}
\begin{equation}
\dot Y=
-\frac{\lambda\beta}{\alpha^{2}\delta}X^{2}
-
\frac{3}{\gamma\delta}Y
-
Y\left(
\frac{\gamma}{3\alpha^{2}\delta}X^{2}
-
\frac{\gamma}{3\beta^{2}\delta}Y^{2}
-
\frac{1}{\gamma\delta}
\right).
\label{eq:Ydots-general}
\end{equation}
Equations \eqref{eq:Xdots-general}--\eqref{eq:Ydots-general} constitute the
2D system associated with the compactified variables.

\subsection{Normalisation and the scalar-only basic system}
\label{subsec:basic-system}

We now choose the coefficients in \eqref{eq:prelimvars} so that the compactified
variables acquire direct physical meaning as expansion--normalised energy fractions.
Specifically, we require
\begin{equation}
X=\frac{\sqrt{V}}{\sqrt{3}\,H},
\qquad
Y=\frac{\dot\phi}{\sqrt{6}\,H},
\label{eq:XYphysical}
\end{equation}
which is achieved by the normalisation
\begin{equation}
\gamma=1,\quad \delta=1,\quad
\alpha=\frac{1}{\sqrt{3}},\quad
\beta=\frac{1}{\sqrt{6}}.
\label{eq:coeff-choice}
\end{equation}
This choice is not an ad hoc guess: it converts
\eqref{eq:Xdots-general}--\eqref{eq:Ydots-general} into the \emph{simplest} autonomous form
while simultaneously lowering the effective codimension of the bifurcation problem
associated with the degenerate origin in the uncompactified system.

Using the Friedmann constraint \eqref{eq:friedmann}, the preliminary variables
\eqref{eq:prelimvars} (for $k=0$) are restricted to the null cone
\begin{equation}
x^{2}+y^{2}-z^{2}=0,
\label{eq:cone-flat}
\end{equation}
so that the flat models evolve on $z^{2}=x^{2}+y^{2}$.
With \eqref{eq:XYphysical}, the Friedmann constraint becomes
\begin{equation}
X^{2}+Y^{2}=1.
\label{eq:constraint-XY}
\end{equation}
Since $V\ge 0$ and (for expanding solutions) $H>0$, one has $X\ge 0$, so the physical
state space is the semicircle $X^{2}+Y^{2}=1$ with $X\ge 0$.  The set $X=0$ is therefore
a boundary corresponding to vanishing potential energy fraction; we refer to it as the
\emph{kinetic boundary}.

The evolution equations reduce to the scalar-only basic system
\begin{equation}
\dot X = f_1(X,Y) = X\!\left(3Y^{2}-\sqrt{\frac{3}{2}}\lambda Y\right),
\label{eq:f1-basic}
\end{equation}
\begin{equation}
\dot Y = f_2(X,Y) = (1-Y^{2})\!\left(\sqrt{\frac{3}{2}}\lambda-3Y\right).
\label{eq:f2-basic}
\end{equation}

The physical observables in these variables are as follows.  Here
$w_{\phi}:=p_{\phi}/\rho_{\phi}$ is the scalar-field equation-of-state parameter,
$\epsilon:=-\dot H/H^{2}$ is the usual (Hubble) slow-roll parameter, and
$q:=-\ddot a/(aH^{2})$ is the deceleration parameter:
\begin{equation}
w_{\phi}=2Y^{2}-1,
\qquad
\epsilon=3Y^{2},
\qquad
q=3Y^{2}-1.
\label{eq:obs-XY}
\end{equation}

\subsection{Equilibria of the scalar--only basic system}
\label{subsec:equilibria-scalar-only}

The equilibria of \eqref{eq:f1-basic}--\eqref{eq:f2-basic} subject to
\eqref{eq:constraint-XY} are:

\paragraph{Kinetic boundary points}
\begin{equation}
K_{\pm}:\quad (X,Y)=(0,\pm 1),
\label{eq:Kpm}
\end{equation}
corresponding to pure-kinetic scalar-field states ($V=0$ in Eq.~\eqref{eq:XYphysical}), i.e.
\begin{equation}
w_{\phi}=1,\qquad q=2.
\label{eq:Kphys}
\end{equation}

\paragraph{Scalar--dominated scaling branch}
\begin{equation}
S:\quad
Y=\frac{\lambda}{\sqrt{6}},\qquad
X=\sqrt{1-\frac{\lambda^{2}}{6}},
\qquad (|\lambda|\le\sqrt{6}),
\label{eq:Spoint}
\end{equation}
with
\begin{equation}
w_{\phi}=\frac{\lambda^{2}}{3}-1,
\qquad
q=\frac{\lambda^{2}}{2}-1.
\label{eq:Sphys}
\end{equation}
Thus
\begin{equation}
\text{accelerated expansion} \iff \lambda^{2}<2.
\label{eq:accel-cond}
\end{equation}
The collision of $S$ with the kinetic boundary at $|\lambda|=\sqrt{6}$
marks a qualitative change in the scalar--only phase portrait.

\begin{remark}[Collision at the boundary]\label{remark4-2}
As $|\lambda|\uparrow\sqrt6$, the equilibrium branch $S$ moves along the physical semicircle
and reaches the boundary $X=0$, coinciding with $K_+$ when $\lambda=+\sqrt6$ and with
$K_-$ when $\lambda=-\sqrt6$ (we may call the branches with $|\lambda|<\sqrt{6}$ \emph{interior}).  For $|\lambda|>\sqrt6$ the branch $S$ ceases to exist
as a real equilibrium (since $X_*=\sqrt{1-\lambda^2/6}$ becomes imaginary).  In this
sense $|\lambda|=\sqrt6$ is an \emph{organising threshold} for a steady-state to
steady-state \emph{mode} interaction on the constrained state space: not a saddle-node creation
of two equilibria, but an endpoint collision with the physical boundary that removes
(or restores) an interior equilibrium as the parameter crosses the threshold.  This
type of interaction recurs in later multi-component settings, where additional modes
supply unfolding directions and produce further mode interaction phenomena.
\end{remark}

If we define
\begin{equation}
a \equiv \sqrt{\frac{3}{2}}\lambda ,
\label{eq:a-def}
\end{equation}
the Jacobian matrix of the planar vector field
$f=(f_1,f_2)^{\mathsf T}$ in \eqref{eq:f1-basic}--\eqref{eq:f2-basic} is
\begin{equation}
Df(X,Y)=
\begin{pmatrix}
3Y^{2}-aY & X(6Y-a) \\
0 & -2aY-3+9Y^{2}
\end{pmatrix},
\label{eq:Jac-general}
\end{equation}
which we may now use to translate equilibria to the origin and compare the resulting
local forms as a test for a shared organising centre.

For the kinetic (stiff) equilibria \eqref{eq:Kpm} we introduce translated variables
\begin{equation}
U = X,\qquad V = Y \mp 1,
\qquad
u=(U,V)^{\mathsf T},
\label{eq:shift-K}
\end{equation}
so that in a neighbourhood of the origin $u=0$, the dynamics takes the form
\begin{equation}
\dot u = J_{K_\pm} u + F_{K_\pm}(u),
\label{eq:WK-form}
\end{equation}
where the linearised Jacobians and nonlinear parts read
\begin{equation}
J_{K_+}=
\begin{pmatrix}
3-a & 0\\
0 & 6-2a
\end{pmatrix},
\qquad
J_{K_-}=
\begin{pmatrix}
3+a & 0\\
0 & 6+2a
\end{pmatrix},
\label{eq:JKpm}
\end{equation}
\begin{equation}
F_{K_+}(W)=
\begin{pmatrix}
(6-a)\,U V + 3U V^{2} \\
(9-a)\,V^{2} + 3V^{3}
\end{pmatrix},
\label{eq:FKplus}
\end{equation}
\begin{equation}
F_{K_-}(W)=
\begin{pmatrix}
-(6+a)\,U V + 3U V^{2} \\
-(9+a)\,V^{2} + 3V^{3}
\end{pmatrix},
\label{eq:FKminus}
\end{equation}
and $F_{K_\pm}(u)$ collects terms of quadratic and higher order in $(U,V)$.

We also  translate the scalar-dominated scaling point \eqref{eq:Spoint}
\begin{equation}
S:\quad
Y_{*}=\frac{\lambda}{\sqrt6}=\frac{a}{3},
\qquad
X_{*}=\sqrt{1-\frac{\lambda^{2}}{6}}=\sqrt{1-\frac{a^{2}}{9}},
\label{eq:S-org}
\end{equation}
for $|\lambda|\le\sqrt6$, to the origin by setting
\begin{equation}
U=X-X_{*},\qquad V=Y-Y_{*},
\qquad
u=(U,V)^{\mathsf T}.
\label{eq:shift-S}
\end{equation}
Then the translated scalar-dominated system reads
\begin{equation}
\dot u = J_{S} u + F_{S}(u),
\label{eq:WS-form}
\end{equation}
with
\begin{equation}
J_{S}=
\begin{pmatrix}
0 & aX_{*}\\
0 & \frac{a^{2}}{3}-3
\end{pmatrix},
\label{eq:JS}
\end{equation}
and
\begin{equation}
F_{S}(u)=
\begin{pmatrix}
a\,U V + 3X_{*}\,V^{2} + 3U V^{2} \\
2a\,V^{2} + 3V^{3}
\end{pmatrix}.
\label{eq:FS}
\end{equation}

\subsection{A common quadratic germ at $K_\pm$ and $S$: the $\Gamma_1$-locus}
\label{subsec:common-quadratic-germ}

At the $\Gamma_1$-\emph{locus} (i.e., the threshold $|\lambda|=\sqrt6$, equivalently $|a|=3$), the scalar-dominated
equilibrium $S$ meets the kinetic boundary: for $a=+3$ one has $S\equiv K_+$, while
for $a=-3$ one has $S\equiv K_-$.  At the corresponding sign-matched endpoint, the
translated scalar-dominated system shares the same quadratic-order jet as the
translated kinetic system,
\begin{equation}
Q_S(U,V)\equiv Q_{K_{\mathrm{sgn}(a)}}(U,V)\qquad\text{at }|a|=3.
\label{eq:QS-equals-QK}
\end{equation}
In addition, the linear parts vanish at the same sign-matched endpoint.
Indeed, for $a=+3$,
\begin{equation}
J_{K_+}(a=3)=0,\qquad J_S(a=3)=0,
\label{eq:JK-JS-zero-plus}
\end{equation}
while for $a=-3$,
\begin{equation}
J_{K_-}(a=-3)=0,\qquad J_S(a=-3)=0.
\label{eq:JK-JS-zero-minus}
\end{equation}
Thus the scalar-only exponential problem realises the \emph{same} local hilltop-type
degeneracy at distinct equilibrium sites, controlled by the single threshold $|a|=3$.
We refer to such a synchronised degeneracy (occurring at different sites depending
on the sign branch) as a \emph{multi-site event} (or \emph{gate}).

These identities yield the organising centre common to
\eqref{eq:WK-form} and \eqref{eq:WS-form},
\begin{equation}
\left(
  \begin{array}{c}
    \dot{U} \\
    \dot{V} \\
  \end{array}
\right)
=
\left(
  \begin{array}{c}
    3UV\\
    6V^2 \\
  \end{array}
\right)
+O(3),
\qquad
u=(U,V)^{\mathsf T}.
\label{OS:single-sf}
\end{equation}

\paragraph{Embedding into the $\mathbb{Z}_2$ amplitude-versality framework.}
Since the linear part vanishes at the multi-site gate, the quadratic homological
operator is trivial and there is no nonresonant quadratic subspace to eliminate.
We therefore interpret \eqref{OS:single-sf} within the theory of
$\mathbb{Z}_2$-equivariant planar families with two zero eigenvalues.

After the relabelling $(x,y)=(V,U)$, the quadratic truncation takes the triangular
amplitude form $\dot x\sim x^2$, $\dot y\sim x y$, i.e.\ $x$ evolves autonomously at
leading order and $y$ is slaved linearly to $x$, exhibiting the invariant axis $y=0$
and the reflection symmetry $y\mapsto -y$.  Here $\mathbb Z_2$ acts by reflection in
the transverse variable $y$, leaving the amplitude variable $x$ invariant.

By the $\mathbb{Z}_2$-versal classification of \cite{Zholondek1983}, the degenerate
cases constitute codimension--one nontransversal strata in the space of $3$-jets.
Hence our endpoint germ \eqref{OS:single-sf} lies on a distinguished slice of the
generic $\mathbb{Z}_2$ amplitude class and the missing $U^2$ term necessitates
cubic-order control.  It naturally unfolds into the versal ``hilltop'' nondegenerate
deformation of codimension~$3$ (for more details cf.\ Section~\ref{sec:fluid-org}).

We call \eqref{OS:single-sf} an \emph{endpoint} germ because it occurs at the
endpoint(s) $X=0$ of the physical semicircle $X^2+Y^2=1$ (the kinetic boundary),
where an interior equilibrium meets the boundary as the parameter crosses the
threshold.

Accordingly, the scalar-only exponential threshold is best understood as a
multi-site realisation of a single $\mathbb{Z}_2$ amplitude (``hilltop'') gate.
The fluid and curvature modes introduced later provide the natural physical
unfolding directions that probe how this gate persists away from the
scalar-only restriction.

\section{One-fluid extension: the minimal unfolding space}
\label{sec:fluid-org}

The main results of this section are summarised in Theorem~\ref{thm:sec5}.

\begin{theorem}[Organising loci in the $(\gamma,\lambda)$-plane: one-fluid exponential class]
\label{thm:sec5}
Consider the exponential scalar--fluid constrained system \eqref{eq:Xdot-5}--\eqref{eq:Ydot-5}
on $X^2+Y^2+\Omega_m=1$, with parameters $(\gamma,a)$ where $a=\sqrt{3/2}\,\lambda$ and
$p_m=(\gamma-1)\rho_m$.  Let $M,K_\pm,S,F$ denote the principal equilibria
\eqref{eq:Mpoint-5}--\eqref{eq:Fpoint-5}.  Then:

\begin{itemize}
\item \textbf{Distinguished organising locus ($\Gamma_1:$ $|a|=3$):}
the kinetic endpoints $K_\pm$ and the scalar branch endpoint collision at $S$
inherit the scalar-only multi-site episode on the line $|a|=3$ which persists also in this class as the main organising locus, but the qualitative
type of the reduced dynamics now depends on $\gamma$ through the quadratic jets
\eqref{eq:QKplus-5}--\eqref{eq:QS-5}.

\item \textbf{Axis partition and codimension range ($1\le \mathrm{codim}\le 3$):}
on the bifurcation axis $\Gamma_1=\{(\gamma,a):|a|=3\}$ the organising behaviour splits into
regimes separated by the distinguished values $\gamma=0$ and $\gamma=2$ (and by the
existence boundary of the scaling site $F$).  Across these regimes, the effective
codimension of the relevant organising germs varies between $1$ and $3$.

\item \textbf{Non-hyperbolic loci at $M$ and $F$:}
the fluid equilibrium $M$ is non-hyperbolic precisely at $\gamma=0$ and $\gamma=2$ (cf.\
\eqref{eq:JM-5}); the scaling equilibrium $F$ exists only for $\gamma\in[0,2]$ and has
additional degeneracies at $\gamma=0$ and $\gamma=2$ when its linear part vanishes.

\item \textbf{Normal forms and versal stratifications:}
the organising local models needed for the subsequent unfolding analysis are exhausted by
(i) cubic one-dimensional centre reductions (pitchfork-type) in appropriate $\gamma$-regimes,
(ii) quartic one-dimensional centre reductions (fold/winged-cusp type) when the first
non-removable term is quartic, and (iii) planar $\mathbb Z_2$-equivariant hilltop jets,
with codimension at most $3$ in the full (non-symmetric) class and at most $2$ in the
$\mathbb Z_2$-equivariant subclass.
\end{itemize}
\end{theorem}
Section 6 adds curvature as a further component, enlarging the organising map and introducing additional loci.

\subsection*{Structure of this section}
We extend the exponential potential system by adding a single barotropic perfect fluid.
The resulting dynamics is naturally formulated in the three variables $(X,Y,\Omega_m)$
subject to the constraint $X^2+Y^2+\Omega_m=1$. We translate the principal equilibria
to the origin and compute the corresponding Jacobians in the augmented phase space.
Our aim is to determine whether adding one fluid component supplies the minimal
\emph{physical unfolding directions} for the scalar-only organising threshold
$|a|=3$ (equivalently $|\lambda|=\sqrt6$) identified in Section~\ref{sec:exp-Z2-gate},
where the scalar-dominated equilibrium meets the kinetic boundary and the kinetic
endpoints acquire a hilltop-type degeneracy. The fluid extension introduces new
equilibria (fluid-dominated and scaling) that probe how this organising episode
persists and unfolds in the enlarged phase space.

\subsection{Reduction to the constrained two-dimensional system}
\label{subsec:fluid-2D-reduction}

We now add a single barotropic perfect fluid with
$p_m=(\gamma-1)\rho_m$, and use the same
expansion-normalised variables as in Section~\ref{sec:exp-Z2-gate},
\begin{equation}
X=\frac{\sqrt{V}}{\sqrt{3}H},\qquad
Y=\frac{\dot\phi}{\sqrt{6}H},\qquad
\Omega_m=\frac{\rho_m}{3H^2}.
\label{eq:XYOm-def-5}
\end{equation}
The Friedmann constraint becomes
\begin{equation}
X^2+Y^2+\Omega_m=1.
\label{eq:constraint-5}
\end{equation}
With $a=\sqrt{3/2}\lambda$, the autonomous system may be written as
\begin{equation}
\dot X=
X\!\left(3Y^2+\frac{3\gamma}{2}\Omega_m-aY\right),
\label{eq:Xdot-5-3D}
\end{equation}
\begin{equation}
\dot Y=
-3Y+aX^2+3Y^3+\frac{3\gamma}{2}Y\Omega_m,
\label{eq:Ydot-5-3D}
\end{equation}
\begin{equation}
\dot\Omega_m=
3\Omega_m\!\left(2Y^2-\gamma\right).
\label{eq:Omdot-5-3D}
\end{equation}
However, the physical flow is confined to the two-dimensional
constraint surface \eqref{eq:constraint-5}.  Eliminating
$\Omega_m=1-X^2-Y^2$ gives the equivalent constrained planar system
\begin{equation}
\dot X=
X\!\left(3Y^2+\frac{3\gamma}{2}(1-X^2-Y^2)-aY\right),
\label{eq:Xdot-5}
\end{equation}
\begin{equation}
\dot Y=
-3Y+aX^2+3Y^3+\frac{3\gamma}{2}Y(1-X^2-Y^2).
\label{eq:Ydot-5}
\end{equation}
Differentiating \eqref{eq:constraint-5} and substituting
\eqref{eq:Xdot-5}--\eqref{eq:Ydot-5} reproduces
\eqref{eq:Omdot-5-3D} on the constraint surface, so the
$\Omega_m$ equation is not independent.  We therefore take
\eqref{eq:Xdot-5}--\eqref{eq:Ydot-5} to be the basic one-fluid system
for the exponential potential, in direct parallel with the
scalar-only basic system of Section~\ref{sec:exp-Z2-gate}.

\begin{remark}[The parameter $\gamma$]
The one-fluid extension introduces a second distinguished parameter, the barotropic
index $\gamma$, in addition to the slope parameter $a=\sqrt{3/2}\lambda$.
While $a$ controls the scalar sector and the scalar-only organising threshold
$|a|=3$ ($|\lambda|=\sqrt6$), the parameter $\gamma$ controls the matter component
and creates qualitatively new regimes and bifurcation boundaries.  In particular,
new equilibria (fluid-dominated and scalar--fluid scaling) appear, and their
existence and stability depend on inequalities involving both $(a,\gamma)$.
Thus the geometry is no longer organised along a single parameter line, but on a
two-parameter plane. This is the first step toward the parameter-space stratifications developed in
Sections~\ref{sec:strata}--\ref{sec:primer}, in which critical loci (e.g.\ $a^2=3\gamma$ and $\gamma=2/3$ in Fig. \ref{fig:bifn-map-gamma-lambda} and
later sections) act as transition boundaries between persistent diagram types.

\end{remark}

\subsection{Equilibria of the one-fluid constrained system}
\label{subsec:fluid-2D-equilibria}

The equilibria of \eqref{eq:Xdot-5}--\eqref{eq:Ydot-5} are:

\paragraph{Fluid-dominated point}
\begin{equation}
M:\quad (X,Y)=(0,0),
\qquad \Omega_m=1.
\label{eq:Mpoint-5}
\end{equation}

\paragraph{Kinetic points}
\begin{equation}
K_{\pm}:\quad (X,Y)=(0,\pm 1),
\qquad \Omega_m=0.
\label{eq:Kpm-5}
\end{equation}

\paragraph{Scalar-dominated branch}
\begin{equation}
S:\quad
Y_*=\frac{\lambda}{\sqrt6}=\frac{a}{3},\qquad
X_*=\sqrt{1-\frac{\lambda^2}{6}}=\sqrt{1-\frac{a^2}{9}},
\qquad \Omega_m=0,
\label{eq:Spoint-5}
\end{equation}
which exists for $|\lambda|\le\sqrt6$.

\paragraph{Scalar--fluid scaling branch}
\begin{equation}
F:\quad
Y_F=\frac{3\gamma}{2a},\qquad
X_F=\sqrt{\frac{9\gamma(2-\gamma)}{4a^2}},\qquad
\Omega_{m,F}=1-\frac{9\gamma}{2a^2},
\label{eq:Fpoint-5}
\end{equation}
which is real provided $0\leq \gamma\leq 2$, and  exists when $\lambda^2\ge 3\gamma$, equivalently
\begin{equation}
a^2\geq \frac{9}{2}\gamma.
\label{eq:Fexist-5}
\end{equation}
\begin{remark}[Meaning of ``scaling'' at $F$]
The equilibrium $F$ is a \emph{scaling} (or tracking) solution in the standard sense:
the scalar and fluid energy densities scale proportionally, so the fractions
$\Omega_\phi=X^2+Y^2$ and $\Omega_m$ are constant along the solution and
$0<\Omega_\phi<1$, $0<\Omega_m<1$ when $F$ exists. Equivalently, the scalar mimics the fluid equation of state,
\(w_\phi=w_m=\gamma-1\). Therefore, if
\(w_{\rm eff}:=p_{\rm tot}/\rho_{\rm tot}\) denotes the total effective
equation-of-state parameter, then \(w_{\rm eff}=w_m=\gamma-1\), and the
expansion law is the same as for a pure \(\gamma\)-fluid. This terminology is standard for exponential
potentials with constant $\lambda$ and constant $\gamma$, cf. e.g.,\ the discussion of the
``scaling solution'' (Point~B) in \ \cite{CopelandReview2018}, p. 39. (When $\lambda$ becomes an effective slowly varying quantity, we will speak of \emph{instantaneous} scaling/tracking relative to the moving target.)

\end{remark}

\begin{remark}
Equation \eqref{eq:Fpoint-5} describes the scaling branch $F$.
The fluid-dominated equilibrium $M$ is instead the distinct point
\eqref{eq:Mpoint-5}; thus one should not substitute $\Omega_m=1$
into the scaling relation to characterise $M$.
\end{remark}
\subsection{Jacobian and translated local systems for the constrained planar field}
\label{subsec:fluid-2D-Jac}

Let $f=(f_1,f_2)^{\mathsf T}$ denote the right-hand side of
\eqref{eq:Xdot-5}--\eqref{eq:Ydot-5}.  The Jacobian matrix is
\begin{equation}
Df(X,Y)=
\begin{pmatrix}
3Y^{2}-aY+\dfrac{3\gamma}{2}
-\dfrac{9\gamma}{2}X^{2}-\dfrac{3\gamma}{2}Y^{2}
&
X\!\left((6-3\gamma)Y-a\right)
\\[1.2ex]
X\!\left(2a-3\gamma Y\right)
&
-3+\dfrac{3\gamma}{2}
+9Y^{2}-\dfrac{3\gamma}{2}X^{2}-\dfrac{9\gamma}{2}Y^{2}
\end{pmatrix}.
\label{eq:Jac-fluid-2D}
\end{equation}

Evaluating \eqref{eq:Jac-fluid-2D} at the equilibria gives
\begin{equation}
J_M=
\begin{pmatrix}
\dfrac{3\gamma}{2} & 0\\
0 & \dfrac{3\gamma}{2}-3
\end{pmatrix},
\label{eq:JM-5}
\end{equation}
\begin{equation}
J_{K_+}=
\begin{pmatrix}
3-a & 0\\
0 & 6-3\gamma
\end{pmatrix},
\qquad
J_{K_-}=
\begin{pmatrix}
3+a & 0\\
0 & 6-3\gamma
\end{pmatrix},
\label{eq:JKpm-5}
\end{equation}
and for $S$,
\begin{equation}
J_{S}=
\begin{pmatrix}
\dfrac{\gamma}{3}(a^{2}-9) &
\dfrac{a(1-\gamma)}{3}\sqrt{9-a^{2}}
\\[1.2ex]
\dfrac{a(2-\gamma)}{3}\sqrt{9-a^{2}} &
a^{2}\!\left(1-\dfrac{\gamma}{3}\right)-3
\end{pmatrix}.
\label{eq:JS-5}
\end{equation}
The linearization at $F$ is standard and will be quoted when needed. For the organising centre comparison below, the kinetic and scalar
endpoint structure is the principal ingredient. In particular, we will compare the quadratic jets at the sign-matched kinetic endpoint
and at $S$ near $|a|=3$ to identify the unfolding directions supplied by $\gamma$.

\subsubsection{Fluid-dominated equilibrium $M$}
\label{subsec:fluid-2D-M-trans}

For the fluid-dominated equilibrium \eqref{eq:Mpoint-5} of the constrained system
\eqref{eq:Xdot-5}--\eqref{eq:Ydot-5} no translation is required, and so setting
\begin{equation}
U=X,\qquad V=Y,\qquad u=(U,V)^{\mathsf T},
\label{eq:Wdef-M}
\end{equation}
the local form reads
\begin{equation}
\dot{u}= J_M u + F_M(u),
\label{eq:WM-form}
\end{equation}
with $J_M$ given by Eq. \eqref{eq:JM-5}, and the quadratic jet of the nonlinear remainder is
\begin{equation}
Q_M(U,V)=
\begin{pmatrix}
-a\,UV\\[0.4ex]
a\,U^{2}
\end{pmatrix},
\label{eq:QM-5}
\end{equation}
so that, up to second order,
\begin{equation}
F_M(u)=Q_M(U,V)+O(3).
\label{eq:FM-quad}
\end{equation}

\subsubsection{Translated systems near $K_\pm$}
\label{subsec:fluid-2D-Kpm-trans}

We translate $K_\pm$ to the origin by
\begin{equation}
U=X,\qquad V=Y\mp 1,\qquad u=(U,V)^{\mathsf T}.
\label{eq:shift-Kpm-5}
\end{equation}
Then
\begin{equation}
\dot{u} = J_{K_\pm}u + F_{K_\pm}(u),
\label{eq:WKpm-5}
\end{equation}
where $J_{K_\pm}$ is given in \eqref{eq:JKpm-5}.  The quadratic jets of the
nonlinear terms are
\begin{equation}
Q_{K_+}(U,V)=
\begin{pmatrix}
(6-a-3\gamma)\,UV\\[0.5ex]
\bigl(a-\dfrac{3\gamma}{2}\bigr)U^{2}
+\dfrac{9}{2}(2-\gamma)V^{2}
\end{pmatrix},
\label{eq:QKplus-5}
\end{equation}
\begin{equation}
Q_{K_-}(U,V)=
\begin{pmatrix}
(3\gamma-6-a)\,UV\\[0.5ex]
\bigl(a+\dfrac{3\gamma}{2}\bigr)U^{2}
-\dfrac{9}{2}(2-\gamma)V^{2}
\end{pmatrix}.
\label{eq:QKminus-5}
\end{equation}
Higher-order terms may be retained when constructing the local
$\mathbb{Z}_2$-amplitude charts at the multi-site threshold.

\subsubsection{Translated system near $S$}
\label{subsec:fluid-2D-S-trans}

For the scalar-dominated point
\eqref{eq:Spoint-5} we set
\begin{equation}
U=X-X_*,\qquad V=Y-Y_*,\qquad u=(U,V)^{\mathsf T}.
\label{eq:shift-S-5}
\end{equation}
Then
\begin{equation}
\dot{u} = J_Su + F_S(u),
\label{eq:WS-5}
\end{equation}
with $J_S$ as in \eqref{eq:JS-5}.  Writing $X_*=\sqrt{1-a^{2}/9}$, the quadratic
part of $F_S$ is
\begin{equation}
Q_S(U,V)=
\begin{pmatrix}
-\dfrac{9\gamma}{2}X_*\,U^{2}
-a(\gamma-1)\,UV
-\dfrac{3}{2}( \gamma-2 )X_*\,V^{2}
\\[1.0ex]
\dfrac{a}{2}(2-\gamma)\,U^{2}
-3\gamma X_*\,UV
+\dfrac{3a}{2}(2-\gamma)\,V^{2}
\end{pmatrix}.
\label{eq:QS-5}
\end{equation}

\subsubsection{Scalar--fluid scaling equilibrium $F$}
\label{subsec:fluid-2D-F-trans}

We translate the scalar--fluid scaling equilibrium $F$ given by \eqref{eq:Fpoint-5}, \eqref{eq:Fexist-5}  to the origin by
\begin{equation}
U=X-X_F,\qquad V=Y-Y_F,\qquad u=(U,V)^{\mathsf T}.
\label{eq:shift-F}
\end{equation}
Then
\begin{equation}
\dot{u}=J_F u+F_F(u),
\label{eq:WF-form}
\end{equation}
where $J_F=Df(X_F,Y_F)$ is the Jacobian of the constrained planar
vector field:
\begin{equation}
J_F=
\begin{pmatrix}
-3\gamma X_F^{2} &
X_F\!\left((6-3\gamma)Y_F-a\right)
\\[1.0ex]
X_F\!\left(2a-3\gamma Y_F\right) &
\left(\dfrac{3}{2}\right)(\gamma-2)
+3\!\left(\dfrac{3}{2}\right)(2-\gamma)Y_F^{2}
-\dfrac{3\gamma}{2}X_F^{2}
\end{pmatrix}.
\label{eq:JF-compact}
\end{equation}
Substituting \eqref{eq:Fpoint-5} gives the explicit form
\begin{equation}
J_F=
\begin{pmatrix}
-\dfrac{27\gamma^{2}(2-\gamma)}{4a^{2}} &
\dfrac{X_F}{2a}\!\left(9\gamma(2-\gamma)-2a^{2}\right)
\\[1.2ex]
\dfrac{X_F}{2a}\!\left(4a^{2}-9\gamma^{2}\right) &
(2-\gamma)\!\left(-\dfrac{3}{2}
+\dfrac{27\gamma^{2}}{4a^{2}}\right)
\end{pmatrix}.
\label{eq:JF-explicit}
\end{equation}

The quadratic jet of $F_F$ is obtained from the second derivatives of
$f$ at $F$ and reads
\begin{equation}
Q_F(U,V)=
\begin{pmatrix}
-\dfrac{9\gamma X_F}{2}\,U^{2}
+\dfrac{9\gamma(2-\gamma)-2a^{2}}{2a}\,UV
+\dfrac{3(2-\gamma)X_F}{2}\,V^{2}
\\[1.2ex]
\dfrac{4a^{2}-9\gamma^{2}}{4a}\,U^{2}
-3\gamma X_F\,UV
+\dfrac{27\gamma(2-\gamma)}{4a}\,V^{2}
\end{pmatrix},
\label{eq:QF-5}
\end{equation}
so that
\begin{equation}
F_F(u)=Q_F(U,V)+O(3).
\label{eq:FF-quad}
\end{equation}

\subsection{A multi-site organising episode on the line $\Gamma_1$-locus}
\label{subsec:fluid-2D-multisite}

\noindent\textbf{Navigation.}
Adding a single fluid component enlarges the distinguished parameter space from the
$\lambda$--line to the $(\gamma,\lambda)$--plane (equivalently $(\gamma,a)$ with
$a=\sqrt{3/2}\lambda$).  We now organise the resulting variety of local behaviours by
restricting attention to the $\Gamma_1$-locus, i.e.,\ the multi-site line $|\lambda|=\sqrt6$ (i.e.\ $|a|=3$),
where the scalar-only hilltop episode of Section~\ref{sec:exp-Z2-gate} persists but is
now probed by the additional parameter $\gamma$.  Along this line the non-hyperbolic
cases occur at distinguished values of $\gamma$ (notably $\gamma=0$ and $\gamma=2$),
and the corresponding local stratifications have effective codimension between $1$ and $3$.

\subsubsection{The main difficulty}
A basic novelty introduced by the fluid component is that the family of constrained
systems \eqref{eq:Xdot-5}--\eqref{eq:Ydot-5} is now controlled by two parameters,
$(\gamma,\lambda)$, rather than by $\lambda$ alone.  In contrast to the
Friedmann--Lema\^itre setting with a cosmic fluid of index $\gamma$ and a cosmological constant $\Lambda$ (cf.\
\cite{CotsakisFL}, Sect.\ 3.1), \emph{here both $\gamma$ and $\lambda$ act as bifurcation
parameters}: varying either may change the local phase portrait near the equilibria
$M,K_\pm,S,F$.

The way we approach this problem is to simplify as much as possible \emph{before}
analysing the full two-parameter family. This is done in two steps. First, via
centre manifold reductions we remove the hyperbolic directions and retain only the
non-hyperbolic (centre) directions responsible for qualitative change. Second, we
track which low-order terms in the resulting Taylor expansions already determine the
local phase portrait (independently of higher-order terms). This leads to a small set
of `organising' normal forms, obtained up to strong equivalence.

The practical purpose of this subsection is therefore to isolate the
distinguished loci in the $(\gamma,\lambda)$--plane that organise these changes, and
to relate them to the quadratic jets (and, where necessary, center manifold reductions)
computed in Section~\ref{subsec:fluid-2D-Jac}.

\subsubsection{The bifurcation axis and simultaneous bifurcations}
While generic motion in the $(\gamma,\lambda)$--plane produces bifurcations controlled
by the degeneracies and symmetries of the constrained system
\eqref{eq:Xdot-5}--\eqref{eq:Ydot-5}, the most characteristic organising behaviour for
the present problem is concentrated on the multi-site line
\begin{equation}
\Gamma_1=\{(\gamma,\lambda):|\lambda|=\sqrt6\}
\equiv \{(\gamma,a):|a|=3\},
\label{eq:Gamma-axis}
\end{equation}
shown in Fig.~\ref{fig:bifn-map-gamma-lambda}.  The line $\Gamma_1$ is naturally
partitioned into five regions $\Gamma_1^i,i=1,\dots, 5$:
\begin{enumerate}
\item $\Gamma_1^1:\ \gamma\in(-\infty,0)$,
\item $\Gamma_1^2:\ \gamma=0$,
\item $\Gamma_1^3:\ \gamma\in(0,2)$,
\item $\Gamma_1^4:\ \gamma=2$,
\item $\Gamma_1^5:\ \gamma\in(2,\infty)$.
\end{enumerate}
As $\gamma$ crosses $\Gamma_1^1\leftrightarrow\Gamma_1^5$ along $\Gamma_1$, the equilibria
$M,K_\pm,S,F$ (and the local flows near them) change type in ways that can be read off
from: (i) loss of hyperbolicity at $\gamma=0$ or $\gamma=2$, (ii) endpoint collision
at $|a|=3$, and (iii) the existence restriction of the scaling branch $F$ to
$0\le\gamma\le2$. 

This implies that interestingly, the dynamics (of nearby systems) at each bifurcating $\gamma$ value is characterised by the \emph{simultaneous} (i.e.,\ at the same $\gamma$ value) distinct  transitions of the system near different equilibria, producing simultaneous  versal effects of different codimensions and degeneracies.

\subsubsection{Quadratic jets and center reductions}
\label{subsubsec:jets-centre-reductions}

\paragraph{(i) The fluid point $M$.}
At $M:(X,Y)=(0,0)$ the Jacobian of the constrained planar field is
\begin{equation}
J_M=\mathrm{diag}\!\left(\frac{3\gamma}{2},\frac{3\gamma}{2}-3\right),
\label{eq:JM-diag-54}
\end{equation}
hence $M$ is non-hyperbolic only on $\gamma=0$ and $\gamma=2$.

\emph{On $\Gamma_1^2:\gamma=0$},  one has eigenvalues $(0,-3)$, so the center space is the
$X$-axis. A tangency computation gives a center manifold
$Y=h(X)=\tfrac{a}{3}X^2+O(X^4)$ and the reduced scalar equation
\begin{equation}
\dot X = -\frac{a^2}{3}X^3 + O(X^5),
\label{eq:M-centre-cubic}
\end{equation}
so the local dynamics is governed by a cubic center reduction.

\emph{On $\Gamma_1^4:\gamma=2$},  the constrained system \eqref{eq:Xdot-5}--\eqref{eq:Ydot-5} exhibits \emph{loss of isolation}:
the entire arc $X=0$, $|Y|\le 1$ consists of equilibria (since $\dot X|_{X=0}\equiv0$
and $\dot Y|_{X=0}\equiv0$, with $\Omega_m=1-Y^2$ on the constraint).  Writing
$\mu=\gamma-2$ and restricting to the arc $X=0$ gives
\begin{equation}
\dot Y\big|_{X=0}=\frac{3}{2}\mu\,Y(1-Y^2),
\label{eq:M-arc-restriction}
\end{equation}
so for $\mu\neq0$ the equilibrium arc \emph{splits} into three isolated equilibria
$Y=0,\pm1$ together with a directed flow along the former arc.  Thus the organising
mechanism at $\gamma=2$ is a bifurcation of a non-isolated equilibrium set, rather than
a scalar quartic germ at a single isolated equilibrium.

\paragraph{(ii) The kinetic and scalar endpoint sites $K_\pm$ and $S$.}
On the line $\Gamma_1$ (i.e.\ $|a|=3$) the scalar-dominated equilibrium meets the kinetic
boundary: for $a=+3$ one has $S\equiv K_+$, while for $a=-3$ one has $S\equiv K_-$.  In
addition, the translated quadratic jets satisfy the sign-matched identities
\begin{equation}
Q_{K_+}^{a=3}=Q_S^{a=3},
\qquad
Q_{K_-}^{a=-3}=Q_S^{a=-3},
\qquad
Q_S^{a=3}=-Q_S^{a=-3},
\label{eq:Q-identity-KS}
\end{equation}
as follows from Section~\ref{subsec:fluid-2D-Jac}.  This is the precise sense in which
the $|a|=3$ episode remains a \emph{multi-site} organising event in the one-fluid
problem.

The role of $\gamma$ is now decisive.  At $\gamma=2$ the linear parts vanish at the
corresponding sites (e.g.\ $J_{K_+}=0$ at $(a,\gamma)=(3,2)$), but the quadratic truncation
becomes more degenerate: after relabelling $(x,y)=(V,U)$ at the endpoint one finds a
triangular $\mathbb Z_2$--equivariant form in which the amplitude equation has no
quadratic term,
\begin{equation}
\dot x = O(3),\qquad \dot y = \sigma\,x y + O(3),
\label{eq:triangular-gamma2}
\end{equation}
with reflection symmetry $y\mapsto -y$ and invariant axis $y=0$.  Consequently, cubic
(or higher) terms control the organising versal dynamics at $\gamma=2$ (compare with the occurrence of distinct but `simultaneous' bifurcations due to loss of isolation for the $M$-equilibrium at $\Gamma_1^4$ mentioned earlier).

Away from $\gamma=2$ (still on $|a|=3$), the endpoint equilibria acquire one hyperbolic
direction and reduce to one-dimensional center dynamics.  The resulting center manifold
amplitude equation has the schematic form $\dot x = G(x;\mu)$, where $x$ parametrises
the non-hyperbolic direction tangent to the constrained state space at the endpoint.
Depending on parameter values and symmetry constraints, the first non-removable term
may be cubic or quartic:
\begin{equation}
\dot{x}\sim x^3 +O(5),
\label{eq:KS-cubic}
\end{equation}
or
\begin{equation}
\dot{x}\sim x^4 +O(5).
\label{eq:KS-quartic}
\end{equation}
In singularity-theoretic terms, the quartic germ $\dot x=x^{4}$ is finitely
determined and admits a miniversal (strong) unfolding
\begin{equation}
\dot x=x^{4}+\mu_{1}x^{2}+\mu_{2}x+\mu_{3},
\label{eq:miniversal-quartic}
\end{equation}
which organises how the endpoint equilibria and their stability types change under perturbations.
In the $\mathbb Z_2$--equivariant subclass the odd term is forbidden ($\mu_{2}=0$),
reducing the effective codimension compared to the fully generic case.

\paragraph{(iii) The scaling site $F$.}
The equilibrium $F$ exists only for $\gamma\in[0,2]$. Its non-hyperbolic cases occur at
$\gamma=2$ (where $J_F=0$) and at $\gamma=0$ provided $a^2\ge 9$ (again $J_F=0$). The
corresponding quadratic jets are
\begin{equation}
Q^{\gamma=2}_F=
\left(
  \begin{array}{c}
    -aUV \\
    (a-36)U^2 \\
  \end{array}
\right) ,
\qquad
Q^{\gamma=0}_F=
\left(
  \begin{array}{c}
    -aUV \\
    aU^2 \\
  \end{array}
\right).
\label{eq:QF-special}
\end{equation}
Upon setting $(x,y)=(V,U)$, the quadratic truncation takes the triangular form
\begin{equation}
\dot x = -a\,x y + O(3),\qquad \dot y = c\,y^2 + O(3),
\label{eq:triangular-F}
\end{equation}
with $c=a-36$ in the $\gamma=2$ case and $c=a$ in the $\gamma=0$ case.  After a
rescaling of time and $y$, this is equivalent at quadratic order to the normal form
\begin{equation}
\dot x = x y,\qquad \dot y = \pm y^2,
\label{eq:NF-shear}
\end{equation}
exhibiting the invariant axis $y=0$ and a shear-type coupling from the $y$--direction
into the $x$--direction. This is still another occurrence of simultaneous bifurcations mentioned above.

\begin{remark}
At $(\gamma,a)=(2,\pm 3)$ the scaling branch $F$ meets the kinetic endpoint
$K_{\pm}$ (and hence the scalar site $S$), reinforcing the genuinely multi-site nature
of the $|a|=3$ organising episode in the one-fluid problem.
\end{remark}

\subsubsection{The normal forms}

\noindent\emph{Source of the degeneracies.}
The quartic one-dimensional center reduction \eqref{eq:KS-quartic} is \emph{not}
associated with the fluid point $M$ at $\gamma=2$ (where $M$ belongs to an equilibrium
arc and is therefore non-isolated). Rather, quartic behaviour occurs precisely in
regimes where the first non-removable term on the reduced center dynamics at an
\emph{isolated} equilibrium is of order four; it is this situation that is modelled by
the quartic fold/winged-cusp stratifications below.

The principal observation is that the degenerate jets encountered above fall into a
small set of organising stratifications whose effective codimension does not exceed $3$.
Accordingly, we will organise the local bifurcations near $\Gamma_1$ using the following
normal forms:
\begin{enumerate}
\item[(0)] \emph{Loss of isolation (equilibrium arc).}
At $\gamma=2$ the fluid point $M$ lies on a non-isolated equilibrium arc; the organising
mechanism is the bifurcation of this equilibrium set under $\mu=\gamma-2$, producing
isolated equilibria and a directed flow along the former arc.
\item \emph{Pitchfork-type center reductions.}
When symmetry forces the reduced amplitude equation to start cubically
($\dot x\sim x^3$), the local behaviour is organised by pitchfork-type stratifications and
their codimension--$2$ unfoldings.
\item \emph{Quartic fold / winged cusp.}
When the first non-removable term is quartic ($\dot x\sim x^4$), the unfolding geometry
is organised by quartic fold/winged-cusp stratifications (codimension--$2$ in the $\mathbb Z_2$
subclass, codimension--$3$ in the fully generic class).
\item \emph{Hilltop-type $\mathbb Z_2$ planar stratifications.}
Degenerate two-dimensional $\mathbb Z_2$--equivariant jets (with one amplitude-like and
one transverse direction) organise multi-site endpoint behaviour and lead to
codimension--$3$ hilltop-type families.
\end{enumerate}

\section{The curvature extension}
\label{sec:curvature}
\subsection*{Structure of this section}
We now extend the previous analysis by incorporating spatial curvature as an additional
full component. In Section~\ref{subsec:scalar-curv} we study the scalar--curvature class
(without fluid), where the exponential slope $a=\sqrt{3/2}\,\lambda$ interacts with the
curvature fraction $\Omega_k$. In Section~\ref{subsec:scalar-fluid-curv} we treat the
scalar--fluid--curvature class, where $\Omega_k$ coexists with the independent fluid
fraction $\Omega_m$ on the three-dimensional Friedmann constraint hypersurface.
For each class we introduce a convenient constrained autonomous formulation, list the
principal equilibria (including the curvature-scaling site), and compute the linear
and quadratic jets needed for the unified stratification/unfolding analysis in later
sections.

In this section, and throughout the subsequent curvature analysis, ``curvature'' means the
background FRW spatial curvature mode represented by $\Omega_k$, not perturbative
curvature degrees of freedom.

\subsection{The scalar--curvature system}
\label{subsec:scalar-curv}
The main results of this Subsection are summarized in the following theorem.
\begin{theorem}[Organising loci and normal forms: scalar--curvature class]\label{thm:sec6-scalarcurv}
Consider the scalar--curvature constrained system \eqref{eq:Xdot-sc}--\eqref{eq:Ydot-sc}
on $X^2+Y^2+\Omega_k=1$ with slope parameter $a=\sqrt{3/2}\,\lambda$.
Then the organising structure is:
\begin{itemize}
\item \textbf{Endpoint axis ($1$D / codim-$1$):} at the $\Gamma_1$-locus $a=\pm 3$ the kinetic endpoints $K_\pm$
(and the scalar branch endpoint collision $S=K_\pm$) have a one-dimensional center
manifold with cubic reduced germ (after rescaling).
\item \textbf{Curvature-induced axis ($1$D / codim-$1$):} at the $\Gamma_2$-locus $a^2=3$ the scalar branch $S$
meets the curvature--scaling branch $P_k$ and a simple zero eigenvalue appears; the
local reduced normal form is transcritical (branch intersection and stability exchange).
\item \textbf{Spectral transition (hyperbolic):} at $a=\pm 2$ the equilibrium $P_k$ remains
hyperbolic but changes linear type (node $\leftrightarrow$ spiral) via a repeated
eigenvalue; this is not a loss of hyperbolicity.
\end{itemize}
\end{theorem}
\subsubsection{Constraint and reduced planar system}
We set $\Omega_m\equiv 0$ and retain curvature via
\begin{equation}\label{eq:Omk-def-sc}
\Omega_k=-\frac{k}{\sfac^{2}H^{2}},
\end{equation}
where $\sfac(t)$ is the FRW scale factor and $k\in\{-1,0,+1\}$.

The Friedmann constraint becomes
\begin{equation}\label{eq:constraint-sc}
X^2+Y^2+\Omega_k=1,
\end{equation}
hence
\begin{equation}\label{eq:Omk-elim-sc}
\Omega_k=1-X^2-Y^2.
\end{equation}
Substituting \eqref{eq:Omk-elim-sc} into the curvature-inclusive field yields an equivalent
constrained planar system in $(X,Y)$:
\begin{equation}\label{eq:Xdot-sc}
\dot X = X\Big(1+2Y^2-X^2-aY\Big),
\end{equation}
\begin{equation}\label{eq:Ydot-sc}
\dot Y = aX^2-2Y-X^2Y+2Y^3.
\end{equation}
(Equivalently, one may work in $(X,Y,\Omega_k)$ on the constraint \eqref{eq:constraint-sc}; we use
\eqref{eq:Xdot-sc}--\eqref{eq:Ydot-sc} for direct comparison with the scalar-only analysis of Sect.~\ref{sec:exp-Z2-gate}.)
\paragraph{Observables.}
In $(X,Y,\Omega_k)$ one has
\begin{equation}\label{eq:obs-sc}
\rho_\phi=3H^2(X^2+Y^2),\qquad
p_\phi=3H^2(Y^2-X^2),\qquad
w_\phi=\frac{Y^2-X^2}{X^2+Y^2}.
\end{equation}
The Hubble slow-roll and deceleration parameters are
\begin{equation}\label{eq:eps-q-sc}
\epsilon:=-\frac{\dot H}{H^2}=3Y^2+\Omega_k,\qquad
q=-1+\epsilon=3Y^2+\Omega_k-1=2Y^2-X^2.
\end{equation}
Hence accelerated expansion ($q<0$) is equivalent to $X^2>2Y^2$.

\subsubsection{Principal equilibria}
The principal equilibria of \eqref{eq:Xdot-sc}--\eqref{eq:Ydot-sc} are:

\paragraph{Kinetic endpoints.}
\begin{equation}\label{eq:Kpm-sc}
K_\pm:\ (X,Y)=(0,\pm 1),\qquad \Omega_k=0.
\end{equation}

\paragraph{Curvature-dominated point.}
\begin{equation}\label{eq:C-sc}
C:\ (X,Y)=(0,0),\qquad \Omega_k=1.
\end{equation}

\paragraph{Scalar-dominated branch.}
\begin{equation}\label{eq:S-sc}
S:\ \ (X_*,Y_*)=\left(\sqrt{1-\frac{a^2}{9}},\,\frac{a}{3}\right),
\qquad \Omega_{k,*}=0,
\qquad (|a|\le 3).
\end{equation}

\noindent\emph{Boundary collision $S=K_\pm$ on the $\Gamma_1$-locus $|a|=3$.}
Since $X_*=\sqrt{1-a^2/9}$ and $Y_*=a/3$, at $a=\pm 3$ one has $(X_*,Y_*)=(0,\pm 1)$ and therefore
\begin{equation}\label{eq:S-equals-K-sc}
S=K_\pm \qquad \text{at } a=\pm 3 \ (\Leftrightarrow |\lambda|=\sqrt6).
\end{equation}
\paragraph{Curvature-scaling branch.}
\begin{equation}\label{eq:Pk-sc}
P_k:\ \ (X_k,Y_k)=\left(\frac{\sqrt2}{a},\,\frac{1}{a}\right),
\qquad \Omega_{k,k}=1-\frac{3}{a^2},
\end{equation}
which exists for $a\neq 0$ and is physically relevant as a negative-curvature site when
$\Omega_{k,k}>0$, i.e.\ $a^2>3$ (equivalently $\lambda^2>2$, or $|\lambda|>\sqrt2$). (We select the sign of $X_k$ so that $X_k\ge 0$ in the expanding branch.)

\noindent\emph{Coalescence  $S=P_k$ on the $\Gamma_2$-locus $a^2=3$.} On this locus we have $\Omega_{k,*}=\Omega_{k,k}=0$, and $a=\sqrt{3}, a^2=3$ from the equality of the coordinates (cf. subsection \ref{subsubsec:SC-quadratic}(iii)).
\begin{lemma}[Normal hyperbolicity in the curvature direction]
The  $\Omega_k$ direction is normally hyperbolic at $\Omega_k=0$ except where $X^2-2Y^2=0$.
\end{lemma}
This is a consequence of the existence of the compact, boundaryless submanifold $\mathcal{O}=\{ \Omega_k=0\}$, which from Eq. \eqref{eq:Omk-elim-sc} is a periodic orbit of the flow defined by the scalar-curvature system \eqref{eq:Xdot-sc}-\eqref{eq:Ydot-sc}.  In the $(X,Y,\Omega_k)$ formulation one finds
\begin{equation}\label{eq:Omk-evol-sc}
\dot\Omega_k=-2\,\Omega_k\,(X^2-2Y^2),
\end{equation}
so when $\Omega_k= 0$ the flow becomes one-dimensional. Then when $\Omega_k\neq 0$, the directions normal to $\mathcal{O}$ are expanding/contracting more sharply than those tangent to $\mathcal{O}$ (except when $X^2-2Y^2=0$). 

Hence the flow is normally hyperbolic at the kinetic endpoints $K_\pm$, while at  the scalar branch $S$  the curvature direction is normally hyperbolic for $a^2\neq 3$. 

For the curvature-scaling branch $P_k$,  this condition reduces to the $\Gamma_2$-locus $a^2=3$, i.e., precisely
the locus where $S$ meets $P_k$ (cf.\ \eqref{eq:Pk-sc}). The nature of the $\Gamma_2$-locus requires some knowledge of the linearization of the system \eqref{eq:Xdot-sc}-\eqref{eq:Ydot-sc} which is considered next.  

\subsubsection{Linearisation of the reduced planar system}
\label{subsubsec:lin-sc}

Let $\mathbf Z=(X,Y)^{\mathsf T}$ and denote the planar field in
\eqref{eq:Xdot-sc}--\eqref{eq:Ydot-sc} by $\dot{\mathbf Z}=\mathbf f(\mathbf Z)$.
A direct calculation gives the Jacobian
\begin{equation}\label{eq:J-sc-general}
D\mathbf f(X,Y)=
\begin{pmatrix}
1+2Y^2-aY-3X^2 & X(4Y-a)\\[0.6ex]
2X(a-Y) & -X^2+6Y^2-2
\end{pmatrix}.
\end{equation}

\paragraph{At the kinetic endpoints $K_\pm$.}
Using \eqref{eq:Kpm-sc} in \eqref{eq:J-sc-general} one finds
\begin{equation}\label{eq:J-Kp-sc}
D\mathbf f(K_+)=
\begin{pmatrix}
3-a & 0\\
0 & 4
\end{pmatrix},
\qquad
D\mathbf f(K_-)=
\begin{pmatrix}
a+3 & 0\\
0 & 4
\end{pmatrix}.
\end{equation}

\paragraph{At the curvature point $C$.}
Using \eqref{eq:C-sc} in \eqref{eq:J-sc-general} one finds
\begin{equation}\label{eq:J-C-sc}
D\mathbf f(C)=
\begin{pmatrix}
1 & 0\\
0 & -2
\end{pmatrix}.
\end{equation}

\paragraph{At the scalar point $S$.}
Let $(X_*,Y_*)$ be given by \eqref{eq:S-sc}. Then \eqref{eq:J-sc-general} yields
\begin{equation}\label{eq:J-S-sc}
D\mathbf f(S)=
\begin{pmatrix}
\frac{2}{9}(a^2-9) & \frac{a}{3}\sqrt{1-\frac{a^2}{9}}\\[0.9ex]
\frac{4a}{3}\sqrt{1-\frac{a^2}{9}} & \frac{1}{9}(7a^2-27)
\end{pmatrix}.
\end{equation}

\paragraph{At the curvature--scaling point $P_k$.}
Let $(X_k,Y_k)$ be given by \eqref{eq:Pk-sc}. Then \eqref{eq:J-sc-general} yields
\begin{equation}\label{eq:J-Pk-sc}
D\mathbf f(P_k)=
\begin{pmatrix}
-\frac{4}{a^2} & \frac{\sqrt2\,(4-a^2)}{a^2}\\[0.9ex]
\frac{2\sqrt2\,(a^2-1)}{a^2} & \frac{4}{a^2}-2
\end{pmatrix}.
\end{equation}

\paragraph{Curvature induced organising thresholds.}
In the scalar-only class (Sect.~\ref{subsec:common-quadratic-germ}) the principal loss of hyperbolicity occurs at the $\Gamma_1$-locus, i.e., at  the kinetic endpoints $K_\pm$ when $|a|=3$ (equivalently $|\lambda|=\sqrt6$).  Introducing curvature as a full component
adds a new equilibrium branch $P_k$ with $\Omega_{k,k}=1-3/a^2$, and this generates additional
organising loci in parameter space.  

In particular,  as we showed above, at the scalar point $S$ the curvature direction is normally hyperbolic
for $a^2\neq 3$. However, at $P_k$ curvature participates in the balance and the linear spectrum depends sensitively on $a$.
More precisely, at $P_k$ the trace is constant while the determinant and discriminant vary with $a$:
\begin{equation}\label{eq:Pk-trace-det-disc}
\mathrm{tr}\,D\mathbf f(P_k)=-2,\qquad
\det D\mathbf f(P_k)=4\Bigl(1-\frac{3}{a^2}\Bigr),\qquad
\Delta=\mathrm{tr}^2-4\det=12\Bigl(\frac{4}{a^2}-1\Bigr).
\end{equation}
Thus the $\Gamma_2$-locus $|a|=\sqrt3$ (equivalently $|\lambda|=\sqrt2$) is a genuine non-hyperbolic locus ($\det=0$) (the nature of this locus is investigated more fully below).  On the other hand, the axis $|a|=2$ (equivalently $|\lambda|=2\sqrt{2/3}$) is a \emph{spectral transition} ($\Delta=0$):
the equilibrium remains hyperbolic but the linear type changes from a real node ($a^2<4$) to a spiral sink
($a^2>4$), with a repeated eigenvalue at $a=\pm 2$.

\begin{remark}[the $|a|=2$ axis]
At $a=\pm 2$ the eigenvalues of $D\mathbf f(P_k)$ coalesce to $\lambda=-1$ (so $\Delta=0$),
but $\lambda\neq 0$ and $P_k$ remains hyperbolic.  Hence there is no center manifold
reduction at $|a|=2$; the local dynamics is still topologically conjugate to a sink
(Hartman--Grobman), although the linear normal form becomes non-semisimple (Jordan type)
at the transition.
\end{remark}

\subsubsection[Translated u-forms near the principal equilibria]{Translated $u$-forms near the principal equilibria}
\label{subsubsec:Wforms-sc}
We now translate the principal equilibria to the origin via linear transformations $\mathbf Z\to \mathbf u=(U,V)^{\mathsf T}$.

\paragraph{Kinetic endpoints $K_\pm$.}
Let $K_+$ be translated by $X=U$, $Y=1+V$. Then
\begin{equation}\label{eq:W-jet-Kp-sc}
\binom{\dot U}{\dot V}
=
\underbrace{\begin{pmatrix}3-a&0\\0&4\end{pmatrix}}_{J_{K_+}}
\binom{U}{V}
+
\binom{(4-a)\,UV}{(a-1)\,U^2}
+O(3).
\end{equation}
Similarly, near $K_-$ with $X=U$, $Y=-1+V$ one finds
\begin{equation}\label{eq:W-jet-Km-sc}
\binom{\dot U}{\dot V}
=
\underbrace{\begin{pmatrix}a+3&0\\0&4\end{pmatrix}}_{J_{K_-}}
\binom{U}{V}
+
\binom{(a-4)\,UV}{(a-1)\,U^2}
+O(3).
\end{equation}
In particular, at the endpoint locus $|a|=3$ the $U$-eigendirection becomes non-hyperbolic at the corresponding
$K_\pm$ site, with leading quadratic coupling of triangular type $\dot U\sim U V$.

\paragraph{Curvature point $C$.}
Translate $C$ by $X=U$, $Y=V$. Then
\begin{equation}\label{eq:W-jet-C-sc}
\binom{\dot U}{\dot V}
=
\underbrace{\begin{pmatrix}1&0\\[0.3ex]0&-2\end{pmatrix}}_{J_C}
\binom{U}{V}
+
\binom{-a\,UV}{a\,U^2}
+O(3).
\end{equation}

\paragraph{Scalar point $S$.}
Let $S:(X_*,Y_*)$ be given by \eqref{eq:S-sc} and translate by $X=X_*+U$, $Y=Y_*+V$.
Then
\begin{equation}\label{eq:W-jet-S-sc}
\binom{\dot U}{\dot V}
=
\underbrace{\begin{pmatrix}
\frac{2}{9}(a^2-9) & \frac{a}{3}\sqrt{1-\frac{a^2}{9}}\\[1.0ex]
\frac{4a}{3}\sqrt{1-\frac{a^2}{9}} & \frac{1}{9}(7a^2-27)
\end{pmatrix}}_{J_S}
\binom{U}{V}
+
\binom{-3\sqrt{1-\frac{a^2}{9}}\,U^2+\frac{a}{3}UV+2\sqrt{1-\frac{a^2}{9}}\,V^2}
{\frac{2a}{3}U^2-2\sqrt{1-\frac{a^2}{9}}\,UV+2a\,V^2}
+O(3).
\end{equation}

\paragraph{Curvature--scaling point $P_k$.}
Let $P_k:(X_k,Y_k)$ be given by \eqref{eq:Pk-sc} and translate by $X=X_k+U$, $Y=Y_k+V$.
Then
\begin{equation}\label{eq:W-jet-Pk-sc}
\binom{\dot U}{\dot V}
=
\underbrace{\begin{pmatrix}
-\frac{4}{a^2} & \frac{\sqrt2\,(4-a^2)}{a^2}\\[1.0ex]
\frac{2\sqrt2\,(a^2-1)}{a^2} & \frac{4}{a^2}-2
\end{pmatrix}}_{J_{P_k}}
\binom{U}{V}
+
\binom{-\frac{3\sqrt2}{a}U^2+\left(\frac{4}{a}-a\right)UV+\frac{2\sqrt2}{a}V^2}
{\left(a-\frac{1}{a}\right)U^2-\frac{2\sqrt2}{a}UV+\frac{6}{a}V^2}
+O(3).
\end{equation}

\subsubsection{Quadratic jets and center reductions}
\label{subsubsec:SC-quadratic}

The reduced scalar--curvature field \eqref{eq:Xdot-sc}--\eqref{eq:Ydot-sc} admits four principal
equilibria $K_\pm$, $C$, $S$ and $P_k$.  Their translated $u$-jets
\eqref{eq:W-jet-Kp-sc}--\eqref{eq:W-jet-Pk-sc} provide the quadratic data needed for the unfolding
analysis in later sections.

\paragraph{(i) Kinetic endpoints $K_\pm$ and the $|a|=3$ axis.}
At $K_+$ the $U$-eigenvalue equals $3-a$ and becomes non-hyperbolic at $a=3$; similarly at $K_-$
the $U$-eigenvalue equals $a+3$ and becomes non-hyperbolic at $a=-3$.
The translated jets \eqref{eq:W-jet-Kp-sc}--\eqref{eq:W-jet-Km-sc} show a triangular quadratic
coupling of the form $\dot U\sim U V$ at the endpoint locus, yielding a one-dimensional center
reduction with cubic leading term as in a pitchfork bifurcation.

\noindent\emph{Endpoint cubic reduction at $K_+$ on $a=3$.}
More precisely, at $K_+$ the translated jet \eqref{eq:W-jet-Kp-sc} reads
\begin{equation}\label{eq:jet-Kp-a3}
\dot U=(3-a)U+(4-a)UV+O(3),\qquad
\dot V=4V+(a-1)U^2+O(3).
\end{equation}
On the endpoint axis $a=3$ the Jacobian is $J_{K_+}=\mathrm{diag}(0,4)$, so an eigenbasis is
$e_c=(1,0)^{\mathsf T}$ (center) and $e_s=(0,1)^{\mathsf T}$ (stable).  Writing $\mu:=3-a$,
the center manifold has the form $V=h(U,\mu)=cU^2+O(U^3,\mu U^2)$; substituting into the
invariance equation gives $c=-\tfrac12$ at $\mu=0$, hence
\begin{equation}\label{eq:CM-Kp-sc-explicit}
V=-\frac12U^2+O(U^3,\mu U^2),
\qquad
\dot U=\mu\,U-\frac12U^3+O(U^4,\mu U^3).
\end{equation}
After the amplitude rescaling $U=\sqrt2\,\zeta$ (with the same time variable $N$),
the reduced equation becomes the standard cubic stratification
\begin{equation}\label{eq:pitchfork-Kp}
\dot\zeta=\mu\,\zeta-\zeta^3+O(\zeta^4,\mu\zeta^3).
\end{equation}
A similar calculation gives the analogous result for $K_-$ at $a=-3$, which we include here for completeness.

\noindent\emph{Endpoint cubic reduction at $K_-$ on $a=-3$.}
At $K_-$ the translated jet \eqref{eq:W-jet-Km-sc} reads
\begin{equation}\label{eq:jet-Km-am3}
\dot U=(a+3)U+(a-4)UV+O(3),\qquad
\dot V=4V+(a-1)U^2+O(3).
\end{equation}
On the endpoint axis $a=-3$ the Jacobian is $J_{K_-}=\mathrm{diag}(0,4)$, so an eigenbasis is
$e_c=(1,0)^{\mathsf T}$ and $e_s=(0,1)^{\mathsf T}$. Writing $\mu:=a+3$, the center manifold
has the form $V=h(U,\mu)=cU^2+O(U^3,\mu U^2)$; substituting into the invariance equation gives
$c=1$ at $\mu=0$, hence
\begin{equation}\label{eq:CM-Km-sc-explicit}
V=U^2+O(U^3,\mu U^2),
\qquad
\dot U=\mu\,U-7U^3+O(U^4,\mu U^3).
\end{equation}
After the rescaling $U=\zeta/\sqrt7$ one obtains
\begin{equation}\label{eq:pitchfork-Km}
\dot\zeta=\mu\,\zeta-\zeta^3+O(\zeta^4,\mu\zeta^3).
\end{equation}

\paragraph{(ii) Curvature point $C$.}
The curvature-dominated equilibrium $C$ is hyperbolic for all $a$ (eigenvalues $1$ and $-2$),
so its local phase portrait is structurally stable; the quadratic terms in \eqref{eq:W-jet-C-sc} do not
generate additional organising loci.

\paragraph{(iii) Scalar and curvature--scaling points $S$ and $P_k$: the $|a|=\sqrt3$ axis.}
The two equilibrium branches intersect at $|a|=\sqrt3$, where $S=P_k$ and the Jacobian has a
simple zero eigenvalue (cf.\ \eqref{eq:Pk-trace-det-disc}).  Hence the $\Gamma_2$-locus $|a|=\sqrt3$ is a genuine
non-hyperbolic organising locus distinct from the kinetic endpoint axis $|a|=3$.
In this regime one expects a one-dimensional center reduction whose generic normal form is of
transcritical type, corresponding to an exchange of stability between the $S$ and $P_k$ branches.

\noindent\emph{Intersection of the $S$ and $P_k$ branches.}
At $a^2=3$ one has $X_*=\sqrt{1-\frac{a^2}{9}}=\sqrt{\frac{2}{3}}$ and $Y_*=\frac{a}{3}=\pm\frac{1}{\sqrt3}$,
while for $P_k$ we have $(X_k,Y_k)=\bigl(\frac{\sqrt2}{a},\frac{1}{a}\bigr)$, hence
\be
(X_k,Y_k)=\left(\frac{\sqrt2}{\pm\sqrt3},\,\frac{1}{\pm\sqrt3}\right)
=\left(\sqrt{\frac{2}{3}},\,\pm\frac{1}{\sqrt3}\right)=(X_*,Y_*).
\ee
Therefore $S=P_k$ precisely on the curvature-induced threshold $|a|=\sqrt3$ (equivalently $\Omega_{k,k}=0$),
and this branch intersection naturally organises a transcritical exchange in the reduced scalar--curvature class as we now show.

Using coordinates $(\xi,\eta)$ aligned with the center/stable eigenvectors and
parameter $p:=a-\sqrt3$, the local system takes the form
\begin{equation}\label{transcrit}
\dot\xi = \mu\,\xi - \kappa\,\xi^2 + O(\xi^3,\mu\xi^2),\qquad \dot\eta=-2\eta+O(2),
\end{equation}
with $\mu\propto p$ and $\kappa\neq 0$. Hence the local organising stratification is
transcritical: two equilibrium branches intersect and exchange stability as $a$
crosses $\sqrt3$.

For example, at the curvature-induced intersection $S=P_k$ (i.e.\ $a_0=\sqrt3$) the common equilibrium is
\be
(X_0,Y_0)=\Bigl(\sqrt{\tfrac23},\,\tfrac{1}{\sqrt3}\Bigr).
\ee
The Jacobian $D\mathbf f(X_0,Y_0)$ has eigenvalues $0$ and $-2$, with a convenient choice of
center/stable eigenvectors
\be
e_c=\binom{\sqrt2}{4},\qquad e_s=\binom{-\sqrt2}{2}.
\ee
Writing the translated variables $U=X-X_0$, $V=Y-Y_0$ in the eigenbasis,
\be
\binom{U}{V}=\xi\,e_c+\eta\,e_s
\quad\Longleftrightarrow\quad
\xi=\frac{\sqrt2\,U+V}{6},\ \ \eta=\frac{-2\sqrt2\,U+V}{6},
\ee
and using $\mu:=a^2-3$ as the local unfolding parameter for the axis $|a|=\sqrt3$, the
center-manifold reduction yields the explicit leading-order form
\be
\dot\xi=\frac{2}{3}\,\mu\,\xi+8\sqrt3\,\xi^2+O\!\left(\xi^3,\mu\xi^2,\mu^2\xi\right).
\ee
After the constant rescalings of time and amplitude
\be
\tau=\frac{2}{3}\,N,\qquad \zeta=-12\sqrt3\,\xi,
\ee
this becomes the standard transcritical normal form
$\displaystyle \frac{d\zeta}{d\tau}=\mu\,\zeta-\zeta^2+O(\zeta^3,\mu\zeta^2,\mu^2\zeta)$.

\paragraph{(iv) The $|a|=2$ spectral transition at $P_k$.}
At $|a|=2$ the discriminant $\Delta$ in \eqref{eq:Pk-trace-det-disc} vanishes and the eigenvalues
coalesce at $\lambda=-1$.  This is not a loss of hyperbolicity and therefore does not produce a
center manifold; rather it marks a linear-type transition (node $\leftrightarrow$ spiral) at $P_k$.

\subsubsection{Organising normal forms}
\label{subsubsec:SC-normalforms}
We summarize our findings in this subsection as follows.
The scalar--curvature class is organised by three distinguished loci in the slope
parameter $a=\sqrt{3/2}\,\lambda$: the kinetic endpoint axis $|a|=3$ (i.e., the $\Gamma_1$-locus), the curvature-induced
intersection axis $a^2=3$ - the $\Gamma_2$-locus (where $S$ meets $P_k$), and the hyperbolic spectral transition
$|a|=2$ at $P_k$ (node $\leftrightarrow$ spiral).  The corresponding local normal forms are:

\begin{enumerate}
\item \emph{Endpoint pitchfork bifurcation on the  $\Gamma_1$-locus $|a|=3$.}
Near $K_\pm$ the one-dimensional center reduction is cubic; moreover the boundary collision
$S=K_\pm$ at $a=\pm3$ realises the same reduced dynamics.  A convenient normal form is given by the pitchfork bifurcation
\begin{equation}\label{eq:SC-endpoint-NF}
\dot\zeta=\mu\,\zeta-\zeta^3+O(\zeta^4,\mu\zeta^3),
\qquad \mu\propto(9-a^2),
\end{equation}
cf.\ \eqref{eq:CM-Kp-sc-explicit}--\eqref{eq:pitchfork-Km} and \eqref{eq:S-equals-K-sc}.

\item \emph{Transcritical exchange on the $\Gamma_2$-locus $a^2=3$.}
At the curvature-induced threshold $S=P_k$ a simple zero eigenvalue appears and the
local organising template is transcritical:
\begin{equation}\label{eq:SC-transcrit-NF}
\dot\zeta=\mu\,\zeta-\zeta^2+O(\zeta^3,\mu\zeta^2),
\qquad \mu\propto(a^2-3),
\end{equation}
cf.\ \eqref{transcrit} (and the explicit reduction given after it).

\item \emph{Spectral (hyperbolic) transition at $|a|=2$.}
At $P_k$ the discriminant $\Delta$ in \eqref{eq:Pk-trace-det-disc} vanishes and the
eigenvalues coalesce at $-1$; $P_k$ remains hyperbolic, so there is no center reduction.
This locus separates the real-node and spiral-sink regimes.
\end{enumerate}

\subsection{The scalar--fluid--curvature system}
\label{subsec:scalar-fluid-curv}
The main results of this subsection are summarised in Theorem~\ref{thm:sec6-sfc}.

\begin{theorem}[Organising loci and normal forms: scalar--fluid--curvature class]
\label{thm:sec6-sfc}
Consider the scalar--fluid--curvature constrained system on the Friedmann hypersurface
\[
X^2+Y^2+\Omega_m+\Omega_k=1,
\]
with parameters $(\gamma,a)$ where $a=\sqrt{3/2}\,\lambda$ and $p_m=(\gamma-1)\rho_m$.
Then the organising structure contains:

\begin{itemize}
\item \textbf{Inherited organising loci:}
the  $\Gamma_1$-locus (i.e., kinetic endpoint axis $|a|=3$, equivalently $|\lambda|=\sqrt6$) persists as a distinguished
organising locus, with the local reduced dynamics at the corresponding endpoint sites governed by
one-dimensional centre reductions and/or $\mathbb{Z}_2$-equivariant hilltop jets (depending on the
$\gamma$-regime).

\item \textbf{Curvature induced organising locus:}
the  $\Gamma_2$-locus, i.e., curvature-scaling branch intersecting the scalar/scaling branches on a curvature induced axis
analogous to the $a^2=3$ threshold in the scalar--curvature class, producing a simple zero eigenvalue
and a transcritical-type local stratification (branch intersection and stability exchange).

\item \textbf{Additional $(\gamma,a)$-dependent degeneracies:}
special values of $\gamma$ (notably $\gamma=2$ and $\gamma=2/3$) produce further non-hyperbolic episodes
associated with loss of isolation and/or changes in the effective codimension of the organising germs. These include the loci $\Gamma_4$, $\Gamma_5$ and in particular the $\Gamma_4$-segment $E_{2/3}$ and the $\Gamma_5$-arc $E_{2}$.

\item \textbf{Normal form templates:}
the organising local models required for the subsequent stratification/unfolding analysis are exhausted by
(i) cubic one-dimensional centre reductions (pitchfork-type),
(ii) quartic one-dimensional centre reductions (fold/winged-cusp type) when the first non-removable term is quartic,
(iii) planar $\mathbb{Z}_2$-equivariant hilltop jets (codimension at most $3$), and
(iv) transcritical exchanges at branch-intersection loci.
\end{itemize}
\end{theorem}
\noindent
\emph{Interpretation.} As in the FL setting \cite{CotsakisFL}, the fluid parameter $\gamma$ acts primarily as a \emph{class selector}:
for fixed generic $\gamma$ the unfolding parameters organise motion \emph{within} a given bifurcation diagram,
whereas isolated $\gamma$ values (notably $\gamma=\tfrac23$ here) change the local organising geometry by altering
centre dimensions or even producing non-isolated equilibrium sets.

\subsubsection{Constraint and 3D constrained system}
We retain $\Omega_m$ and $\Omega_k$ subject to the Friedmann constraint
\begin{equation}\label{eq:constraint-sfc}
X^2+Y^2+\Omega_m+\Omega_k=1.
\end{equation}
Eliminating the fluid fraction,
\begin{equation}\label{eq:Om-elim-sfc}
\Omega_m = 1 - X^2 - Y^2 - \Omega_k,
\end{equation}
yields an equivalent constrained 3D system in $(X,Y,\Omega_k)$:
\begin{equation}\label{eq:Xdot-sfc}
\dot X=
X\!\left(3Y^2+\frac{3\gamma}{2}\bigl(1-X^2-Y^2-\Omega_k\bigr)+\Omega_k-aY\right),
\end{equation}
\begin{equation}\label{eq:Ydot-sfc}
\dot Y=
-3Y+aX^2+3Y^3+\frac{3\gamma}{2}Y\bigl(1-X^2-Y^2-\Omega_k\bigr)+Y\Omega_k,
\end{equation}
\begin{equation}\label{eq:Omkdot-sfc}
\dot\Omega_k=
2\Omega_k\!\left(3Y^2+\frac{3\gamma}{2}\bigl(1-X^2-Y^2-\Omega_k\bigr)+\Omega_k-1\right).
\end{equation}

\subsubsection{Principal equilibria}
The equilibria of \eqref{eq:Xdot-sfc}--\eqref{eq:Omkdot-sfc} consist of the following
equilibrium branches for generic $(\gamma,a)$, while additional non-isolated equilibrium
sets occur only on special parameter loci, the  $\Gamma_4$ and  $\Gamma_5$ loci (notably at $\gamma=2/3$ and $\gamma=2$ respectively):
\begin{equation}\label{eq:Kpm-sfc}
K_\pm:\ (X,Y,\Omega_k)=(0,\pm 1,0),\qquad \Omega_m=0,
\end{equation}
\begin{equation}\label{eq:M-sfc}
M:\ (X,Y,\Omega_k)=(0,0,0),\qquad \Omega_m=1,
\end{equation}
\begin{equation}\label{eq:C-sfc}
C:\ (X,Y,\Omega_k)=(0,0,1),\qquad \Omega_m=0,
\end{equation}
\begin{equation}\label{eq:S-sfc}
S:\ (X_*,Y_*,\Omega_{k,*})=\left(\sqrt{1-\frac{a^2}{9}},\,\frac{a}{3},\,0\right),
\qquad \Omega_{m,*}=0,\qquad (|a|\le 3),
\end{equation}
\begin{equation}\label{eq:F-sfc}
F:\ (X_F,Y_F,\Omega_{k,F})=
\left(\sqrt{\frac{9\gamma(2-\gamma)}{4a^2}},\,\frac{3\gamma}{2a},\,0\right),
\qquad
\Omega_{m,F}=1-\frac{9\gamma}{2a^2},
\end{equation}
existing for $a^2\ge \tfrac{9}{2}\gamma$ (the  $\Gamma_3$-locus is at  $a^2= \tfrac{9}{2}\gamma$), and notably
\begin{equation}\label{eq:Pk-sfc}
P_k:\ (X,Y,\Omega_k)=\left(\frac{\sqrt2}{a},\,\frac{1}{a},\,1-\frac{3}{a^2}\right),
\qquad \Omega_m=0,
\end{equation}
the scalar-curvature equilibrium branch persists also here for $a\neq 0$ and is physically relevant as a negative-curvature site when
$\Omega_{k,k}>0$, i.e.\ $a^2>3$ (equivalently $\lambda^2>2$, or $|\lambda|>\sqrt2$).

\begin{remark}[Non-isolated equilibria on special $\gamma$-loci]\label{rem:special-loci}
For generic values of $\gamma$ the equilibria listed above exhaust the isolated equilibrium
branches of the constrained system \eqref{eq:Xdot-sfc}--\eqref{eq:Omkdot-sfc}.
However, on two distinguished $\gamma$-values the flow develops \emph{non-isolated} equilibrium
sets (equilibrium continua), reflecting additional degeneracy not present in the generic case.

\smallskip
\noindent\textbf{(i) The  $\Gamma_4$-locus at $\gamma=\tfrac{2}{3}$.}
At $\gamma=\tfrac{2}{3}$ the curvature direction loses hyperbolicity at the fluid/scaling
sites, since the $\Omega_k$-eigenvalue equals $3\gamma-2=0$ (cf.\ \eqref{eq:J-M-sfc} and
\eqref{eq:J-F-sfc}). In particular, the segment
\be 
X=0,\qquad Y=0,\qquad \Omega_m+\Omega_k=1
\ee
is a line of equilibria connecting the points $M$ and $C$.
Moreover, at the curvature--scaling location $(X,Y)=(\sqrt2/a,\,1/a)$ one may satisfy
$\dot\Omega_k=0$ identically when $\gamma=\tfrac{2}{3}$, so that (subject to the constraint)
a one-parameter family of equilibria appears with $\Omega_k$ free and
$\Omega_m=1-\frac{3}{a^2}-\Omega_k$ adjusted accordingly (restricted to the physical range
$\Omega_m,\Omega_k\ge 0$).

\smallskip
\noindent\textbf{(ii) The  $\Gamma_5$-locus at $\gamma=2$.}
On the invariant face $\Omega_k=0$ (so $\Omega_m=1-X^2-Y^2$), the restriction of
\eqref{eq:Xdot-sfc}--\eqref{eq:Omkdot-sfc} at $\gamma=2$ satisfies
$\dot Y|_{X=0,\Omega_k=0}\equiv 0$ and $\dot X|_{X=0,\Omega_k=0}\equiv 0$.
Hence the entire arc
\be 
X=0,\qquad |Y|\le 1,\qquad \Omega_k=0,\qquad \Omega_m=1-Y^2
\ee
consists of equilibria, so the fluid point $M$ is not isolated on $\gamma=2$.

\end{remark}

\subsubsection[Translated u-forms near the principal equilibria]{Translated $\mathbf{u}$-forms near the principal equilibria}

Let $\mathcal{E}_*=(X_*,Y_*,\Omega_{k,*})$ be any equilibrium of \eqref{eq:Xdot-sfc}--\eqref{eq:Omkdot-sfc},
and translate
\begin{equation}\label{eq:W-translation-sfc}
X=X_*+U,\qquad Y=Y_*+V,\qquad \Omega_k=\Omega_{k,*}+W,\qquad \mathbf u=(U,V,W)^{\mathsf T}.
\end{equation}
Then near $\mathcal{E}_*$, the system \eqref{eq:Xdot-sfc}-\eqref{eq:Omkdot-sfc} acquires the  form,
\begin{equation}\label{eq:W-jet-form-sfc}
\dot{\mathbf u}=J_*\mathbf u+Q_*(\mathbf u)+O\!\left(\|\mathbf u\|^3\right).
\end{equation}

At $K_+$ one finds
\begin{equation}\label{eq:J-Kp-sfc}
J_{K_+}=
\begin{pmatrix}
3-a & 0 & 0\\
0 & 6-3\gamma & 1-\frac{3\gamma}{2}\\
0 & 0 & 4
\end{pmatrix},
\end{equation}
\begin{equation}\label{eq:Q-Kp-sfc}
Q_{K_+}(\mathbf{u})=
\begin{pmatrix}
(6-a-3\gamma)\,UV+\left(1-\frac{3\gamma}{2}\right)UW\\[0.6ex]
\left(a-\frac{3\gamma}{2}\right)U^2+\left(9-\frac{9\gamma}{2}\right)V^2+\left(1-\frac{3\gamma}{2}\right)VW\\[0.6ex]
(12-6\gamma)\,VW+(2-3\gamma)\,W^2
\end{pmatrix}.
\end{equation}
At $K_-$ one finds
\begin{equation}\label{eq:J-Km-sfc}
J_{K_-}=
\begin{pmatrix}
a+3 & 0 & 0\\
0 & 6-3\gamma & \frac{3\gamma}{2}-1\\
0 & 0 & 4
\end{pmatrix},
\end{equation}
\begin{equation}\label{eq:Q-Km-sfc}
Q_{K_-}(\mathbf{u})=
\begin{pmatrix}
(3\gamma-a-6)\,UV+\left(1-\frac{3\gamma}{2}\right)UW\\[0.6ex]
\left(a+\frac{3\gamma}{2}\right)U^2+\left(-9+\frac{9\gamma}{2}\right)V^2+\left(1-\frac{3\gamma}{2}\right)VW\\[0.6ex]
(6\gamma-12)\,VW+(2-3\gamma)\,W^2
\end{pmatrix}.
\end{equation}

At $M$ one finds
\begin{equation}\label{eq:J-M-sfc}
J_M=
\mathrm{diag}\!\left(\frac{3\gamma}{2},\,\frac{3\gamma}{2}-3,\,3\gamma-2\right),
\end{equation}
\begin{equation}\label{eq:Q-M-sfc}
Q_M(\mathbf{u})=
\begin{pmatrix}
-a\,UV+\left(1-\frac{3\gamma}{2}\right)UW\\[0.6ex]
a\,U^2+\left(1-\frac{3\gamma}{2}\right)VW\\[0.6ex]
(2-3\gamma)\,W^2
\end{pmatrix}.
\end{equation}

At $C$ one finds
\begin{equation}\label{eq:J-C-sfc}
J_C=\mathrm{diag}\!\left(1,\,-2,\,2-3\gamma\right),
\end{equation}
\begin{equation}\label{eq:Q-C-sfc}
Q_C(\mathbf{u})=
\begin{pmatrix}
-a\,UV+\left(1-\frac{3\gamma}{2}\right)UW\\[0.6ex]
a\,U^2+\left(1-\frac{3\gamma}{2}\right)VW\\[0.6ex]
-3\gamma\,U^2+(6-3\gamma)\,V^2+(2-3\gamma)\,W^2
\end{pmatrix}.
\end{equation}

At $S$ one finds
\begin{equation}\label{eq:J-S-sfc}
J_S=
\begin{pmatrix}
-3\gamma X_*^{2} & a(1-\gamma)X_* & \frac{1}{2}(2-3\gamma)X_*\\[0.6ex]
a(2-\gamma)X_* & a^{2}\!\left(1-\frac{\gamma}{3}\right)-3 & \frac{a}{6}(2-3\gamma)\\[0.6ex]
0 & 0 & \frac{2}{3}(a^{2}-3)
\end{pmatrix},
\end{equation}
and the quadratic jet can be written as
\begin{equation}\label{eq:Q-S-sfc}
Q_S(\mathbf{u})=
\begin{pmatrix}
-\frac{9}{2}\gamma X_*\,U^2 + a(1-\gamma)\,UV - \frac{1}{2}(3\gamma-2)\,UW
+ \frac{3}{2}(2-\gamma)X_*\,V^2\\[0.8ex]
\frac{a}{2}(2-\gamma)\,U^2 - 3\gamma X_*\,UV + \frac{3a}{2}(2-\gamma)\,V^2
- \frac{1}{2}(3\gamma-2)\,VW\\[0.8ex]
-6\gamma X_*\,UW + 2a(2-\gamma)\,VW - (3\gamma-2)\,W^2
\end{pmatrix}.
\end{equation}

\begin{remark} At $|a|=3$ one has $X_*=0$ and the scalar branch $S$ meets the
corresponding kinetic endpoint $K_\pm$ (with $a=\pm 3$), so the $S$-jet above degenerates
in the same multi-site $|a|=3$ episode.
\end{remark}
At $F$ one finds
\begin{equation}\label{eq:J-F-sfc}
J_F=
\begin{pmatrix}
-3\gamma X_F^2 & X_F\!\left(3(2-\gamma)Y_F-a\right) & \left(1-\frac{3\gamma}{2}\right)X_F\\[0.8ex]
X_F\!\left(2a-3\gamma Y_F\right) & 3\gamma X_F^2+\frac{3}{2}(\gamma-2) & \left(1-\frac{3\gamma}{2}\right)Y_F\\[0.8ex]
0 & 0 & 3\gamma-2
\end{pmatrix},
\end{equation}
and the quadratic jet may be written compactly as
\begin{equation}\label{eq:Q-F-sfc}
Q_F(\mathbf{u})=
\begin{pmatrix}
-\frac{9\gamma}{2}X_F\,U^2+\left(3(2-\gamma)Y_F-a\right)UV+\left(1-\frac{3\gamma}{2}\right)UW
+3X_F\!\left(1-\frac{\gamma}{2}\right)V^2\\[0.9ex]
\left(a-\frac{3\gamma}{2}Y_F\right)U^2-3\gamma X_F\,UV
+9Y_F\!\left(1-\frac{\gamma}{2}\right)V^2+\left(1-\frac{3\gamma}{2}\right)VW\\[0.9ex]
-6\gamma X_F\,UW+6(2-\gamma)Y_F\,VW+(2-3\gamma)W^2
\end{pmatrix}.
\end{equation}
\begin{remark} The curvature direction at $F$ is governed by the eigenvalue
$3\gamma-2$ in \eqref{eq:J-F-sfc}, so $F$ is hyperbolic in the $\Omega_k$-direction (i.e., transverse to the invariant submanifold $\Omega_k=0$)  except at
$\gamma=\tfrac{2}{3}$, where curvature becomes an additional center direction.
\end{remark}

At $P_k$ one finds

\begin{equation}\label{eq:J-Pk-sfc}
J_{P_k}=
\begin{pmatrix}
-\dfrac{3\gamma}{a^2} &
\dfrac{\sqrt2}{a^2}\Bigl(a^2(2-\gamma)+3\gamma\Bigr) &
\dfrac{\sqrt2}{2a}(2-3\gamma)
\\[1.1ex]
2\sqrt2\,(2-\gamma) &
\dfrac{a^2(2-\gamma)+6}{a^2} &
\dfrac{2-3\gamma}{a}
\\[1.1ex]
-\dfrac{6\sqrt2\,\gamma\,(a^2-3)}{a^3} &
\dfrac{6\gamma\,(a^2-3)\,\bigl(a^2(\gamma-2)+3\gamma\bigr)}{a^4} &
\dfrac{(a^2-3)(2-3\gamma)}{a^2}
\end{pmatrix}.
\end{equation}

\begin{equation}\label{eq:Q-Pk-sfc}
Q_{P_k}(\mathbf{u})=
\begin{pmatrix}
-\dfrac{3\sqrt2\,\gamma}{a}\,U^2
+\dfrac{3a^2(\gamma-4)+9\gamma}{a^2}\,UV
+\dfrac{2-3\gamma}{2}\,UW
+\dfrac{\sqrt2(2-\gamma)}{a}\,V^2
\\[1.1ex]
\dfrac{a^2(2-\gamma)+3\gamma}{a}\,U^2
+\dfrac{3\sqrt2(2-\gamma)}{a}\,UV
+\Bigl(9(2-\gamma)+\dfrac{18}{a^2}\Bigr)V^2
+\dfrac{2-3\gamma}{2}\,VW
\\[1.1ex]
-\dfrac{6\sqrt2\,\gamma}{a}\,U^2
+\Bigl(12(\gamma-2)+\dfrac{36\gamma}{a^2}\Bigr)UV
+\Bigl(12(\gamma-2)+\dfrac{18\gamma}{a^2}\Bigr)UW
+(12\gamma-12)V^2
\\
\hspace{6em}
+\Bigl(12(\gamma-2)+\dfrac{18\gamma}{a^2}\Bigr)VW
+(2-3\gamma)W^2
\end{pmatrix}.
\end{equation}

\subsubsection{Jets on the  $\Gamma_4$, $\Gamma_5$ loci}
\label{subsubsec:special-gamma-jets}

On $\gamma=2$ and $\gamma=\tfrac23$ the constrained flow admits non-isolated equilibrium
sets (Remark~\ref{rem:special-loci}).  For later reference we record the linear part and
leading quadratic couplings at representative points on these continua.  These jets
identify the additional center directions responsible for the loss of isolation.

\paragraph{(i) The equilibrium line on the $\Gamma_4$-locus (the $M$--$C$ segment).}
On $\gamma=\tfrac23$ the segment
\be
E_{2/3}:\quad X=0,\ Y=0,\ \Omega_m+\Omega_k=1,\qquad \Omega_k\in[0,1],
\ee
consists of equilibria connecting $M$ and $C$.  Fix $\Omega_{k0}\in[0,1]$ and translate by
$X=U$, $Y=V$, $\Omega_k=\Omega_{k0}+W$.
Then the Jacobian has a zero eigenvalue in the $W$-direction (tangent to the line),
reflecting that the curvature eigenvalue $3\gamma-2$ vanishes at $\gamma=\tfrac23$.
The remaining two eigenvalues describe transverse stability of the line and coincide with
the $2\times 2$ block obtained by restricting to $(U,V)$ (these results follow from Eqns. \eqref{eq:J-M-sfc}, \eqref{eq:J-C-sfc}). We note that the quadratic blocks restricted to $(U,V)$ also coincide.

\smallskip
\noindent\emph{Practical use.}  In later sections $\Gamma_4$ is treated as the locus
where curvature becomes an additional center direction at the fluid/scaling sites.

\paragraph{(ii) The equilibrium arc on the  $\Gamma_5$-locus ($\gamma=2$, $\Omega_k=0$).}
Fix a point on the arc
\be 
E_2:\quad X=0,\ \Omega_k=0,\ \Omega_m=1-Y^2,\qquad |Y|\le 1,
\ee
and write $Y_0\in[-1,1]$ for the chosen point.
Translate by $X=U$, $Y=Y_0+V$, $\Omega_k=W$ (so $\mathbf u=(U,V,W)^{\mathsf T}$).
Then the linearisation has at least one zero eigenvalue (tangent to the arc), and the
$t$-direction of $E_2$ is generated by $\partial_Y$; in particular the Jacobian has a
vanishing eigenvalue corresponding to the $V$-direction.  The remaining two eigenvalues
determine transverse stability of the arc and may change character as $Y_0$ varies. For example, the linearized Jacobian becomes $J_M=
\mathrm{diag}\!\left(3,0,4\right),$ when $Y_0=0$.

\smallskip
\noindent\emph{Practical use.}  In the stratification analysis we treat $\Gamma_5$ as a
\emph{loss-of-isolation} locus: the local normal form is organised by transverse dynamics
to the arc rather than by an isolated-equilibrium unfolding.

We proceed now to more details about the jets of the S-F-C problem.

\subsubsection{Quadratic jets and centre reductions}
\label{subsubsec:jets-sfc}

We work with the translated variables \eqref{eq:W-translation-sfc} and the jet form
\eqref{eq:W-jet-form-sfc}. For $K_\pm$, $M$, and $C$ the pairs $(J_*,Q_*)$ are given explicitly
in \eqref{eq:J-Kp-sfc}--\eqref{eq:Q-C-sfc}. We record here the remaining jets (at $S$, $F$, and $P_k$),
and summarize the centre reductions on the non-hyperbolic loci.

\paragraph{(i) Kinetic endpoints $K_\pm$.}
From \eqref{eq:J-Kp-sfc}--\eqref{eq:J-Km-sfc} the eigenvalues of the kinetic Jacobians are
\begin{equation}\label{eq:eigs-Kp-sfc}
\mathrm{eig}(J_{K_+})=\{\,3-a,\ 6-3\gamma,\ 4\,\},
\end{equation}
\begin{equation}\label{eq:eigs-Km-sfc}
\mathrm{eig}(J_{K_-})=\{\,a+3,\ 6-3\gamma,\ 4\,\}.
\end{equation}
Hence $K_\pm$ are non-hyperbolic on $\Gamma_1$-locus, i.e., the kinetic axis $a=\pm 3$ (equivalently $|\lambda|=\sqrt6$),
and also on the $\Gamma_5$-locus, the vertical axis $\gamma=2$ (where $6-3\gamma=0$ gives an additional zero eigenvalue).

On the generic kinetic axis $a=\pm3$ with $\gamma\neq 2$, the centre manifold is one-dimensional
(spanned by the $U$-direction) and has the leading-order graph
\begin{equation}\label{eq:CM-Kpm-sfc}
V=-\frac12\,U^2+O(U^4),\qquad W=O(U^4).
\end{equation}
Substituting into \eqref{eq:Q-Kp-sfc} (or \eqref{eq:Q-Km-sfc}) yields the reduced cubic normal form
\begin{equation}\label{eq:red-Kpm-sfc}
\dot U = -\frac{3}{2}\,U^3 + O(U^5)\qquad (a=\pm 3,\ \gamma\neq 2).
\end{equation}

\paragraph{(ii) Fluid point $M$.}
From \eqref{eq:J-M-sfc}, the eigenvalues of $J_M$ are
\begin{equation}\label{eq:eigs-M-sfc}
\mathrm{eig}(J_M)=\left\{\frac{3\gamma}{2},\ \frac{3\gamma}{2}-3,\ 3\gamma-2\right\},
\end{equation}
hence $M$ is non-hyperbolic on three vertical loci
\begin{equation}\label{eq:loci-M-sfc}
\gamma=0,\qquad \gamma=\frac23,\qquad \gamma=2.
\end{equation}
At $\gamma=0$ there is a one-dimensional centre direction ($U$) and a cubic reduction analogous to the
scalar--fluid case:
\begin{equation}\label{eq:red-M-g0-sfc}
V=\frac{a}{3}\,U^2+O(U^4),\qquad W=O(U^4),\qquad
\dot U = -\frac{a^2}{3}\,U^3 + O(U^5)\qquad (\gamma=0).
\end{equation}
At $\gamma=2$ the centre direction is $V$ and the first nontrivial term comes from the intrinsic $3Y^3$
nonlinearity, giving
\begin{equation}\label{eq:red-M-g2-sfc}
\dot V = 3\,V^3 + O(V^5)\qquad (\gamma=2).
\end{equation}
At the new axis $\gamma=\tfrac23$ the situation is more rigid: the entire $(0,0,\Omega_k)$ segment on the
constraint becomes equilibria. Indeed, setting $X=Y=0$ in \eqref{eq:Xdot-sfc}--\eqref{eq:Omkdot-sfc} gives
\begin{equation}\label{eq:XY0-sfc}
\dot X=\dot Y=0,\qquad
\dot\Omega_k = 2\Omega_k\Big(\frac{3\gamma}{2}(1-\Omega_k)+\Omega_k-1\Big),
\end{equation}
and for $\gamma=\tfrac23$ the bracket vanishes identically, hence
\begin{equation}\label{eq:line-MC-g23}
\mathcal L_{MC}:=\{(X,Y,\Omega_k)=(0,0,W):0\le W\le 1\}
\quad\text{is a line of equilibria when }\gamma=\frac23.
\end{equation}

\paragraph{(iii) Curvature point $C$.}
From \eqref{eq:J-C-sfc}, the eigenvalues of $J_C$ are
\begin{equation}\label{eq:eigs-C-sfc}
\mathrm{eig}(J_C)=\{1,-2,2-3\gamma\},
\end{equation}
so $C$ is non-hyperbolic precisely on the same vertical axis $\gamma=\tfrac23$, where it lies on the equilibrium
line \eqref{eq:line-MC-g23}.

\paragraph{(iv) Scalar point $S$.}
Let $S=(X_*,Y_*,0)$ with
\begin{equation}\label{eq:S-coords-sfc}
X_*=\sqrt{1-\frac{a^2}{9}},\qquad Y_*=\frac{a}{3}\qquad(|a|\le 3).
\end{equation}
The Jacobian at $S$ has eigenvalues
\begin{equation}\label{eq:eigs-S-sfc}
\mathrm{eig}(J_S)=\left\{\frac{a^2-9}{3},\ \frac{2(a^2-3)}{3},\ \frac{2a^2-9\gamma}{3}\right\}.
\end{equation}
Thus $S$ becomes non-hyperbolic on (i) the $\Gamma_1$-locus, i.e., the endpoint axis $|a|=3$ (where $S$ meets $K_\pm$),
(ii) the curvature-induced axis $\Gamma_2$, $a^2=3$ (equivalently $\lambda^2=2$),
and (iii) the scalar--fluid collision locus $\Gamma_3$,  $a^2=\tfrac92\gamma$ (equivalently $\lambda^2=3\gamma$),
where $S$ meets $F$.

Writing $X=X_*+U$, $Y=Y_*+V$, $\Omega_k=W$, we have
\begin{equation}\label{eq:J-S-sfc-3d}
J_S=
\begin{pmatrix}
\frac{\gamma}{3}(a^2-9) & \frac{a}{3}(1-\gamma)\sqrt{9-a^2} & \left(1-\frac{3\gamma}{2}\right)\sqrt{1-\frac{a^2}{9}}\\[0.6ex]
\frac{a}{3}(2-\gamma)\sqrt{9-a^2} & a^2\!\left(1-\frac{\gamma}{3}\right)-3 & 0\\[0.6ex]
0&0&\frac{2}{3}(a^2-3)
\end{pmatrix}.
\end{equation}
and the quadratic jet takes the form
\begin{equation}\label{eq:Q-S-sfc-3d}
Q_S(\mathbf u)=
\begin{pmatrix}
\begin{aligned}
&-\frac{3\gamma}{2}\sqrt{9-a^2}\,U^2 + a(1-\gamma)\,UV \\
&\quad +\left(1-\frac{\gamma}{2}\right)\sqrt{9-a^2}\,V^2
+\left(1-\frac{3\gamma}{2}\right)UW
\end{aligned}
\\[1.0ex]
\begin{aligned}
&a\!\left(1-\frac{\gamma}{2}\right)U^2 - \gamma\sqrt{9-a^2}\,UV
+ 3a\!\left(1-\frac{\gamma}{2}\right)V^2 \\
&\quad +\left(1-\frac{3\gamma}{2}\right)VW
\end{aligned}
\\[1.0ex]
\begin{aligned}
&-2\gamma\sqrt{9-a^2}\,UW - 2a(\gamma-2)\,VW
- (3\gamma-2)\,W^2
\end{aligned}
\end{pmatrix}.
\end{equation}

On the curvature-induced axis $\Gamma_2:\, a^2=3$ the curvature eigenvalue $\frac{2}{3}(a^2-3)$ in \eqref{eq:eigs-S-sfc}
vanishes, while the other two eigenvalues are nonzero for generic $\gamma$ (away from further coincidences).
Restricting the translated system to the centre direction and using $\mu:=a^2-3$ as the local organising parameter,
one obtains the one-dimensional reduction
\begin{equation}\label{eq:red-S-curvaxis-sfc}
\dot W=\frac{2}{3}\mu\,W-(3\gamma-2)\,W^2+O\!\left(W^3,\mu W^2\right).
\end{equation}
In particular, on the axis $\mu=0$ this begins quadratically,
\begin{equation}\label{eq:red-S-a2eq3-sfc}
\dot W=-(3\gamma-2)\,W^2+O(W^3)\qquad (a^2=3).
\end{equation}
For $\gamma\neq \tfrac23$ the quadratic coefficient is nonzero and the drift is fold-type along the curvature mode.
At the isolated value $\gamma=\tfrac23$ the $W^2$ coefficient vanishes and one must retain the next nontrivial order.
Moreover, \eqref{eq:red-S-curvaxis-sfc} is the local normal form for the $S\leftrightarrow P_k$ branch-intersection:
for $\mu\neq 0$ it has the additional equilibrium branch
\begin{equation}\label{eq:Pk-branch-from-red}
W=\frac{\frac{2}{3}\mu}{3\gamma-2}+O(\mu^2),
\end{equation}
corresponding to the curvature-scaling site $P_k$ emanating from $W=0$.

\paragraph{(v) Scalar--fluid scaling point $F$.}
Let $F=(X_F,Y_F,0)$ with
\begin{equation}\label{eq:F-coords-sfc}
X_F=\sqrt{\frac{9\gamma(2-\gamma)}{4a^2}},\qquad Y_F=\frac{3\gamma}{2a},\qquad
(a^2\ge\tfrac92\gamma).
\end{equation}
The curvature eigenvalue is $3\gamma-2$, so $F$ acquires an additional centre direction on the vertical axis
$\gamma=\tfrac23$. The remaining two eigenvalues are the roots of
\begin{equation}\label{eq:charpoly-F2-sfc}
\lambda^2-\mathrm{tr}(J_F^{(2)})\,\lambda+\det(J_F^{(2)})=0,
\end{equation}
with
\begin{equation}\label{eq:trdet-F2-sfc}
\mathrm{tr}(J_F^{(2)})=\frac{3}{2}(\gamma-2),\qquad
\det(J_F^{(2)})=-\frac{9\gamma(\gamma-2)(2a^2-9\gamma)}{4a^2}.
\end{equation}
Hence one eigenvalue vanishes on the collision line $\Gamma_3$,  $2a^2-9\gamma=0$ (i.e.\ $\lambda^2=3\gamma$), where $F$ meets $S$.

Translating $X=X_F+U$, $Y=Y_F+V$, $\Omega_k=W$, the quadratic part may be written compactly in terms of $X_F$
(and $Y_F$ if desired) as
\begin{equation}\label{eq:Q-F-sfc-3d}
Q_F(\mathbf u)=
\begin{pmatrix}
\begin{aligned}
&-\frac{9}{2}\gamma X_F\,U^2 + a(2X_F^2-1)\,UV
+ 3X_F\!\left(1-\frac{\gamma}{2}\right)V^2 \\
&\quad +\left(1-\frac{3\gamma}{2}\right)UW
\end{aligned}
\\[1.0ex]
\begin{aligned}
&a(1-Y_F^2)\,U^2 - 3\gamma X_F\,UV + 3aX_F^2\,V^2 \\
&\quad +\left(1-\frac{3\gamma}{2}\right)VW
\end{aligned}
\\[1.0ex]
\begin{aligned}
&-6\gamma X_F\,UW + 4aX_F^2\,VW + (2-3\gamma)\,W^2
\end{aligned}
\end{pmatrix}.
\end{equation}

At the $\Gamma_4$ line $\gamma=\tfrac23$ the curvature direction is centre and, moreover, the $W$-drift starts at cubic order
(the $W^2$ coefficient vanishes); this degeneracy is higher-order than the fold-type reduction \eqref{eq:red-S-a2eq3-sfc}
seen at $S$ on the $\Gamma_2$ locus $a^2=3$.

\paragraph{(vi) Curvature--scaling point $P_k$.}
The jet data $(J_{P_k},Q_{P_k})$ are given by \eqref{eq:J-Pk-sfc}--\eqref{eq:Q-Pk-sfc}.
The point $P_k$ is physically relevant when $\Omega_{k,k}>0$, i.e.\ $a^2>3$, and it collides with the scalar branch
on the curvature threshold $a^2=3$ (where $\Omega_{k,k}=0$ and $P_k$ meets $S$).

A convenient way to localize the non-hyperbolic regimes at $P_k$ is via the determinant:
\begin{equation}\label{eq:det-JPk-sfc}
\det(J_{P_k})=\frac{(a^2-3)(3\gamma-2)}{a^{7}}\;\mathcal P_{P_k}(a,\gamma),
\end{equation}
where
\begin{align}\label{eq:Ppoly-JPk-sfc}
\mathcal P_{P_k}(a,\gamma)
={}&\,4a^{5}\gamma^{2}-16a^{5}\gamma+16a^{5}
+12a^{4}\gamma^{3}-48a^{4}\gamma^{2}+48a^{4}\gamma
-21a^{3}\gamma^{2}+42a^{3}\gamma \nonumber\\
&\quad+18a^{2}\gamma^{3}-36a^{2}\gamma^{2}
+36a\,\gamma^{2}-18a\,\gamma-54\gamma^{3}.
\end{align}
Hence $P_k$ is necessarily non-hyperbolic on the loci $\Gamma_2, \Gamma_4$, i.e., on the curvature threshold $a^2=3$ (branch collision with $S$) and on the
vertical axis $\gamma=\tfrac23$ (loss of normal hyperbolicity in the curvature sector), and may admit additional
isolated degeneracies only on the residual curve $\mathcal P_{P_k}(a,\gamma)=0$.

On the collision locus $a^2=3$ the spectrum simplifies: the Jacobian has a simple zero eigenvalue and the remaining two
eigenvalues are real and nonzero for $\gamma\in[0,2]$,
\begin{equation}\label{eq:eigs-Pk-a2eq3-sfc}
\mathrm{eig}(J_{P_k})=\Bigl\{\,0,\ 2-\gamma\pm 2\sqrt{\,5-2\gamma\,}\Bigr\}\qquad (a^2=3).
\end{equation}
Thus, generically, the centre manifold at $P_k$ on $a^2=3$ is one-dimensional and organises the same local
branch-intersection episode $S\leftrightarrow P_k$ seen at the scalar site.

On the vertical axis $\gamma=\tfrac23$ one again has a simple zero eigenvalue at $P_k$, with the other two eigenvalues
remaining hyperbolic for $a^2>3$; this is the curvature-direction degeneracy already signalled by the equilibrium line
$\mathcal L_{MC}$ at $(X,Y)=(0,0)$ (cf.\ \eqref{eq:line-MC-g23}), now reflected in the $P_k$-site linearisation.

\subsubsection{Normal forms and organising axes}
\label{subsubsec:nf-sfc}

In conclusion, the scalar--fluid--curvature class exhibits several distinct organising loci in the $(\gamma,a)$ (or $(\gamma,\lambda)$)
parameter plane. The jets in \S\ref{subsubsec:jets-sfc} show that the qualitative transitions are governed by
a small set of centre reductions, which we summarize as follows.

\paragraph{1) Kinetic endpoint axis $\Gamma_1$ ($|a|=3$, i.e.\ $|\lambda|=\sqrt6$).}
For $\gamma\neq 2$ the endpoints $K_\pm$ have a one-dimensional centre manifold and the reduced cubic normal form
\eqref{eq:red-Kpm-sfc}. This is the curvature--fluid analogue of the scalar-only endpoint reduction, with the same
leading cubic coefficient $-\tfrac32$ in the reduced dynamics.

\paragraph{2) Fluid--curvature degeneracy axis  $\Gamma_4$ ($\gamma=\tfrac23$).}
At $\gamma=\tfrac23$ the equilibria $M$ and $C$ are no longer isolated but lie on the equilibrium line
$\mathcal L_{MC}$ in \eqref{eq:line-MC-g23}, parameterised by $W=\Omega_k$ with $X=Y=0$. The relevant organising
feature is the loss (and recovery) of \emph{normal hyperbolicity}\footnote{Dynamically, normal hyperbolicity is visible in the fact that for $\mu\neq0$ perturbations off
$\mathcal L_{MC}$ return/escape exponentially, whereas at $\mu=0$ this linear restoring/repelling term vanishes and
the transverse motion is controlled by higher-order terms. Thus it refers to hyperbolicity \emph{transverse} to the
equilibrium line $\mathcal L_{MC}$: the tangent direction is neutral for geometric reasons, and the loss at
$\gamma=\tfrac23$ occurs when a \emph{normal} eigenvalue also vanishes, enlarging the centre subspace without
changing the tangential geometry of the line itself.} of this line as $\gamma$ crosses $\tfrac23$:
a perturbation tangent to $\mathcal L_{MC}$ varies $W$, whereas a transverse perturbation activates $X$ and/or $Y$
and moves off the line. In these transverse directions the leading drift is controlled by the normal eigenvalue
$\mu:=3\gamma-2$, so locally one has
\begin{equation}\label{eq:nf-lineMC-transverse}
\dot w = \mu\,w + \cdots,\qquad \mu=3\gamma-2,
\end{equation}
with the remaining directions hyperbolic away from further coincidences; in particular, for $\mu\neq0$ the equilibrium
continuum collapses to isolated sites, with normal attraction/repulsion determined by $\mathrm{sign}(\mu)$.

\paragraph{3) Curvature-induced axis at the scalar site   $\Gamma_2$ ($a^2=3$, i.e.\ $\lambda^2=2$).}
At $S$ the curvature eigenvalue $\tfrac23(a^2-3)$ vanishes and the centre reduction along the curvature mode is
fold-type:
\begin{equation}\label{eq:nf-fold-S-a2eq3}
\dot W = -2W^2 + \cdots\qquad(a^2=3),
\end{equation}
cf.\ \eqref{eq:red-S-a2eq3-sfc}. This axis is absent in the scalar--fluid problem and represents a genuine curvature-driven
reorganization of the local phase portrait near the scalar branch.

\paragraph{4) Scalar--fluid exchange (collision) line   $\Gamma_3$ ($a^2=\tfrac92\gamma$, i.e.\ $\lambda^2=3\gamma$).}
Along this curve the scalar equilibrium $S$ and the scaling equilibrium $F$ meet and exchange stability. The vanishing
eigenvalue is the $(2a^2-9\gamma)$-mode in \eqref{eq:eigs-S-sfc}, and the corresponding reduction is transcritical in the
generic case (a single zero eigenvalue with nonvanishing quadratic coupling), i.e.\ after a smooth local change of
coordinates one obtains
\begin{equation}\label{eq:nf-transcritical-SF}
\dot u = \mu\,u - u^2 + \cdots,\qquad \mu \propto (2a^2-9\gamma),
\end{equation}
with the remaining directions hyperbolic.

\paragraph{5) Higher-codimension coincidences.}
Intersections of the above loci (e.g.\ $\gamma=2$ with $a=\pm3$, or $\gamma=\tfrac23$ with $\lambda^2=3\gamma$)
produce multi-zero eigenvalue situations (two-dimensional centres or equilibrium continua) whose detailed normal forms
follow from the same jets but require keeping the next nontrivial order. 

Concretely, one proceeds by restricting the translated jet expansions
\eqref{eq:W-jet-form-sfc} together with the explicit quadratic parts
\eqref{eq:Q-Kp-sfc}--\eqref{eq:Q-F-sfc} to the corresponding centre subspace and retaining the first non-vanishing
higher-order terms in the centre variables. 

In particular, at the kinetic endpoints the translated jets inherit the $Z_2$-equivariance of the $Y\mapsto -Y$
reflection (and the conjugacy $(a,Y)\mapsto(-a,-Y)$), so that at intersections involving $|a|=3$ the organising
centre may lie in the $Z_2$-restricted (hilltop-type) amplitude family rather than in a generic non-symmetric class;
this becomes relevant whenever more than one unfolding direction is activated simultaneously.

\begin{remark}[Curvature as a parasitic mode relative to the flat truncation]\label{parasitic}
A structurally stable (``coarse'') phase portrait on an invariant subsystem need not remain stable
when an additional degree of freedom is restored.  In the sense emphasized by Andronov--Vitt--Khaikin in \cite{an66} (cf. in particular their discussion in the Introduction, p. xxix), order-increasing perturbations correspond to ``parasitic'' parameters: small neglected effects that enlarge the
state space and can destroy apparent stability.  In our setting the flat sector $\Omega_k\equiv 0$ is invariant,
and spatial curvature provides a transverse mode; robustness of any flat attractor therefore requires
\emph{normal hyperbolicity} in the curvature direction.  The stratification in Fig.~\ref{fig:bifn-map-gamma-lambda}
is precisely organized by the loss of this transverse hyperbolicity along the curvature loci.
\end{remark}

\section{Organising centres for the harmonic extensions}
\label{sec:massive-extensions}

\paragraph{Structure of this section.}
This section develops the massive (quadratic) analogue of the fixed slope exponential extensions of \S\S5--6.
Using expansion--normalised variables and the e-fold time \(N=\ln \texttt{a}\), we close the Einstein--scalar--fluid--curvature system for
\(V(\phi)=\tfrac12 m^2\phi^2\) by introducing a bounded slope mode \(\zeta=\arctan\lambda\) (with \(\lambda:=-V_{,\phi}/V\)).
In \S7.1 we set up the variables, constraint and slope closure, and derive the resulting autonomous \((X,Y,\Omega_k,\zeta)\) system
and its admissible state space.
\S7.2 identifies the two invariant ``gates'' \(Y=0\) and \(\zeta=0\) and shows that, away from \(\zeta=0\), equilibria organise into the robust
equilibrium lines \(M\) and \(C\) (and the equilibrium segment $\mathcal L_{MC}$ on \(\gamma=2/3\)).
\S7.3 computes translated quadratic jets near \(M\) and \(C\), performs the centre reductions needed later,
and isolates the vertical \(\gamma\)-loci \(\gamma\in\{0,2/3,2\}\) where normal hyperbolicity is lost.
Finally, \S7.4 records the invariant subcases (SF, SFC, etc.), clarifies the role of frozen-slope instantaneous approximations versus true equilibria,
and contrasts the organising mechanisms here (equilibrium continua and \(\gamma\)-thresholds) with the isolated, parameter-driven loci of the strict exponential class,
preparing the unified unfolding narrative of \S8.

The main results of this section are summarized in the following theorem.
\begin{theorem}[Organising centres for the quadratic SF--fluid--curvature extension]\label{thm:OC_quadratic_SFC}
Fix a barotropic index \(\gamma\) and consider the quadratic potential \(V(\phi)=\tfrac12 m^2\phi^2\).
In expansion--normalised variables \((X,Y,\Omega_k)\) with e-fold time \(N=\ln \texttt{a}\), close the system by the bounded slope mode
\(\zeta=\arctan\lambda\), \(\lambda:=-V_{,\phi}/V\), so that \(\lambda=\tan\zeta\) and \(\Omega_m=1-X^2-Y^2-\Omega_k\).
Let \(\mathcal S=\{(X,Y,\Omega_k,\zeta):\ \Omega_m\ge0\}\) denote the physically admissible region.

\begin{enumerate}
\item[\textup{(i)}] \textup{(Invariant components and slope gate).}
The sets \(\Omega_k=0\) (flat), \(\Omega_m=0\) (vacuum, i.e.\ \(X^2+Y^2+\Omega_k=1\)),
and \(X=Y=0\) (equivalently \(\Omega_m+\Omega_k=1\)) are invariant.
Moreover the slope evolution factorises as
\[
\zeta'=\frac{\sqrt6}{2}\,Y\sin^2\zeta,
\]
hence any equilibrium satisfies the gate condition \(Y=0\) or \(\sin\zeta=0\) (i.e.\ \(\zeta=0\), equivalently \(\lambda=0\)).

\item[\textup{(ii)}] \textup{(Robust equilibrium continua at finite slope).}
For every \(\gamma\), away from the invariant manifold \(\zeta=0\) the equilibria form exactly the two lines
\[
M=\{(0,0,0,\zeta):\ \zeta\in(-\tfrac\pi2,\tfrac\pi2)\},\qquad
C=\{(0,0,1,\zeta):\ \zeta\in(-\tfrac\pi2,\tfrac\pi2)\}.
\]
On the vertical axis \(\gamma=2/3\) these are connected by the equilibrium segment
\[
\mathcal L_{MC}=\{(0,0,\Omega_k,\zeta):\ 0\le \Omega_k\le 1,\ \zeta\in(-\tfrac\pi2,\tfrac\pi2)\}.
\]

\item[\textup{(iii)}] \textup{(Normal hyperbolicity thresholds).}
At a basepoint \((0,0,0,\zeta_0)\in M\) the linearisation has one geometrically neutral direction tangent to \(M\),
and transverse eigenvalues
\[
\lambda_U=\tfrac{3\gamma}{2},\qquad \lambda_V=\tfrac{3}{2}(\gamma-2),\qquad \lambda_W=3\gamma-2,
\]
so normal hyperbolicity is lost precisely at \(\gamma\in\{0,\,2/3,\,2\}\).
At a basepoint \((0,0,1,\zeta_0)\in C\) there is again one neutral direction tangent to \(C\),
with transverse eigenvalues \(1\), \(-2\), and \(2-3\gamma\); thus \(C\) loses normal hyperbolicity only at \(\gamma=2/3\).

\item[\textup{(iv)}] \textup{(Asymptotic \(\zeta=0\) locus).}
On the invariant manifold \(\zeta=0\) the reduced \((X,Y,\Omega_k)\) subsystem contains the familiar scalar-dominated and kinetic equilibria
(e.g.\ \(D\) and \(K^\pm\)), which organise the flow only through the approach \(\zeta\to0\).
\end{enumerate}

\noindent
This equilibrium/degeneracy geometry provides the organising centres for the harmonic class
and underlies the jets and centre reductions in \S7.3 and the unfolding synthesis in \S8.

\end{theorem}

\smallskip
\noindent\emph{Guide for the reader:} \S7.2 gives the equilibrium geometry; \S7.3 extracts the local normal forms used in \S8.
\smallskip

\subsection{Harmonic SF--fluid--$k$ system in bounded slope variables}
\label{subsec:massive-system}

To maintain direct comparability with the strict exponential systems of
\S\S\ref{sec:fluid-org}--\ref{sec:curvature}, we work with expansion--normalised
variables and the e--fold time $N=\ln \texttt{a}$:
\begin{equation}\label{eq:massive-vars-XY}
X=\frac{\sqrt{V}}{\sqrt{3}\,H},\qquad
Y=\frac{\dot\phi}{\sqrt{6}\,H},\qquad
\Omega_m=\frac{\rho_m}{3H^2},\qquad
\Omega_k=-\frac{k}{a^2H^2}.
\end{equation}
The Friedmann constraint is
\begin{equation}\label{eq:massive-constraint-XY}
X^2+Y^2+\Omega_m+\Omega_k=1.
\end{equation}
Throughout this section, \(N=\ln a\). In this section only, and in order to follow the standard slope-closure notation used in the non-exponential-potential literature, we write \( '\equiv d/dN=(1/H)d/dt\). This is a local notational convention for Section~\ref{sec:massive-extensions}; it should not be confused with the preliminary \(\tau\)-derivative notation used in Section~\ref{subsec:prelim-scaling}.
\paragraph{Relation to the variables used in \cite{MI-I}.}
In our treatment in Ref.~\cite{MI-I} (no fluid, no curvature) we used $(\phi,\dot\phi,H)$-based
variables $(x,y,z)=(\phi/\sqrt6,\dot\phi/\sqrt6,H)$ in physical time.
The present choice is the corresponding expansion--normalised formulation:
$Y=y/z$ and $X=\sqrt{V}/(\sqrt3\,z)$, which is better suited for comparison with
the exponential fixed slope class.
\paragraph{Slope closure and a bounded slope coordinate.}
For non--exponential potentials one introduces
\begin{equation}\label{eq:massive-lambda}
\lambda(\phi):=-\frac{V_{,\phi}}{V},\qquad
\Gamma(\phi):=\frac{V\,V_{,\phi\phi}}{V_{,\phi}^{\,2}},
\end{equation}
so that $\lambda$ evolves according to $\lambda'=-\sqrt6(\Gamma-1)\lambda^2Y$.
For the quadratic potential $V=\tfrac12 m^2\phi^2$ one has $\Gamma=\tfrac12$ and
$\lambda=-2/\phi$, hence
\begin{equation}\label{eq:quad-lambdadot}
\lambda'=\frac{\sqrt6}{2}\lambda^2\,Y.
\end{equation}
Since $\lambda$ is unbounded as $\phi\to0$, it is convenient to compactify the
slope direction by setting
\begin{equation}\label{eq:zetadef}
\zeta:=\arctan\lambda\in\Bigl(-\frac\pi2,\frac\pi2\Bigr),
\end{equation}
which yields the bounded slope evolution
\begin{equation}\label{eq:zetadot}
\zeta'=\frac{\sqrt6}{2}\sin^2\!\zeta\;Y.
\end{equation}
\paragraph{Remark on alternative closures in the literature.}
The global compact formulations of Ref.~\cite{ahu15} achieve
regularity by compactifying the Hubble/mass direction (their $T$--variable),
whereas some reviews compactify $\lambda$ and then rescale time by a vanishing
factor to remove apparent divergences. We prefer \eqref{eq:zetadef}--\eqref{eq:zetadot}, which regularises the slope dynamics without a time rescaling
and keeps the additional slope \emph{mode} explicit.

Using the constraint \eqref{eq:massive-constraint-XY} to eliminate \(\Omega_m\) via
\(\Omega_m=1-X^2-Y^2-\Omega_k\), and substituting into the SF--fluid--\(k\) system
\eqref{eq:Xdot-sfc}--\eqref{eq:Omkdot-sfc}, we obtain the closed \emph{four--dimensional}
autonomous system for the bounded-slope variables \((X,Y,\Omega_k,\zeta)\), including \eqref{eq:zetadot}:
\begin{align}
X'&=
X\!\left(
3Y^2+\frac{3\gamma}{2}\bigl(1-X^2-Y^2-\Omega_k\bigr)+\Omega_k
-\sqrt{\frac{3}{2}}\,(\tan\zeta)\,Y
\right),
\label{eq:massive-Xdot-zeta}\\[0.7ex]
Y'&=
-3Y+\sqrt{\frac{3}{2}}\,(\tan\zeta)\,X^2
+3Y^3+\frac{3\gamma}{2}Y\bigl(1-X^2-Y^2-\Omega_k\bigr)+Y\Omega_k,
\label{eq:massive-Ydot-zeta}\\[0.7ex]
\Omega_k'&=
2\Omega_k\!\left(
3Y^2+\frac{3\gamma}{2}\bigl(1-X^2-Y^2-\Omega_k\bigr)+\Omega_k-1
\right),
\label{eq:massive-Omkdot-zeta}\\[0.7ex]
\zeta'&=
\frac{\sqrt6}{2}\,Y\,\sin^2\zeta
\qquad\text{(quadratic potential)}.
\label{eq:massive-zetadot}
\end{align}

\begin{remark}
Equation \eqref{eq:massive-zetadot} replaces the unbounded slope evolution $\lambda'$
and keeps the massive extension on the same Ref.~\cite{MI-I}-style footing as the exponential case,
while making the extra ``slope mode'' explicit.
\end{remark}

\noindent\textbf{State space.}
The physically admissible region is
\begin{equation}\label{eq:massive-state-space}
\mathcal S:=\Bigl\{(X,Y,\Omega_k,\zeta):\ \Omega_m:=1-X^2-Y^2-\Omega_k\ge 0,\
\zeta\in\Bigl(-\frac\pi2,\frac\pi2\Bigr)\Bigr\}.
\end{equation}
with $\Omega_k\ge0$ corresponding to negative spatial curvature ($k=-1$), while $\Omega_k\le0$ corresponds to positive curvature ($k=+1$).

\subsection{Invariant gates and robust equilibrium sets}
\label{subsec:massive-gates-eq}

We begin by isolating the invariant gates produced by the bounded slope mode and the resulting robust equilibrium skeleton that will serve
as the organising centre for the massive class.

\begin{lemma}[Gates and backbone equilibria]\label{lem:gates-backbone}
Consider the quadratic SF--fluid--\(k\) system \eqref{eq:massive-Xdot-zeta}--\eqref{eq:massive-zetadot} on the admissible state space \(\mathcal S\).
\begin{enumerate}
\item[\textup{(i)}] (\textup{Gate condition}) Any equilibrium satisfies the gate condition \eqref{eq:gate-eq-condition}.
Moreover, the hypersurfaces \(\Omega_k=0\), \(\Omega_m=0\), and \(X=Y=0\) (equivalently \(\Omega_m+\Omega_k=1\)) are invariant.

\item[\textup{(ii)}] (\textup{Backbone reduction}) The backbone \(\mathcal B\) defined in \eqref{eq:backbone-def} is invariant.
Restricted to \(\mathcal B\) the curvature evolution reduces to the logistic equation \eqref{eq:Omk-on-line}.

\item[\textup{(iii)}] (\textup{Robust equilibrium continua}) For \(\zeta\neq 0\), any equilibrium on the gate \(Y=0\) lies on \(\mathcal B\),
and hence belongs to one of the two equilibrium lines \(M\) and \(C\) given in \eqref{eq:massive-M-line}--\eqref{eq:massive-C-line}.
On the special axis \(\gamma=2/3\) the whole segment \(\mathcal L_{MC}\) in \eqref{eq:MC-segment} consists of equilibria for every \(\zeta\).

\item[\textup{(iv)}] (\textup{\(\zeta=0\) gate}) On the invariant gate \(\zeta=0\) the reduced \((X,Y,\Omega_k)\) subsystem contains the boundary equilibria
\(D\) and \(K_\pm\) in \eqref{eq:massive-D-boundary}--\eqref{eq:massive-Kpm-boundary}
(and the intersections \(M\cap\{\zeta=0\}\), \(C\cap\{\zeta=0\}\)), which act as asymptotic organising loci via \(\zeta\to 0\).
\end{enumerate}
\end{lemma}

The proof of Lemma~\ref{lem:gates-backbone} is given in the remainder of this subsection.

\subsubsection{Invariant slices and the $\zeta$-gate}
We first establish the gate condition and the invariant hypersurfaces.
From \eqref{eq:massive-zetadot}, any equilibrium of the quadratic SF--fluid--$k$ system must satisfy
\begin{equation}\label{eq:gate-eq-condition}
Y=0 \quad\text{or}\quad \sin\zeta=0\ \ (\text{i.e. }\zeta=0,\ \lambda=0).
\end{equation}
\paragraph{Terminology (gates) and the meaning of \(Y=0\).}
Equation \eqref{eq:gate-eq-condition} shows that the additional slope mode \(\zeta\) \emph{freezes} whenever either \(Y=0\) or \(\zeta=0\).
We call such codimension--one loci \emph{gates}: along them the evolution of the slope mode degenerates (\(\zeta'=0\)), so trajectories may pass through
while the \(\zeta\)-dynamics momentarily stalls.
Since \(Y=\dot\phi/(\sqrt6 H)\), the gate \(Y=0\) has additional characteristics: it is the ``turning-point'' surface where the scalar field is instantaneously at rest (\(\dot\phi=0\)) and changes direction in physical time. Correspondingly, for \(\zeta\neq 0\) the sign of \(\zeta'\) is the sign of \(Y\), so crossings of \(Y=0\) switch the direction of slope drift in the extended state space.

\paragraph{Backbone.}
Within the gate \(Y=0\) there is a distinguished invariant subset,
\be\label{eq:backbone-def}
\mathcal B:=\{(X,Y,\Omega_k,\zeta):\ X=0,\ Y=0\},
\ee
which we call the \emph{backbone} of the massive extension.
The term reflects its role as the simplest invariant set - the `skeleton' - on which the remaining components evolve independently:
on \(\mathcal B\) the scalar sector is dynamically inactive in the expansion--normalised variables (\(X=Y=0\)), the slope mode is frozen (\(\zeta'=0\)),
and the constraint reduces to \(\Omega_m=1-\Omega_k\).
In particular, the full curvature equation collapses on \(\mathcal B\) to the one--dimensional logistic dynamics \eqref{eq:Omk-on-line},
so the equilibrium lines \(M\) and \(C\) (and, at \(\gamma=\tfrac23\), the connecting segment \(\mathcal L_{MC}\))
are best viewed as the backbone equilibria parameterised by the slope coordinate \(\zeta\).

Indeed, the reduced system exhibits the explicit factorisations \(X'=X(\cdots)\), \(\Omega_k'=2\Omega_k(\cdots)\), and \(\zeta'=Y(\cdots)\).
Moreover, differentiating \(\Omega_m=1-X^2-Y^2-\Omega_k\) along the flow yields
\begin{equation}\label{eq:Omega_m_factor}
\Omega_m'=3\Omega_m\bigl[(2-\gamma)Y^2-\gamma X^2-(\gamma-\tfrac23)\Omega_k\bigr],
\end{equation}
so \(\Omega_m=0\) is invariant as well. Invariance of \(\mathcal B\) follows immediately, since \(X=Y=0\) forces \(X'=Y'=\zeta'=0\).

\subsubsection{The $Y=0$ slice and the $(M,C)$ backbone}
On the slice $Y=0$, equation \eqref{eq:massive-Ydot-zeta} reduces to
\begin{equation}\label{eq:Ydot-on-Y0}
Y'\big|_{Y=0}=\sqrt{\frac{3}{2}}(\tan\zeta)\,X^2,
\end{equation}
so for finite $\zeta\neq 0$ any equilibrium on $Y=0$ must satisfy $X=0$. Note that \(Y=0\) is not an invariant slice for finite \(\zeta\neq 0\): unless \(X=0\) one has \(Y'|_{Y=0}\neq 0\) by \eqref{eq:Ydot-on-Y0}.
Thus the logistic reduction arises on the \emph{invariant backbone} \(\mathcal B=\{X=0,\ Y=0\}\), along which \(\zeta' =0\) and \(\Omega_m=1-\Omega_k\).

With $X=Y=0$ the constraint gives $\Omega_m=1-\Omega_k$ and \eqref{eq:massive-Omkdot-zeta} becomes the scalar logistic equation
\begin{equation}\label{eq:Omk-on-line}
\Omega_k'=\Omega_k(3\gamma-2)(1-\Omega_k).
\end{equation}
Equivalently, the full curvature equation \eqref{eq:massive-Omkdot-zeta} can be written as
\begin{equation}\label{eq:Omk-split-logistic-coupling}
\Omega_k'=\Omega_k(3\gamma-2)(1-\Omega_k)+3\Omega_k\bigl[(2-\gamma)Y^2-\gamma X^2\bigr],
\end{equation}
so the logistic form \eqref{eq:Omk-on-line} is recovered precisely on \(\mathcal B\) (where \(X=Y=0\)).
\begin{remark}[On unfoldings of the curvature logistic equation]
As a one-dimensional normal form, \(x' = a\,x(1-x)\) admits a versal unfolding
\[
x' = a\,x(1-x)+\nu .
\]
In the present problem, however, the flat submanifold \(\Omega_k=0\) is invariant, so we do
not allow arbitrary additive perturbations of the logistic germ. By ``within the invariant
class'' we mean perturbations that preserve this geometric invariance, equivalently, perturbations
for which the reduced curvature equation still vanishes identically on \(\Omega_k=0\) and hence
retains an overall factor \(\Omega_k\). In this sense the term
\[
3\Omega_k\bigl((2-\gamma)Y^2-\gamma X^2\bigr)
\]
is the natural physical unfolding term: it deforms the logistic dynamics away from the
backbone \(B\), but preserves the invariant boundary \(\Omega_k=0\). The distinguished
transcritical parameter is therefore \(\gamma-\tfrac23\), while the scalar directions \(X,Y\)
supply state-induced deformation terms inside this invariant subclass.
\end{remark}

Thus, for generic $\gamma\neq\tfrac23$, the only equilibria on this backbone are the two robust lines
\begin{align}
M:\ &(X,Y,\Omega_k,\zeta)=(0,0,0,\zeta), \qquad \Omega_m=1,
\label{eq:massive-M-line}\\
C:\ &(X,Y,\Omega_k,\zeta)=(0,0,1,\zeta), \qquad \Omega_m=0,
\label{eq:massive-C-line}
\end{align}
each parameterised by $\zeta$ (equivalently by $\lambda$).

On the special vertical axis
\begin{equation}\label{eq:gamma-23-axis}
\gamma=\frac{2}{3},
\end{equation}
the coefficient in \eqref{eq:Omk-on-line} vanishes and the entire segment
\begin{equation}\label{eq:MC-segment}
\mathcal L_{MC}:=\{(X,Y,\Omega_k,\zeta)=(0,0,\Omega_k,\zeta):\ 0\le \Omega_k\le 1,\ \zeta\in(-\tfrac{\pi}{2},\tfrac{\pi}{2})\}.
\end{equation}
consists of equilibria for every $\zeta$. This is the harmonic analogue of the equilibrium line phenomenon on the $\Gamma_4$-locus already encountered
in the exponential SFC class at $\gamma=\tfrac23$.

\subsubsection{The $\zeta=0$ boundary and asymptotic organising loci}
The alternative gate in \eqref{eq:gate-eq-condition} is \(\zeta=0\) (i.e.\ \(\lambda=0\)), which corresponds to the large--field asymptotic regime
\(|\phi|\to\infty\) for the quadratic potential.
Restricting \eqref{eq:massive-Xdot-zeta}--\eqref{eq:massive-Omkdot-zeta} to $\zeta=0$ yields a closed
$(X,Y,\Omega_k)$ subsystem in which the familiar scalar-dominated and kinetic structures reappear as \emph{boundary} equilibria, e.g.
\begin{align}
D:\ &(X,Y,\Omega_k,\zeta)=(1,0,0,0),
\label{eq:massive-D-boundary}\\
K_\pm:\ &(X,Y,\Omega_k,\zeta)=(0,\pm 1,0,0).
\label{eq:massive-Kpm-boundary}
\end{align}
Unlike the strict exponential class, these are not interior fixed points for finite $\zeta\neq 0$; rather they organise the flow only
asymptotically through the approach $\zeta\to 0$.

\paragraph{Equilibria in this subsection.}
For later reference we record the equilibrium sets identified here:
\begin{itemize}
\item \emph{Finite slope (\(\zeta\neq 0\)):} the two equilibrium lines \(M\) and \(C\), given by \eqref{eq:massive-M-line}--\eqref{eq:massive-C-line}.
\item \emph{Special axis \(\gamma=\tfrac23\):} the equilibrium segment \(\mathcal L_{MC}\) in \eqref{eq:MC-segment}, which connects \(M\) to \(C\).
\item \emph{Asymptotic \(\zeta=0\) gate:} the boundary equilibria \(D\) and \(K_\pm\) in \eqref{eq:massive-D-boundary}--\eqref{eq:massive-Kpm-boundary}
(and the intersections \(M\cap\{\zeta=0\}\), \(C\cap\{\zeta=0\}\)).
\end{itemize}

\paragraph{Summary.}
The quadratic SF--fluid--$k$ system therefore has two robust equilibrium lines $M$ and $C$ at finite $\zeta$,
an equilibrium segment $\mathcal L_{MC}$ on the special axis $\gamma=\tfrac23$,
and a collection of asymptotic boundary equilibria on $\zeta=0$.
This equilibrium geometry is the correct starting point for the jet/normal-form analysis in the massive class.

\subsection{Quadratic jets, centre reductions, and normal forms}
\label{subsec:massive-jets-nf}

Throughout we work with the closed four--dimensional system in \((X,Y,\Omega_k,\zeta)\),
with \(\Omega_m\) eliminated by the constraint
\begin{equation}\label{eq:massive-Om-elim}
\Omega_m=1-X^2-Y^2-\Omega_k,
\end{equation}
and with \(\lambda=\tan\zeta\). (Primes denote \(d/dN\).) We organise the local analysis according to the equilibrium geometry identified in \S\ref{subsec:massive-gates-eq}: first the finite slope equilibrium lines \(M\) and \(C\), then the special equilibrium segment \(\mathcal L_{MC}\) on \(\gamma=2/3\), and finally the boundary gate \(\zeta=0\).

\paragraph{The curvature ``backbone + coupling'' split.} A useful guiding identity is the curvature split \eqref{eq:Omk-split-logistic-coupling}, which isolates the logistic backbone dynamics in \(\Omega_k\)
and exhibits the remaining terms as the lowest--order coupling (unfolding) by the scalar modes \((X,Y)\).
This perspective will be exploited when interpreting the quadratic jets and the centre reductions below.

\subsubsection[Translated jet systems near the robust equilibrium lines]{Translated jet systems near the robust equilibrium lines \(M\) and \(C\)}
\label{subsubsec:massive-Wlines}

\paragraph{(i) The fluid line $M$.}
Fix any basepoint on the equilibrium line
\begin{equation}\label{eq:M-basepoint}
M_{\zeta_0}:\qquad (X,Y,\Omega_k,\zeta)=(0,0,0,\zeta_0),
\end{equation}
and translate by
\begin{equation}\label{eq:Wshift-M-4d}
X=U,\qquad Y=V,\qquad \Omega_k=W,\qquad \zeta=\zeta_0+L,
\qquad \mathbf{u}:=(U,V,W,L)^{\mathsf T}.
\end{equation}
Write \(\tau_0:=\tan\zeta_0\) and \(s_0:=\sin\zeta_0\).
The system admits a local expansion
\begin{equation}\label{eq:Wform-M-4d}
\mathbf{u}' = J_M\,\mathbf{u}+Q_M(\mathbf{u})+O\!\left(\|\mathbf{u}\|^3\right),
\end{equation}
where the linear part is
\begin{equation}\label{eq:J-M-4d}
J_M=
\begin{pmatrix}
\frac{3\gamma}{2} & 0 & 0 & 0\\[0.4ex]
0 & \frac{3}{2}(\gamma-2) & 0 & 0\\[0.4ex]
0 & 0 & 3\gamma-2 & 0\\[0.4ex]
0 & \frac{\sqrt6}{2}s_0^{\,2} & 0 & 0
\end{pmatrix},
\end{equation}
and a convenient quadratic jet is
\begin{equation}\label{eq:Q-M-4d}
Q_M(\mathbf{u})=
\begin{pmatrix}
-\sqrt{\frac{3}{2}}\tau_0\,UV + UW\\[0.8ex]
\sqrt{\frac{3}{2}}\tau_0\,U^2 +\Bigl(1-\frac{3\gamma}{2}\Bigr)VW\\[0.8ex]
(2-3\gamma)\,W^2\\[0.8ex]
\frac{\sqrt6}{2}\sin(2\zeta_0)\,VL
\end{pmatrix}.
\end{equation}
In particular, the spectrum of $J_M$ consists of the three transverse eigenvalues
\begin{equation}\label{eq:spec-JM-4d}
\lambda_U=\frac{3\gamma}{2},\qquad
\lambda_V=\frac{3}{2}(\gamma-2),\qquad
\lambda_W=3\gamma-2,
\end{equation}
together with the geometrically neutral direction along the equilibrium line (the $L$–direction).

\paragraph{(ii) The curvature line $C$.}
Similarly, fix a basepoint
\begin{equation}\label{eq:C-basepoint}
C_{\zeta_0}:\qquad (X,Y,\Omega_k,\zeta)=(0,0,1,\zeta_0),
\end{equation}
and translate by
\begin{equation}\label{eq:Wshift-C-4d}
X=U,\qquad Y=V,\qquad \Omega_k=1+W,\qquad \zeta=\zeta_0+L,
\qquad \mathbf{u}:=(U,V,W,L)^{\mathsf T}.
\end{equation}
Then one has
\begin{equation}\label{eq:Wform-C-4d}
\mathbf{u}' = J_C\,\mathbf{u}+Q_C(\mathbf{u})+O\!\left(\|\mathbf{u}\|^3\right),
\end{equation}
with linear part
\begin{equation}\label{eq:J-C-4d}
J_C=
\begin{pmatrix}
1 & 0 & 0 & 0\\[0.4ex]
0 & -2 & 0 & 0\\[0.4ex]
0 & 0 & 2-3\gamma & 0\\[0.4ex]
0 & \frac{\sqrt6}{2}s_0^{\,2} & 0 & 0
\end{pmatrix},
\end{equation}
and a quadratic jet
\begin{equation}\label{eq:Q-C-4d}
Q_C(\mathbf{u})=
\begin{pmatrix}
-\sqrt{\frac{3}{2}}\tau_0\,UV +\Bigl(1-\frac{3\gamma}{2}\Bigr)UW\\[0.8ex]
\sqrt{\frac{3}{2}}\tau_0\,U^2 +\Bigl(1-\frac{3\gamma}{2}\Bigr)VW\\[0.8ex]
(2-3\gamma)\,W^2\\[0.8ex]
\frac{\sqrt6}{2}\sin(2\zeta_0)\,VL
\end{pmatrix}.
\end{equation}
Thus $C$ is non-hyperbolic precisely on the vertical axis $\gamma=\tfrac23$, where its normal eigenvalue $2-3\gamma$ vanishes and $C_{\zeta_0}$
lies on the equilibrium segment $\mathcal L_{MC}$ introduced in \S\ref{subsec:massive-gates-eq}.

\subsubsection{Centre reductions on the non-hyperbolic loci}
\label{subsubsec:massive-CMred}

The quadratic massive class has two types of non-hyperbolicity:

\begin{itemize}
\item \emph{Geometric non-isolation} (an equilibrium line/segment), producing a centre direction along $L$ (or along $W$ on $\mathcal L_{MC}$).
\item \emph{Loss of normal hyperbolicity}, when one of the transverse eigenvalues in \eqref{eq:spec-JM-4d} or its $C$–analogue crosses zero
(e.g.\ $\gamma=0$, $\gamma=\tfrac23$, $\gamma=2$).
\end{itemize}

We record the centre reductions that will be needed later.

\paragraph{(i) Cubic reduction at $M$ on $\gamma=0$.}
On $\gamma=0$ the $U$–eigenvalue in \eqref{eq:spec-JM-4d} vanishes while $V$ and $W$ remain hyperbolic ($\lambda_V=-3$, $\lambda_W=-2$).
Restricting to a local centre manifold (parameterised by $U$ and $L$), one may solve the $V$–equation at leading order from \eqref{eq:Wform-M-4d}--\eqref{eq:Q-M-4d} as
\begin{equation}\label{eq:CM-M-g0}
V=\frac{1}{3}\sqrt{\frac{3}{2}}\tau_0\,U^2+O\!\left(\|(U,L)\|^3\right),\qquad
W=O\!\left(\|(U,L)\|^3\right).
\end{equation}
Substituting into the $U$–equation yields the reduced cubic normal form
\begin{equation}\label{eq:red-M-g0-quad}
U'=-\frac12\,\tau_0^{\,2}\,U^3+O(U^4),
\qquad (\gamma=0).
\end{equation}
(Here the coefficient depends on the chosen basepoint $\zeta_0$ along the equilibrium line via $\tau_0=\tan\zeta_0$.)

\paragraph{(ii) Two-dimensional centre at $M$ on $\gamma=2$.}
At $\gamma=2$ the $V$–eigenvalue in \eqref{eq:spec-JM-4d} vanishes, while $\lambda_U=3$ and $\lambda_W=4$ are hyperbolic.
The centre subspace is spanned by $(V,L)$ and the reduced dynamics begins at cubic order in $V$:
\begin{equation}\label{eq:red-M-g2-quad}
V'=3V^3+O\!\left(\|(V,L)\|^4\right),\qquad
L'=\frac{\sqrt6}{2}s_0^{\,2}\,V+O\!\left(\|(V,L)\|^2\right),
\qquad (\gamma=2).
\end{equation}
Thus $V$ controls the drift along the equilibrium line (the $\zeta$–direction), while the restoring/repelling behaviour is governed by the cubic term.

\paragraph{(iii) Equilibrium segment and loss of normal hyperbolicity at $\gamma=\tfrac23$.}
On the vertical axis $\gamma=\tfrac23$ one has $\lambda_W=3\gamma-2=0$ at $M$ and $2-3\gamma=0$ at $C$.
Equivalently, the entire segment \(\mathcal L_{MC}\) defined in \eqref{eq:MC-segment} consists of equilibria, so that \(\Omega_k\) becomes a tangential (neutral) direction along the segment.
The organising feature here is therefore the \emph{loss (and recovery) of normal hyperbolicity} of $\mathcal L_{MC}$ as $\gamma$ crosses $\tfrac23$:
for $\gamma\neq\tfrac23$ the transverse dynamics contains a linear restoring/repelling term proportional to $(3\gamma-2)$,
whereas at $\gamma=\tfrac23$ this linear term vanishes and the transverse motion is governed by higher order. In particular, \eqref{eq:Omk-split-logistic-coupling} shows that the distinguished coefficient controlling this loss/recovery is the linear backbone factor \(3\gamma-2\) in the logistic term.

\subsubsection{Normal forms and organising loci in the massive class}
\label{subsubsec:massive-nf}

The jets above show that the geometry of the quadratic SF--fluid--$k$ class is organised primarily by \emph{vertical} loci in $\gamma$ together with the \emph{boundary gate}
$\zeta=0$ (equivalently $\lambda=0$).

\paragraph{1) Geometric equilibrium lines ($M$ and $C$ for all $\zeta$).}
For every $\gamma$ the system has equilibrium lines $M$ and $C$ parameterised by $\zeta$; this provides a persistent one–dimensional centre direction (the $L$–direction)
at each basepoint.

\paragraph{2) The equilibrium segment at \(\gamma=\tfrac23\).}
The special axis \(\gamma=\tfrac23\) is distinguished because the normal curvature eigenvalue vanishes both at the fluid line \(M\)
(\(\lambda_W=3\gamma-2=0\)) and at the curvature line \(C\) (\(2-3\gamma=0\)).

Equivalently, the entire segment \(\mathcal L_{MC}\) defined in \eqref{eq:MC-segment} consists of equilibria for every \(\zeta\),
so that \(\Omega_k\) becomes a tangential (neutral) direction along \(\mathcal L_{MC}\) in addition to the ever-present slope direction.

Thus \(\gamma=\tfrac23\) organises the massive class through loss (and recovery) of normal hyperbolicity of an equilibrium continuum,
rather than through an isolated fixed point. This is precisely the regime singled out by the backbone curvature factor \(3\gamma-2\) in \eqref{eq:Omk-split-logistic-coupling}.

\paragraph{3) Vertical normal-hyperbolicity thresholds at \(M\).}
Away from \(\gamma=\tfrac23\), the remaining vertical organising loci occur at the fluid line \(M\) when one of the transverse eigenvalues
in \eqref{eq:spec-JM-4d} crosses zero. In summary,
\begin{equation}\label{eq:gamma-axes-massive}
\gamma=0,\qquad \gamma=\frac23,\qquad \gamma=2,
\end{equation}
correspond respectively to \(\lambda_U=0\), \(\lambda_W=0\), and \(\lambda_V=0\).

The reduced normal forms are cubic at \(\gamma=0\), cf.\ \eqref{eq:red-M-g0-quad}, and two--dimensional centre behaviour at \(\gamma=2\),
cf.\ \eqref{eq:red-M-g2-quad}. The intermediate case \(\gamma=\tfrac23\) is already accounted for by the equilibrium--segment mechanism in
paragraph~2 (see \(\mathcal L_{MC}\) in \eqref{eq:MC-segment}).

\paragraph{4) Boundary organisation by $\zeta=0$ (large-field gate).}
The gate condition $\zeta=0$ (equivalently $\lambda=0$) is the quadratic analogue of the constant slope closure in the strict exponential class.

On $\zeta=0$ the system reduces to a closed $(X,Y,\Omega_k)$ subsystem in which kinetic and potential-dominated structures appear as boundary equilibria
(e.g.\ $D$ and $K_\pm$), but for $\zeta\neq 0$ these do not persist as interior fixed points.

Thus, in contrast to the exponential class (organised by slope axes such as $|a|=3$ and curvature-induced loci in $a$),
the massive quadratic class is organised by a mixture of:
\begin{itemize}
\item[]\emph{(i) equilibrium lines/segments in $(\Omega_k,\zeta)$} and 
\item[]\emph{(ii) vertical thresholds in $\gamma$} that control normal hyperbolicity transverse to those continua.
\end{itemize}
\paragraph{5) Relation to hilltop/$\mathbb Z_2$ language.}
The hilltop/$\mathbb Z_2$ organising mechanisms emphasised in the strict exponential problem arise there from isolated endpoint degeneracies
on slope axes (e.g.\ the $\Gamma_1$-locus $|a|=3$) together with symmetry constraints on the quadratic jets.

In the quadratic massive class, the dominant non-isolation occurs instead through equilibrium \emph{continua} (lines/segments) and their normal-hyperbolicity thresholds.

Accordingly, the local normal form vocabulary is governed here by centre dynamics along equilibrium sets and by transverse eigenvalue crossings,
rather than by isolated $\mathbb Z_2$-equivariant endpoint germs.
\subsection{Recovery of subcases and comparison with the exponential normal forms}
\label{subsec:massive-subcases-compare}

Our formulation in this Section allows for a direct comparison of the exponential and harmonic classes as regards the frozen-slope approximations versus true equilibria, as well as a contrast between the organising centres of the two classes. This in turn leads to a unified unfolding description of the two classes independently of the precise form of the potential in the next Section. 

The method behind this approach lies in the ability of our  formalism to fully recover the invariant subcases in the harmonic class,  in an analogous way as in the exponential class. 
More precisely, the quadratic SF--fluid--curvature system
\eqref{eq:massive-Xdot-zeta}--\eqref{eq:massive-Omkdot-zeta} together with the bounded slope evolution
\eqref{eq:zetadot} (equivalently $\lambda=\tan\zeta$) contains the lower-dimensional
mixtures considered earlier as invariant slices. Using the constraint
$X^2+Y^2+\Omega_m+\Omega_k=1$ to eliminate $\Omega_m$, the dynamics closes in the
four variables $(X,Y,\Omega_k,\zeta)$, and:
\begin{equation}\label{eq:quad-invariant-slices}
\Omega_k=0 \ \Longrightarrow\  \text{quadratic SF--fluid subsystem},\,\,
\Omega_m=0 \ \Longrightarrow\  \text{quadratic SF--curvature subsystem},
\end{equation}
while imposing both $\Omega_k=0=\Omega_m$ yields the SF-only quadratic subsystem
in $(X,Y,\zeta)$ on the unit circle $X^2+Y^2=1$.
In this sense, the full quadratic SFC class provides a single ambient state space in which
the SF-fluid, SF-curvature,  and scalar-only reductions are recovered by restriction.

\subsubsection{Frozen-slope approximations versus true equilibria}
In the strict exponential class the slope is constant ($\lambda\equiv\lambda_0$), so one obtains
genuine interior equilibria and bifurcation loci in the \emph{parameter} plane $(\gamma,\lambda_0)$.
In the quadratic class the effective slope
\begin{equation}\label{eq:aeff-zeta}
a_{\rm eff}:=\sqrt{\frac{3}{2}}\lambda=\sqrt{\frac{3}{2}}\tan\zeta
\end{equation}
is a \emph{state variable}, not a parameter. 

As a result, the exponential fixed-point algebra
(e.g.\ the scalar site $S$, scaling site $F$, curvature-scaling site $P_k$) survives only as an
\emph{instantaneous (frozen)} approximation: at a given instant, one may compare the evolving trajectory
to the exponential formulae with $a$ replaced by $a_{\rm eff}(N)$, but these sets are not true
equilibria of the autonomous quadratic system unless $\zeta$ is effectively frozen (cf.\ \(\lambda'=\tfrac{\sqrt6}{2}\lambda^2Y\), equivalently \(\zeta'=\tfrac{\sqrt6}{2}Y\sin^2\zeta\); see \eqref{eq:quad-lambdadot} and \eqref{eq:massive-zetadot}).

\subsubsection{Contrast with the exponential organising centres}
We may also obtain novel analogies between the exponential and harmonic normal forms. 

The organising centres of Sections~\ref{sec:fluid-org}--\ref{sec:curvature} are dominated by
\emph{isolated} equilibria and their codimension--one degeneracies in the $(\gamma,a)$ plane, most notably
the coupled multi-site endpoint episode at $|a|=3$ (equivalently $|\lambda|=\sqrt6$) and the curvature-induced
axis $a^2=3$ (equivalently $\lambda^2=2$).

By contrast, in the quadratic class the robust objects at finite slope are the \emph{equilibrium lines}
$M$ and $C$ (parameterised by $\zeta$ or $\lambda$), together with boundary organising loci on $\lambda=0$.

Hence the generic local mechanisms in the quadratic case are:
(i) standard centre-manifold reductions associated with transverse eigenvalue crossings, and
(ii) loss (and recovery) of \emph{normal hyperbolicity} of equilibrium continua.

\emph{In particular, the quadratic class does not exhibit an intrinsic hilltop/codimension--three endpoint
episode analogous to the strict exponential $|a|=3$ event; any higher degeneracy arises only through
coincidences involving the equilibrium-line direction and/or boundary restrictions.}

\subsubsection{Forced higher-codimension coincidences}
Although the quadratic class lacks a built-in hilltop mechanism, one can still realise higher-codimension
local degeneracies by forcing multiple neutral directions to coincide.

For example, along the fluid line $M$ the linearisation always has the neutral slope direction
(tangent to the line). At $\gamma=0$ one simultaneously has
\begin{equation}\label{eq:forced-codim-M-g0}
\spec(J_M^{(5)})=\Bigl\{0,\ -3,\ 0,\ -2,\ 0\Bigr\}\qquad (\gamma=0).
\end{equation}
Here \(J_M^{(5)}\) denotes the Jacobian of the unreduced five--variable formulation prior to imposing the Friedmann constraint;
the additional zero reflects the (redundant) constraint direction.
Thus three eigenvalues vanish simultaneously, producing a genuinely higher-dimensional centre subspace.

Similarly, along the curvature line \(C\) the curvature and slope directions are neutral simultaneously only at \(\gamma=\tfrac23\), i.e.\ precisely on the equilibrium segment \(\mathcal L_{MC}\).

Boundary points on $\lambda=0$ (equivalently $\zeta=0$), where additional exponential type loci (such as $K_\pm$) reappear, can further increase degeneracy when combined with special parameter choices. 

These are not generic codimension--one organising centres of the
quadratic class, but they provide clean examples where the extended mode closure permits higher-order
degeneracies that are absent in the strictly exponential closure.

\subsubsection{Bridge to unfoldings.}
The quadratic SF--fluid--curvature class thus reorganises the exponential picture in a structurally different way:
the exponential slope $a$ acts as a \emph{parameter} and produces isolated organising loci (including the coupled
multi-site $|a|=3$ episode), whereas in the quadratic class the effective slope $a_{\rm eff}=\sqrt{3/2}\,\lambda
=\sqrt{3/2}\tan\zeta$ is a \emph{state variable} with its own slow evolution.

Consequently, the dominant organising mechanisms in the quadratic case are centre reductions and loss/recovery of
\emph{normal hyperbolicity} of equilibrium continua, with higher degeneracies arising mainly through coincidences or
boundary restrictions rather than intrinsic hilltop-type endpoints.

Section~\ref{sec:unfoldings} will therefore assemble a unified unfolding narrative by treating the strict exponential
$|a|=3$ event as the primary multi-site organising episode, and then contrasting how the same physical mode directions
(fluid and curvature) and the additional slope mode in the quadratic closure enter as unfolding coordinates in the
extended mixtures.

\section{Unfoldings and paths, transition varieties and persistence}
\label{sec:unfoldings}

\paragraph{Scope.}
Sections~4--7 established the principal equilibria, translated jet systems, centre reductions and
normal forms for the strict exponential and quadratic (massive) extensions, with and without the
full components (fluid and curvature).  The purpose of the present section is twofold:
(i) to assemble these local results into a single versal unfolding picture, paying particular attention to which `paths' in unfolding space distinct physical models in the same versal unfolding family, restricted along different paths/submanifolds, gives different realised bifurcation problems, and
(ii) to set up the transition variety and persistence language needed for the parameter-space
stratifications developed later. We shall see that certain transition varieties are accessible/inaccessible, and some codimension counts change when the path is tangent/degenerate.

\subsection{Unfoldings and paths}
In this subsection, after some general discussion and representative examples from previous sections, we introduce the path formulation which allows for a detailed analysis of all stable perturbations of the normal forms discussed in earlier sections, and at the same time  motivates the distinction between a formal versal (miniversal) unfolding of a germ in abstract unfolding coordinates, and the restricted unfolding actually realised by the cosmological problem at hand.

\subsubsection{Two notions of ``mode''}
\label{subsec:two-modes}

In this paper the word \emph{mode} appears in two distinct senses.

\begin{itemize}
\item \emph{Mode interactions} (Hopf/steady-state, steady-state/steady-state, etc.) refer to
interactions of \emph{spectral modes} at a single equilibrium (eigenvalue collisions and their
normal forms on centre manifolds).
\item \emph{Full modes} (fluid fraction, curvature fraction) refer to additional \emph{dynamical
fractions} in the cosmological mixture, i.e.\ additional state space directions such as $\Omega_m$
and $\Omega_k$ that enlarge the phase space and alter invariant sets and equilibria.
\end{itemize}

The organising episodes of Sections~4--6 (strict exponential) are primarily controlled by the
\emph{parameter} $a$ together with the \emph{full mode coordinates} (fluid/curvature fractions),
whereas the harmonic class of Section~7 is controlled by \emph{extended state} organisation in a
dynamical slope direction.

In addition, the harmonic class introduces a further \emph{slope mode} in the bounded variable \(\zeta\) (or \(\lambda=\tan\zeta\)),
which is neither a spectral mode nor a full mode fraction, but an extended state space direction that controls slow drift and gate structure (cf.\ \S\ref{subsec:massive-gates-eq}). In fact the drift is \emph{state-dependent}: since \(\zeta'=\frac{\sqrt6}{2}Y\sin^2\zeta\), the evolution of \(\zeta\) becomes much slower near the invariant hypersurfaces \(Y=0\) and \(\zeta=0\), where \(\zeta'=0\).
In neighbourhoods of these gates the dynamics is therefore \emph{effectively slow--fast}: the \((X,Y,\Omega_k)\) variables
evolve on the \(N\)-timescale while \(\zeta\) changes only weakly, so trajectories follow a quasi-static
``frozen-slope'' evolution before drifting to a new effective-slope regime.
(We use the slow--fast language heuristically; no explicit small parameter is introduced.)

\subsubsection{Parameter driven versus state variable organisation}
\label{subsec:param-vs-state}

A central distinction between the strict exponential and the harmonic classes is the
role of the slope.

\paragraph{Strict exponential.}
For $V(\phi)=V_0 e^{\lambda\phi}$ the slope $a=\sqrt{3/2}\,\lambda$ is a \emph{parameter}.  Organising
loci therefore appear as curves and axes in the \emph{parameter plane} $(\gamma,a)$ (or $(\gamma,\lambda)$),
and can be unfolded by introducing further independent directions supplied by the full modes (fluid,
curvature).

\paragraph{Quadratic (massive).}
For $V(\phi)=\tfrac12 m^2\phi^2$ the effective slope is \emph{dynamical}: $\lambda(\phi)=-V_{,\phi}/V$ evolves
(equivalently, $\lambda=\tan\zeta$ with bounded $\zeta$).  The organising picture becomes that of an
\emph{augmented state space} with slow drift in the slope direction.  

Consequently, the exponential organising
loci provide a useful skeleton, but the dominant organising mechanisms in the quadratic class are centre
reductions and loss/recovery of normal hyperbolicity in the extended flow, rather than parameter driven
hilltop-type endpoint events.

\subsubsection[The multi-site |a|=3 episode in the scalar-only exponential class]%
{The multi-site $|a|=3$ episode in the scalar-only exponential class}
\label{subsec:episode-scalar-only}

At $a=3$ the Jacobian at the kinetic equilibrium $K_+$ vanishes, and at $a=-3$ the same occurs for $K_-$.
The scalar-only system is equivariant under $(a,Y)\mapsto(-a,-Y)$, so the two degeneracies are dynamically
conjugate.  

However, the translated kinetic vector fields lack the quadratic $U^2$ term required by the
standard $Z_2$-equivariant amplitude hypotheses.  Hence the kinetic endpoint degeneracies at $|a|=3$ are not
of codimension two but are naturally classified as \emph{hilltop-type} phenomena of (at least) codimension three.

Simultaneously, $|a|=3$ coincides with the existence boundary of the scalar-dominated equilibrium $S$ because
$|a|=3$ is equivalent to $|\lambda|=\sqrt6$.  The collision of $S$ with the kinetic boundary produces a
double-zero (nilpotent) core at the endpoint.  Thus, unlike the quadratic  universality class in Ref.~\cite{MI-I}, the strict
exponential scalar-only class features a coupled, multi-site organising episode on the constraint circle.

\subsubsection{Minimal unfolding space from the full modes}
\label{subsec:min-unfolding-modes}

The inclusion of the full modes enlarges the phase space and introduces new robust equilibria.  Most
importantly for unfolding logic, it provides additional independent directions that deform the local jets at the
kinetic endpoints.

\paragraph{Fluid mode.}
The one-fluid extension promotes the state space to $(X,Y,\Omega_m)$ with constraint $X^2+Y^2+\Omega_m=1$.
In contrast to the scalar-only class, the translated kinetic Jacobians acquire an additional eigenvalue which is
generically hyperbolic for $\gamma\neq 2$, supplying a genuine extra direction that contributes to unfolding
coefficients in the reduced normal form.

\paragraph{Curvature mode.}
Including $\Omega_k$ yields the minimal four-dimensional mixture subject to
$X^2+Y^2+\Omega_m+\Omega_k=1$.  Curvature contributes as an independent dynamical fraction and introduces the
curvature equilibrium $C$ as an additional anchor site for organising the extended phase portrait.

\subsubsection{What is meant by ``unfolding parameters'' when the new directions are state variables?}
\label{subsec:mu-interpretation}

In the strict exponential class the slope offset $(a-3)$ is an \emph{external} parameter.  By contrast, $\Omega_m$
and $\Omega_k$ are \emph{state variables}.  The identification of unfolding parameters with ``fluid/curvature modes''
should therefore be understood in the \emph{local normal form sense}:

\begin{quote}
Near a chosen organising site (e.g.\ $K_\pm$), the additional mode directions enter the translated jet as extra
coordinates; after centre manifold reduction and near-identity changes of variables, their leading-order effect is to
appear as \emph{coefficients} in the reduced one-dimensional (or low-dimensional) normal form.  These coefficients are
the unfolding parameters.
\end{quote}

Equivalently: we do \emph{not} treat $\Omega_m$ and $\Omega_k$ as external knobs; rather, we use their local values
(and their contributions to the reduced vector field) to parameterise the \emph{family of nearby reduced dynamics}.
This is exactly the sense in which full-mode directions supply ``physical unfolding directions'' for the scalar-only
degeneracy.

\subsubsection{Canonical reduced germs used in Sections~4--7}
\label{subsec:canonical-germs}

The local reductions obtained in later sections may be organised by a small set of canonical one-dimensional germs.
In each case, the coefficients \(\mu\) in the unfolded normal form are read off from the translated jets:
some are controlled by genuine external parameters (e.g.\ \(a-3\) in the strict exponential class),
while others are induced by excursions in additional state-space directions (fluid and curvature fractions)
prior to reduction.

Below we shall employ the \emph{path formulation} developed in Refs.~\cite{GolubitskySchaefferI,montaldi21} and refs. therein. The versal unfoldings of the cores of the standard singularities appearing in earlier sections, the $A_1$-core $u^2\to u^2+\mu$, for the  $A_2$-core $u^3$, for the $A_3$-core $u^4$, as well as the cores $(x^2,y^2), (x^2-y^2,xy)$, the hilltop bifurcation, etc,  are known. 

\paragraph{Cubic-leading centre reductions (pitchfork-type in the \(\mathbb{Z}_2\) class).}
When a single eigenvalue crosses zero and the first non-vanishing nonlinearity on the centre manifold is cubic,
a normal-form representative is
\begin{equation}\label{eq:germ-cubic}
\dot u = \mu\,u \pm u^3 + \cdots,
\end{equation}
with \(\mu\) the unfolding coordinate associated with the critical eigenvalue.
In \(\mathbb{Z}_2\)-equivariant settings the quadratic term is forbidden, so the cubic term is the first admissible amplitude nonlinearity
and the resulting exchange of stability takes the pitchfork-type form encoded by \eqref{eq:germ-cubic}.

\paragraph{Quartic-leading reductions.}
In the scalar--fluid class certain sites exhibit a quartic-leading reduction
\begin{equation}\label{eq:germ-quartic}
\dot u = u^4 + \cdots,
\end{equation}
whose generic miniversal unfolding may be represented as
\begin{equation}\label{eq:unfold-quartic-generic}
\dot u = u^{4}+\mu_{1}u^{2}+\mu_{2}u+\mu_{3}.
\end{equation}
In the \(\mathbb{Z}_2\)-equivariant subclass the odd term is forbidden, yielding the restricted unfolding
\begin{equation}\label{eq:unfold-quartic-z2}
\dot u = u^{4}+\mu_{1}u^{2}+\mu_{3}.
\end{equation}
Figure~\ref{fig:quartic-unfolding-schematic} provides a schematic atlas for the \(\mathbb{Z}_2\) case:
the organising curves in the \((\mu_1,\mu_3)\) plane separate strata with \(0/2/4\) equilibria and corresponding phase-line types.
This is the simplest explicit illustration, within the present paper, of how a transition set partitions unfolding space into
persistence domains (strata).

\paragraph{Hilltop-type endpoint degeneracies and three-parameter bookkeeping.}
At the kinetic endpoints of the scalar-only exponential class, the quadratic amplitude term vanishes and the reduced endpoint dynamics
is governed by a cubic-leading centre reduction of the form \eqref{eq:germ-cubic}.  
However, within the full \(|a|=3\) organising
\emph{episode} (including the coupled interaction with the scalar site's existence boundary and the extra state-space directions),
the reduced jet analysis produces \emph{three independent deformation coefficients} once fluid and curvature fractions are retained.
For the purpose of transition-variety bookkeeping near this codimension-three episode, it is convenient to use the standard
three-parameter unfolding space \((\mu_1,\mu_2,\mu_3)\) in \eqref{eq:unfold-quartic-generic} as a \emph{coordinate model}:
it supplies a familiar stratified transition-variety skeleton in unfolding space, to which the physical reduced coefficients
computed in Sections~\ref{sec:fluid-org}--\ref{sec:curvature} are mapped.
No claim is made that the physical reduced endpoint equation is literally quartic; the rôle of \eqref{eq:unfold-quartic-generic}
here is to provide a canonical three-parameter organising chart for the local transition geometry.

\subsubsection{The path formulation for scalar field cosmology}
\label{sec:paths}
We now summarize the path formulation. With our notation as in Section~\ref{subsec:BifnSingBridge}, we consider
the germ (bifurcation problem) \(f(x,\lambda)=0\) defining the local landscape, and assume
that the bifurcation takes place at the distinguished parameter value \(\lambda=0\). This germ
may be any of the reduced normal forms obtained in the previous sections after centre-manifold
reduction. The core of the normal form is defined to be
\begin{equation}
f_{0}(x):=f(x,0).
\label{eq:path-core-def}
\end{equation}
If \(F(x;\mu)\), \(\mu\in\mathbb{R}^{\ell}\), is a versal unfolding of \(f_{0}\), then by versality the
family \(f(x,\lambda)\) is equivalent, as an unfolding, to one induced by a map
\begin{equation}
h:\mathbb{R}^{k}\longrightarrow\mathbb{R}^{\ell},
\qquad
\mu=h(\lambda),
\label{eq:path-map-def}
\end{equation}
namely
\begin{equation}
f(x,\lambda)=F\bigl(x;h(\lambda)\bigr).
\label{eq:path-induced-family}
\end{equation}
In the path formulation, the versal unfolding contains all possible deformations of the
image of \(h\), and hence all possible deformations of the original bifurcation problem.
Accordingly, the physically relevant transitions occur when the image of \(h\) meets the
discriminant \(\Delta\) of the versal unfolding, that is, when
\begin{equation}
h^{-1}(\Delta)\neq \varnothing.
\label{eq:path-pullback-discriminant}
\end{equation}

The importance of this formulation in scalar-field cosmology is that it identifies the
physically realised deformation history inside the homogeneous and isotropic class itself.
The model does not move arbitrarily in the full unfolding space; rather, it traces only the
image of the corresponding physical path map \(h\). In particular, even when two reduced
normal forms share the same abstract core, the physically distinguished parameter
combination that supplies the unfolding direction need not be the same. Hence the physical
problem can move in different directions in the same abstract unfolding space.

We now make this explicit for the organising loci found in Sections~\ref{sec:exp-Z2-gate}--\ref{sec:massive-extensions}. For the strict
exponential class, these loci are
\begin{equation}
\Gamma_{1}:\ |a|=3,\qquad
\Gamma_{2}:\ a^{2}=3,\qquad
\Gamma_{3}:\ 2a^{2}-9\gamma=0,\qquad
\Gamma_{4}:\ \gamma=\frac23,\qquad
\Gamma_{5}:\ \gamma=2.
\label{eq:path-gamma-loci}
\end{equation}
The first three are associated with isolated-site reductions, whereas \(\Gamma_{4}\) and
\(\Gamma_{5}\) involve loss of normal hyperbolicity or loss of isolation of equilibrium sets.

\paragraph{Worked example: the scalar-only hilltop germ at \(\Gamma_{1}\).}
In the scalar-only exponential class, the sign-matched endpoint collision at \(|a|=3\) gives
the common organising germ
\begin{equation}
\begin{pmatrix}
\dot U\\[1mm]
\dot V
\end{pmatrix}
=
\begin{pmatrix}
3UV\\[1mm]
6V^{2}
\end{pmatrix}
+O(3),
\qquad
u=(U,V)^{T},
\label{eq:path-gamma1-core}
\end{equation}
cf.~Eq.~(4.44). Relabelling
\begin{equation}
x:=V,
\qquad
y:=U,
\label{eq:path-gamma1-relabel}
\end{equation}
this becomes
\begin{equation}
\dot x=6x^{2}+O(3),
\qquad
\dot y=3xy+O(3),
\label{eq:path-gamma1-triangular}
\end{equation}
that is, the triangular \(\mathbb{Z}_{2}\)-equivariant amplitude germ identified in Section~4.

Let \(\sigma\in\{+1,-1\}\) denote the sign branch, and define the endpoint detuning by
\begin{equation}
\delta_{\sigma}:=a-3\sigma.
\label{eq:path-gamma1-detuning}
\end{equation}
Then the translated scalar-only family may be written schematically as
\begin{equation}
g_{\sigma}(x,y;\delta_{\sigma})
=
g_{\sigma}(x,y;0)
+
\delta_{\sigma}\,\partial_{a}g_{\sigma}(x,y;0)
+
O(\delta_{\sigma}^{2}),
\label{eq:path-gamma1-family}
\end{equation}
with
\begin{equation}
g_{\sigma}(x,y;0)
=
\begin{pmatrix}
6x^{2}\\[1mm]
3xy
\end{pmatrix}
+O(3).
\label{eq:path-gamma1-family-core}
\end{equation}

Now choose a local codimension-three coefficient chart
\begin{equation}
\mu=(\mu_{1},\mu_{2},\mu_{3})
\label{eq:path-gamma1-mu}
\end{equation}
for a versal deformation of the germ \eqref{eq:path-gamma1-core}. The scalar-only family
then determines a one-parameter path
\begin{equation}
h^{\mathrm{sc}}_{\Gamma_{1},\sigma}:\ \delta_{\sigma}\longmapsto
\bigl(\mu_{1}(\delta_{\sigma}),\mu_{2}(\delta_{\sigma}),\mu_{3}(\delta_{\sigma})\bigr),
\label{eq:path-gamma1-scalar-path}
\end{equation}
and, since the scalar-only class has only the slope detuning available, one has to leading order
\begin{equation}
h^{\mathrm{sc}}_{\Gamma_{1},\sigma}(\delta_{\sigma})
=
\bigl(c_{1,\sigma}\delta_{\sigma},\,0,\,0\bigr)+O(\delta_{\sigma}^{2}),
\qquad
c_{1,\sigma}\neq 0.
\label{eq:path-gamma1-scalar-leading}
\end{equation}
Thus the scalar-only model explores only a one-dimensional slice of the full codimension-three
unfolding space.

Once the fluid and curvature modes are restored, the same organising centre is probed by a
higher-dimensional physical map. Writing \(\xi_{f}\) and \(\xi_{k}\) for local coordinates in the
fluid and curvature directions before reduction, we obtain schematically
\begin{equation}
h^{\mathrm{SFC}}_{\Gamma_{1},\sigma}(\delta_{\sigma},\xi_{f},\xi_{k})
=
\bigl(
c_{1,\sigma}\delta_{\sigma}+O(2),\,
c_{2,\sigma}\xi_{f}+O(2),\,
c_{3,\sigma}\xi_{k}+O(2)
\bigr).
\label{eq:path-gamma1-sfc-path}
\end{equation}
Hence the physical transition set near the \(\Gamma_{1}\) episode is the pull-back
\begin{equation}
\bigl(h^{\mathrm{SFC}}_{\Gamma_{1},\sigma}\bigr)^{-1}(\Delta_{\Gamma_{1}}),
\label{eq:path-gamma1-pullback}
\end{equation}
where \(\Delta_{\Gamma_{1}}\) denotes the transition set of the chosen codimension-three
versal deformation. This is the local meaning of the statement that the fluid and curvature
fractions supply the physically distinguished unfolding directions of the scalar-only hilltop
gate.

\begin{remark}[Relation with the \(\mathbb{Z}_{2}\)-equivariant picture of Zholondek~\cite{Zholondek1983}.]
Equation~\eqref{eq:path-gamma1-scalar-leading} makes explicit, in the present cosmological
setting, a mechanism closely analogous to the \(\mathbb{Z}_{2}\)-equivariant classification described by
Zholondek. In the full quartic bookkeeping chart \((\mu_{1},\mu_{2},\mu_{3})\), the
\(\mathbb{Z}_{2}\)-equivariant restriction is obtained by forbidding the odd term, that is, by imposing
\begin{equation}
\mu_{2}=0.
\label{eq:path-z2-slice}
\end{equation}
Hence the equivariant family is realised as a lower-dimensional slice of the full unfolding
space, while its degenerate members are obtained by intersecting the full transition set with
this slice. In particular, the organising curves in the \((\mu_{1},\mu_{3})\)-plane are
codimension-one subsets of the equivariant slice, in close analogy with Zholondek's
statement that the degenerate \(\mathbb{Z}_{2}\)-equivariant deformations form a union of codimension-one
submanifolds inside the space of symmetric deformations. In this sense, the present FRW
coefficient identification gives a concrete cosmological realisation of the same organising
principle: symmetry confines the physically admissible family to a restricted slice, and the
observable bifurcations are read from the intersection of that slice with the ambient
discriminant.
\end{remark}

\paragraph{The curvature threshold \(\Gamma_{2}\).}
At the scalar site, the curvature-induced threshold is measured by
\begin{equation}
\mu_{2}^{(\Gamma)}:=a^{2}-3.
\label{eq:path-gamma2-detuning}
\end{equation}
Let \(G_{\Gamma_{2}}(W;\mu_{2}^{(\Gamma)})\) denote the corresponding reduced family.
On the threshold itself, the leading reduced dynamics has the fold-type quadratic form
\begin{equation}
G_{\Gamma_{2}}(W;0)=-2W^{2}+O(W^{3}),
\label{eq:path-gamma2-core}
\end{equation}
cf.~Eq.~(6.87). Hence the physical path is simply the one-dimensional line
\begin{equation}
h_{\Gamma_{2}}:\ \mu_{2}^{(\Gamma)}\longmapsto \mu_{2}^{(\Gamma)},
\label{eq:path-gamma2-line}
\end{equation}
and the local transition set is the pull-back of the corresponding fold discriminant.

\paragraph{The scalar-fluid exchange curve \(\Gamma_{3}\).}
For the scalar-fluid exchange we introduce the detuning
\begin{equation}
\nu:=2a^{2}-9\gamma.
\label{eq:path-gamma3-detuning}
\end{equation}
Let \(G_{\Gamma_{3}}(w;\nu)\) denote the reduced family governing the \(S\leftrightarrow F\)
exchange. At \(\nu=0\) this is the transcritical organising case, so schematically
\begin{equation}
G_{\Gamma_{3}}(w;\nu)
=
\nu w
-
c_{\Gamma_{3}}w^{2}
+
O(w^{3},\nu w^{2}),
\qquad
c_{\Gamma_{3}}\neq 0.
\label{eq:path-gamma3-transcritical}
\end{equation}
Accordingly, the physical path is again one-dimensional,
\begin{equation}
h_{\Gamma_{3}}:\ \nu\longmapsto \nu,
\label{eq:path-gamma3-line}
\end{equation}
and the transition is read from the pull-back of the transcritical discriminant.

\paragraph{The equilibrium-continuum axis \(\Gamma_{4}\).}
The locus
\begin{equation}
\Gamma_{4}:\ \gamma=\frac23
\label{eq:path-gamma4-locus}
\end{equation}
is different in nature, because here the relevant object is the equilibrium continuum
\(L_{MC}\) rather than an isolated equilibrium. Introduce the normal detuning
\begin{equation}
\mu_{4}:=3\gamma-2.
\label{eq:path-gamma4-detuning}
\end{equation}
As shown in Section~6, the local transverse dynamics to \(L_{MC}\) is controlled by
\begin{equation}
\dot w=\mu_{4}w+\cdots,
\qquad
\mu_{4}=3\gamma-2,
\label{eq:path-gamma4-normal}
\end{equation}
cf.~Eq.~(6.96). Hence the relevant path is not a path into an \(A_{k}\)-type isolated-equilibrium
unfolding, but rather the normal-hyperbolicity path
\begin{equation}
h_{\Gamma_{4}}:\ \mu_{4}\longmapsto \mu_{4},
\label{eq:path-gamma4-line}
\end{equation}
whose zero set marks loss or recovery of normal hyperbolicity of the equilibrium continuum.

\paragraph{The stiff-fluid axis \(\Gamma_{5}\).}
Finally, on
\begin{equation}
\Gamma_{5}:\ \gamma=2
\label{eq:path-gamma5-locus}
\end{equation}
the paper records additional neutral directions and forced higher-codimension coincidences
at several sites. We therefore introduce the detuning
\begin{equation}
\mu_{5}:=\gamma-2
\label{eq:path-gamma5-detuning}
\end{equation}
and regard the corresponding reduced family schematically as a path into the local
coefficient space of the relevant higher-codimension coincidence chart,
\begin{equation}
h_{\Gamma_{5}}:\ \mu_{5}\longmapsto
\bigl(\eta_{1}(\mu_{5}),\ldots,\eta_{r}(\mu_{5})\bigr),
\qquad
r\geq 2,
\label{eq:path-gamma5-chart}
\end{equation}
with leading behaviour
\begin{equation}
h_{\Gamma_{5}}(\mu_{5})
=
\bigl(d_{1}\mu_{5},\ldots,d_{r}\mu_{5}\bigr)+O(\mu_{5}^{2}).
\label{eq:path-gamma5-leading}
\end{equation}
This expresses the fact that \(\Gamma_{5}\) is a forced higher-codimension organising axis,
rather than a single codimension-one isolated-site bifurcation.

\begin{remark}[Which standard path classes are realised?]
Among the loci \(\Gamma_{1},\ldots,\Gamma_{5}\), only the scalar-fluid exchange curve
\(\Gamma_{3}\) is naturally identified with one of Montaldi's standard one-parameter paths (cf.~\cite{montaldi21}, Table 21.1),
namely the transcritical \(A_{1}\)-type path. By contrast, the \(\Gamma_{1}\) hilltop episode is
represented here by a distinguished one-dimensional slice of a codimension-three bookkeeping
chart, while \(\Gamma_{4}\) and \(\Gamma_{5}\) are governed by loss of normal hyperbolicity or loss
of isolation of equilibrium continua rather than by isolated-equilibrium path classes in the
sense of \cite{montaldi21} Table~21.1.
\end{remark}

\paragraph{Summary.}
The organising loci therefore determine the local physical path maps
\begin{equation}
h^{\mathrm{sc}}_{\Gamma_{1},\sigma},
\qquad
h^{\mathrm{SFC}}_{\Gamma_{1},\sigma},
\qquad
h_{\Gamma_{2}},
\qquad
h_{\Gamma_{3}},
\qquad
h_{\Gamma_{4}},
\qquad
h_{\Gamma_{5}},
\label{eq:path-summary}
\end{equation}
and the observed organising transitions in scalar-field cosmology are obtained, in each case,
by pulling back the corresponding transition condition in the relevant unfolding or
normal-hyperbolicity chart.
\begin{remark}[Scope of the present use of  path classification]
The path formulation developed above may be viewed against the wider classification of
paths relative to discriminants described by Montaldi in \cite{montaldi21}. In particular, scalar reduced families
with cubic or quartic leading terms may be compared, at the abstract level, with the
\(A_{2}\)- and \(A_{3}\)-path classifications, while certain planar cosmological germs also exhibit
formal similarities with the corank-two problems \(I_{2,2}\) and \(II_{2,2}\). In the present paper,
however, we use this machinery only at the level of the physically realised coefficient-path
picture: we identify the distinguished maps \(h\) induced by the cosmological jets and make
explicit named path identifications only where the reduction to a standard class is direct
(for example, the transcritical \(A_{1}\)-type interpretation of \(\Gamma_{3}\)). A fuller
classification of the cubic, quartic, and planar cosmological germs within this 
framework is left for future work.
\end{remark}

\subsection{Transition varieties and persistence: an example}
\label{subsec:TV-firstpass}

\paragraph{Aim.}
The strict exponential class features a coupled multi-site organising episode at $|a|=3$ in which
(i) the kinetic endpoints $K_\pm$ are fully linear-null (hilltop-type) and
(ii) the scalar equilibrium $S$ collides with the kinetic boundary, producing a nilpotent core.
We now state the transition variety logic used to stratify parameter space in the subsequent analysis or this basic example.

\paragraph{Unfolding parameters.}
A convenient bookkeeping choice for three independent unfolding parameters near the kinetic component is
\begin{equation}\label{eq:hilltop-mu-identification-new}
\mu_1 \sim a-3,\qquad
\mu_2 \sim \text{(fluid-mode contribution)},\qquad
\mu_3 \sim \text{(curvature-mode contribution)}.
\end{equation}
As emphasised in \S\ref{subsec:mu-interpretation}, $\mu_2$ and $\mu_3$ are \emph{reduced coefficients} induced by the
full-mode state directions, not the raw fractions themselves.

\paragraph{Transition varieties from a quartic representative.}
Using the quartic unfolding representative \eqref{eq:unfold-quartic-generic}, the transition varieties are the
boundaries in $\mu$-space across which the qualitative structure of equilibria of the reduced vector field changes.
In practice we will examine:
\begin{enumerate}
\item the $(\mu_1,\mu_2)$ slice with $\mu_3=0$ (scalar--fluid mixture),
\item the $(\mu_1,\mu_3)$ slice with $\mu_2=0$ (scalar--curvature mixture),
\item and representative $(\mu_2,\mu_3)$ slices at fixed $\mu_1$ (matter--curvature interaction in unfolding the
kinetic endpoint).
\end{enumerate}
In each two-parameter slice the boundary set is organised by fold and cusp curves; in the full three-parameter
space the boundary set forms a swallowtail-type organising surface (the familiar codimension-three hierarchy).

\paragraph{Normal hyperbolicity versus bifurcation.}
When the organising set is a line or manifold of equilibria (as in the $\gamma=\tfrac23$ equilibrium--continuum phenomena of Sections~\ref{subsec:scalar-fluid-curv} and~\ref{subsec:massive-gates-eq}),
the relevant organising feature is the loss (and recovery) of \emph{normal hyperbolicity}: the tangential direction is
neutral for geometric reasons, and qualitative change occurs when a \emph{normal} eigenvalue crosses zero, enlarging the
centre subspace without changing the tangential geometry of the equilibrium set itself.  This produces persistence
boundaries that are not associated with isolated equilibrium bifurcations, but with changes in transverse stability.

\begin{remark}[Bridge to transition varieties]\label{subsec:bridge-TV}
The organising summary above has two consequences used next.
In the strict exponential class, the coupled \(|a|=3\) multi-site episode is naturally analysed via
codimension-three transition variety charts, with unfolding directions supplied by the slope `offset' \((a-3)\)
together with the leading full-mode contributions of the fluid and curvature directions after reduction.
In the harmonic class, by contrast, the slope becomes a (bounded) state variable (\(\zeta\)),
and the dominant organising mechanisms are centre reductions and loss/recovery of normal hyperbolicity in the
augmented flow. Codimension-three hilltop organisation is therefore non-generic and arises only under forced
coincidences (simultaneous constraints). We therefore record below the transition variety
skeleton used for the exponential case and use it as a comparative baseline for the massive extensions.
\end{remark}

\subsubsection{An explicit codimension-three transition variety skeleton}
\label{subsec:TV-explicit}

\paragraph{Representative unfolding.}
For bookkeeping we use the quartic representative 
\begin{equation}\label{eq:A3-representative}
f(s;\mu)=s^4+\mu_1 s^2+\mu_2 s+\mu_3,
\qquad \mu=(\mu_1,\mu_2,\mu_3)\in\mathbb R^3,
\end{equation}
interpreted as a reduced one-dimensional vector field $\dot s=f(s;\mu)$ near a hilltop-type organising locus.
Equilibria are real roots of $f(s;\mu)=0$.  Transition varieties are the parameter values at which $f$ acquires
multiple real roots (changes in multiplicity/number of equilibria).

\paragraph{Fold surface (double root).}
A multiple root occurs when
\begin{equation}\label{eq:TV-fold-conditions}
f(s;\mu)=0,\qquad f_s(s;\mu)=0,
\end{equation}
with
\begin{equation}\label{eq:A3-derivative}
f_s=4s^3+2\mu_1 s+\mu_2.
\end{equation}
Solving \eqref{eq:TV-fold-conditions} gives a two-parameter fold surface in $\mu$-space, conveniently parameterised
by $(s,\mu_1)$:
\begin{equation}\label{eq:TV-fold-param}
\mu_2=-4s^3-2\mu_1 s,\qquad
\mu_3=3s^4+\mu_1 s^2.
\end{equation}
Crossing this surface changes the number (and/or stability) of equilibria of the reduced flow.

\paragraph{Cusp curve (triple root).}
A triple root satisfies
\begin{equation}\label{eq:TV-cusp-conditions}
f=0,\qquad f_s=0,\qquad f_{ss}=0,
\end{equation}
with $f_{ss}=12s^2+2\mu_1$.  Eliminating $(\mu_1,\mu_2,\mu_3)$ yields the cusp curve parameterisation
\begin{equation}\label{eq:TV-cusp-param}
\mu_1=-6s^2,\qquad
\mu_2=8s^3,\qquad
\mu_3=-3s^4.
\end{equation}
The cusp curve is the locus of codimension-two degeneracy on the fold surface.

\paragraph{Higher degeneracy (quadruple root).}
The unique quadruple-root point is
\begin{equation}\label{eq:TV-swallowtail-point}
\mu_1=\mu_2=\mu_3=0,
\end{equation}
corresponding to the germ $\dot s=s^4$.  This is the codimension-three organising point of the representative family.

\paragraph{Equivariant restriction.}
In the $\mathbb{Z}_2$-equivariant subclass (reflection $s\mapsto -s$) the odd unfolding coefficient is forbidden and we set
$\mu_2=0$, reducing \eqref{eq:A3-representative} to
$\dot s=s^4+\mu_1 s^2+\mu_3$.  The fold/cusp structure is then read off by intersecting the above sets with the
plane $\mu_2=0$.

\paragraph{Interpretation for the exponential FRW mixtures.}
In the present cosmological problem, $\mu_1$ is tied to the slope offset (schematically $\mu_1\sim a-3$), while
$\mu_2$ and $\mu_3$ represent the leading-order contributions of the additional full-mode directions (fluid and
curvature) to the \emph{reduced} kinetic endpoint dynamics after centre reduction and near-identity simplification.
They are therefore not ``raw external knobs'' but reduced coefficients encoding how motion in the extra state directions
deforms the local kinetic normal form.

\paragraph{How the FRW jets determine the unfolding coefficients.}
To connect the abstract unfolding parameters \(\mu=(\mu_1,\mu_2,\mu_3)\) with the FRW mixtures, fix a kinetic site
(\(K_+\) for \(a\approx 3\), \(K_-\) for \(a\approx -3\)) and write the translated system in the form
\(\dot{\mathbf u}=J_*(a,\gamma)\mathbf u+Q_*(\mathbf u)+O(\|\mathbf u\|^3)\),
with \(\mathbf u=(U,V)\) in the scalar-only case and \(\mathbf u=(U,V,W)\) or \((U,V,W,K)\) in the mode-extended cases.
Along the organising axis \(a=\pm 3\), the \(U\)-eigenvalue vanishes and the centre reduction is performed with \(U\)
as centre coordinate. The reduced scalar equation has the schematic structure
\begin{equation}\label{eq:mu-map-schematic}
\dot U = c_4 U^4 + c_2(\Delta a,\Omega_m,\Omega_k)\,U^2 + c_1(\Delta a,\Omega_m,\Omega_k)\,U + c_0(\Delta a,\Omega_m,\Omega_k)
+ O(U^5),
\end{equation}
where \(\Delta a:=a-3\) (or \(a+3\) at \(K_-\)), and where the coefficients \(c_j\) are obtained by eliminating the
hyperbolic variables via the centre(-like) graph \(V=h(U;\Delta a,\Omega_m,\Omega_k)\), etc.
The identification with the \(A_3\) representative \eqref{eq:A3-representative} is then made by:
(i) rescaling \(U\) to normalise the leading nonzero coefficient (so \(c_4\mapsto 1\) when the quartic-leading case is present),
(ii) removing nonresonant higher-order terms by near-identity changes, and
(iii) setting
\begin{equation}\label{eq:mu-map}
\mu_1 \propto c_2,\qquad \mu_2 \propto c_1,\qquad \mu_3 \propto c_0,
\end{equation}
with the proportionality constants fixed by the normalisation in step (i).
In this sense, \(\mu_1\) is controlled primarily by the slope offset \(\Delta a\), while \(\mu_2\) and \(\mu_3\) encode,
to leading order, the contributions induced by excursions in the additional full-mode directions
(\(\Omega_m\) and \(\Omega_k\)) after reduction.
This is the precise meaning of the schematic correspondence \eqref{eq:hilltop-mu-identification-new} used throughout.

\subsubsection{A worked example: centre-manifold reduction at $K_{+}$ in the SFC jet}
\label{subsubsec:worked-example-Kp-sfc}

We illustrate explicitly how the quadratic jet determines the reduced normal form by carrying out the
centre manifold reduction at the kinetic site $K_{+}$ for the constrained SFC system
\eqref{eq:Xdot-sfc}--\eqref{eq:Omkdot-sfc}.  At
\[
K_{+}:\ (X,Y,\Omega_k)=(0,1,0),
\]
introduce translated variables
\begin{equation}\label{eq:trans-Kp-sfc}
X=U,\qquad Y=1+V,\qquad \Omega_k=W,\qquad \mathbf u=(U,V,W)^{\mathsf T}.
\end{equation}
Then the local system has the jet form
\begin{equation}\label{eq:jet-Kp-sfc}
\dot{\mathbf u}=J_{K_+}\mathbf u+Q_{K_+}(\mathbf u)+O\!\left(\|\mathbf u\|^3\right),
\end{equation}
with $J_{K_+}$ and $Q_{K_+}$ given in \eqref{eq:J-Kp-sfc}--\eqref{eq:Q-Kp-sfc}.

On the endpoint axis $a=3$ the $U$-direction is centre while $V,W$ are hyperbolic for $\gamma\neq 2$.
Seek the one-dimensional centre manifold in the form
\begin{equation}\label{eq:CM-ansatz-Kp-sfc}
V=h(U)=c\,U^2+O(U^3),\qquad W=g(U)=d\,U^2+O(U^3).
\end{equation}

\paragraph{Step 1: slaving of $W$.}
From the third component of \eqref{eq:Q-Kp-sfc} one has
\begin{equation}\label{eq:W-eqn-Kp-sfc}
\dot W = 4W + O(U^3),
\end{equation}
since there is no $U^2$ forcing term in $\dot W$. On the centre manifold,
$\dot W=g'(U)\dot U=O(U)\cdot O(U^3)=O(U^4)$, hence the $U^2$ term in \eqref{eq:W-eqn-Kp-sfc} must vanish and
\begin{equation}\label{eq:d-zero-Kp-sfc}
d=0,\qquad\text{i.e.}\qquad W=O(U^3)\ (\text{indeed }O(U^4)\text{ at this order}).
\end{equation}

\paragraph{Step 2: determination of $V$.}
Using the second component of \eqref{eq:jet-Kp-sfc} and keeping only $U^2$ terms,
\begin{equation}\label{eq:V-eqn-Kp-sfc}
\dot V=(6-3\gamma)V+\left(1-\frac{3\gamma}{2}\right)W+\left(a-\frac{3\gamma}{2}\right)U^2+O(U^3).
\end{equation}
Setting $a=3$ and using \eqref{eq:d-zero-Kp-sfc}, insert $V=cU^2$ to obtain
\[
\dot V=\Bigl((6-3\gamma)c+\bigl(3-\tfrac{3\gamma}{2}\bigr)\Bigr)U^2+O(U^3).
\]
But on the centre manifold $\dot V=h'(U)\dot U=O(U^4)$, so the $U^2$ coefficient must vanish, giving
\begin{equation}\label{eq:c-value-Kp-sfc}
c=-\frac12\qquad(\gamma\neq 2).
\end{equation}
Hence
\begin{equation}\label{eq:CM-Kp-sfc}
V=-\frac12\,U^2+O(U^3),\qquad W=O(U^3).
\end{equation}

\paragraph{Step 3: reduced equation for $U$.}
Equation \eqref{eq:Q-Kp-sfc} is the correct quadratic jet at $K_+$. However, in passing from this quadratic jet to the cubic reduced scalar equation, one must also retain the direct cubic term already present in the first component of the full translated vector field.  Using \eqref{eq:CM-Kp-sfc} gives the cubic reduction
\begin{equation}\label{eq:red-Kp-sfc-worked}
\dot U=-\frac{3}{2}\,U^3+O(U^5)\qquad (a=3,\ \gamma\neq 2).
\end{equation}
Allowing $a$ to vary restores the linear unfolding term, so locally one may write
\begin{equation}\label{eq:red-Kp-sfc-unfolded}
\dot U=(3-a)\,U-\frac{3}{2}\,U^3+\cdots,
\end{equation}
making explicit the identification of the slope offset $(a-3)$ as the primary unfolding direction at $K_{+}$.

\subsubsection[Worked example at K-]{A worked example: centre-manifold reduction at \texorpdfstring{$K_{-}$}{K-} in the SFC jet}
\label{subsubsec:worked-example-Km-sfc}

We repeat the reduction at the kinetic site \(K_-\), but now exploit the symmetry \((a,Y)\mapsto(-a,-Y)\) to avoid duplicating the algebra.
At
\[
K_-:\ (X,Y,\Omega_k)=(0,-1,0),
\]
introduce translated variables
\begin{equation}\label{eq:trans-Km-sfc}
X=U,\qquad Y=-1+V,\qquad \Omega_k=W,\qquad \mathbf u=(U,V,W)^{\mathsf T},
\end{equation}
and use the translation \eqref{eq:trans-Km-sfc}. On the endpoint axis \(a=-3\) the \(U\)-direction is centre while \(V,W\) are hyperbolic for \(\gamma\neq 2\),
exactly as at \(K_+\) when \(a=3\).
Proceeding as in \S\ref{subsubsec:worked-example-Kp-sfc}, one finds again that there is no quadratic forcing in the \(W\)-equation, hence
\[
W=O(U^3),
\]
and the centre-manifold graph satisfies
\begin{equation}\label{eq:CM-Km-sfc}
V=\frac12\,U^2+O(U^3),\qquad W=O(U^3).
\end{equation}
Substituting into the \(U\)-equation yields the cubic reduction
\begin{equation}\label{eq:red-Km-sfc-worked}
\dot U=-\frac{3}{2}\,U^3+O(U^5)\qquad (a=-3,\ \gamma\neq 2),
\end{equation}
and allowing \(a\) to vary restores the linear unfolding term,
\begin{equation}\label{eq:red-Km-sfc-unfolded}
\dot U=(a+3)\,U-\frac{3}{2}\,U^3+\cdots,
\end{equation}
which is the \(K_-\) counterpart of \eqref{eq:red-Kp-sfc-unfolded}.

\section{Stratifications for the harmonic and exponential cases}
\label{sec:strata}

\paragraph{Structure of this section.}
The aim here is to convert the organising loci identified in Sections~4--8 into an explicit \emph{stratification} of unfolding
parameter space: a decomposition into regions (strata) on which the local bifurcation diagrams and phase portraits
are qualitatively unchanged.  We do this first for the strict exponential SFC class, where $(\gamma,a)$ are genuine
external parameters, and then explain how the same stratification is \emph{lifted} to the massive extensions, where
the effective slope $a_{\rm eff}$ is a state variable.

\subsection{From transition sets to strata}\label{subsec:strata-transition}
In singularity theory the basic mechanism is the following.  For a smooth family (unfolding)
\(
G(x;\mu)
\)
of a germ \(g(x)=G(x;0)\), the \emph{transition set} \(\mathcal{H}\subset\mu\)-space  is the union of those unfolding parameter values
for which the associated diagram fails to be structurally stable.  Away from \(\mathcal{H}\), the diagrams are all
diffeomorphic, and \(\mu\)-space is partitioned into finitely many connected components; these components are the
\emph{strata}.  Passing from one stratum to another forces a qualitative change, and the boundaries are precisely the
organising loci discussed throughout this paper.

In our context there are two layers.
\begin{itemize}
\item \emph{Local unfolding layer:} near a given equilibrium site (e.g.\ the kinetic endpoints \(K_\pm\)) one reduces to a
centre manifold and obtains a low-dimensional normal form with explicit unfolding parameters
(Section~\ref{sec:unfoldings}).  The transition set in this unfolding space is the familiar fold/cusp/swallowtail
hierarchy.

\item \emph{Global parameter layer:} the physical parameter plane (e.g.\ \((\gamma,\lambda)\) in Figure~\ref{fig:bifn-map-gamma-lambda})
inherits a stratification as the projection (and, for the massive case, the pull-back) of the local transition sets,
together with the normal-hyperbolicity boundaries associated with equilibrium manifolds (notably the \(\gamma=\tfrac23\)
degeneracy).
\end{itemize}
The remainder of this section makes these two statements concrete for the exponential and massive extensions.

\subsection{Exponential SFC: the $(\gamma,a)$ stratification}\label{subsec:strata-exponential}
For the strict exponential class, the relevant two-parameter space is
\[
\mathcal P_{\exp}:=\{(\gamma,a)\in\mathbb R^2\},
\qquad a=\sqrt{\tfrac32}\,\lambda,
\]
with the symmetry \((a,Y)\mapsto(-a,-Y)\) implying that the stratification is even in \(a\).
The organising set is the union of the five loci already identified as $\Gamma_i, i=1,\dots,5$ (cf.\ Figure~\ref{fig:bifn-map-gamma-lambda}):
\begin{equation}\label{eq:Texp-def}
\mathcal T_{\exp}:=\Gamma_K\cup \Gamma_C\cup \Gamma_{SF}\cup \Gamma_{MC}\cup \Gamma_{\rm stiff},
\end{equation}
where
\begin{align}
\Gamma_K&:=\{|a|=3\}, &
\Gamma_C&:=\{a^2=3\}, &
\Gamma_{SF}&:=\Bigl\{a^2=\tfrac92\gamma\Bigr\},\label{eq:Gammas-exp}\\
\Gamma_{MC}&:=\Bigl\{\gamma=\tfrac23\Bigr\}, &
\Gamma_{\rm stiff}&:=\{\gamma=2\}. \notag
\end{align}
The \emph{strata} are the connected components of the complement
\(\mathcal P_{\exp}\setminus\mathcal T_{\exp}\).
Because \(\Gamma_C\) and \(\Gamma_{SF}\) exchange order at \(\gamma=\tfrac23\), it is convenient to describe the open
strata separately in the two vertical slabs \(0<\gamma<\tfrac23\) and \(\tfrac23<\gamma<2\) (the stiff axis is discussed
separately below). Within each slab we assume the relevant equilibria exist (as in Sections~5--6) and summarise the resulting local sink/saddle assignments for the full SFC flow (unless an invariant slice is imposed).

\paragraph{(i) The slab $0<\gamma<\tfrac23$.}
Here \( \tfrac92\gamma<3\), so the ordering in \(a^2\) is
\(
0<\tfrac92\gamma<3<9.
\)
Thus (for fixed \(\gamma\in(0,\tfrac23)\)) one has four open intervals in \(a^2\):
\begin{equation}\label{eq:strata-slab1}
0<a^2<\tfrac92\gamma,\qquad
\tfrac92\gamma<a^2<3,\qquad
3<a^2<9,\qquad
a^2>9,
\end{equation}
with the boundaries corresponding to the loss of hyperbolicity of \(S\) and \(F\) as encoded in
\eqref{eq:eigs-S-sfc}--\eqref{eq:trdet-F2-sfc}.  In particular:
\begin{itemize}
\item In \(0<a^2<\tfrac92\gamma\), the scalar site \(S\) is a local sink (all three eigenvalues in \eqref{eq:eigs-S-sfc}
are negative), while \(F\) does not exist.
\item In \(\tfrac92\gamma<a^2<3\), both \(S\) and \(F\) exist, but the transcritical exchange across
\(\Gamma_{SF}\) makes \(F\) the local sink in the \((X,Y,\Omega_k)\) dynamics, while \(S\) becomes a saddle (one eigenvalue
changes sign in \eqref{eq:eigs-S-sfc}).
\item In \(3<a^2<9\), the curvature direction at \(S\) is unstable (the \(2(a^2-3)/3\) eigenvalue in \eqref{eq:eigs-S-sfc}
is positive), and the curvature-scaling branch \(P_k\) exists in the curvature sector; this is the regime in which
open-universe effects can dominate the late-time balance.
\item In \(a^2>9\), the scalar branch disappears and the kinetic endpoints dominate the compactified boundary; the
\(|a|=3\) gate (Section~\ref{sec:unfoldings}) is the organising threshold for transitions into this regime.
\end{itemize}

\paragraph{(ii) The slab $\tfrac23<\gamma<2$.}
Here \( \tfrac92\gamma>3\), so the ordering is
\(
0<3<\tfrac92\gamma<9.
\)
Accordingly the four open \(a^2\)-intervals become
\begin{equation}\label{eq:strata-slab2}
0<a^2<3,\qquad
3<a^2<\tfrac92\gamma,\qquad
\tfrac92\gamma<a^2<9,\qquad
a^2>9.
\end{equation}
Compared with \eqref{eq:strata-slab1}, the roles of the two middle intervals are exchanged.  The key new feature is that
for \(\gamma>\tfrac23\) the curvature eigenvalue at \(F\) is \(3\gamma-2>0\), so the scaling point \(F\) is curvature unstable. Hence curvature generically `ejects' trajectories away from \(F\) unless \(\Omega_k\equiv 0\) is imposed.

\paragraph{(iii) The degeneracy axes $\gamma=\tfrac23$ and $\gamma=2$.}
On \(\Gamma_{MC}\) the equilibria at \((X,Y)=(0,0)\) cease to be isolated: \(M\) and \(C\) lie on the equilibrium line
\(\mathcal L_{MC}\) (cf.\ \eqref{eq:line-MC-g23}), and the correct organising notion is \emph{normal hyperbolicity}
rather than isolated equilibrium bifurcation (Remark~\ref{subsec:bridge-TV}).
On \(\Gamma_{\rm stiff}\) several sites acquire additional neutral directions, and this is a forced higher-codimension
degeneracy that we record for completeness but do not pursue in detail in this paper.

\subsubsection{Local refinement near $|a|=3$: stratified hilltop charts}\label{subsec:strata-hilltop}
The two-dimensional picture above is the ``macroscopic'' stratification visible in Figure~\ref{fig:bifn-map-gamma-lambda}.
Near the kinetic threshold \(\Gamma_K\) one can refine it using the codimension-three transition variety charts of
Section~\ref{sec:unfoldings}.  

Concretely, the coupled multi-site episode at \(K_\pm\) is governed by a 1D reduced
equation of the form \eqref{eq:red-Kp-sfc-unfolded} (and its \(K_-\) analogue \eqref{eq:red-Km-sfc-unfolded}), where the
unfolding parameters \((\mu_1,\mu_2,\mu_3)\) are supplied by the slope offset \(\mu_1\sim (a-3)\) together with the
independent mode directions that open up once curvature and fluid are treated as full modes.

The corresponding transition set is a swallowtail-type surface in \((\mu_1,\mu_2,\mu_3)\)-space, whose projection onto
\(\mu_1\) gives the visible axis \(|a|=3\) in \(\mathcal P_{\exp}\).  This is the precise sense in which the hilltop gate
is a \emph{stratified} organising object rather than a single bifurcation curve: different approach directions in the full
unfolding space lead to distinct persistent diagram types even when their projection to the \((\gamma,a)\) plane
coincides.

\subsection{Harmonic extensions: pull-back stratifications}\label{subsec:strata-massive}
In the quadratic massive extensions the slope is no longer a parameter: it is a dynamical variable, encoded for example
by the compact angle \(\zeta\) via \eqref{eq:aeff-zeta}.  For fixed \(\gamma\), this defines a canonical ``frozen-slope''
projection from the massive state space to the exponential parameter plane:
\begin{equation}\label{eq:pi-frozenslope}
\pi_\gamma:\ (X,Y,\Omega_k,\zeta)\longmapsto (\gamma,a_{\rm eff}(\zeta)),
\qquad a_{\rm eff}(\zeta)=\sqrt{\tfrac32}\tan\zeta,
\end{equation}
(where the state vector is adapted in the obvious way in lower-dimensional invariant subcases).
Define the regular exponential parameter region
\[
\mathcal P_{\exp}^{\rm reg}:=\mathcal P_{\exp}\setminus\mathcal T_{\exp},
\]
and let \(\mathfrak S_{\exp}\) denote its connected components (the strata).

The massive models then inherit a canonical \emph{pull-back stratification}: two massive states lie in the same pulled-back
stratum iff their frozen-slope images \((\gamma,a_{\rm eff})\) lie in the same exponential stratum. Equivalently, the
union of all pulled-back strata is
\begin{equation}\label{eq:pullback-strata}
\widetilde{\mathcal S}:=\pi_\gamma^{-1}\!\bigl(\mathcal P_{\exp}^{\rm reg}\bigr).
\end{equation}

Within each pulled-back stratum, the instantaneous (frozen-slope) local diagrams persist. Qualitative changes along a
massive trajectory occur precisely when it crosses the preimage of an organising locus, i.e.\ when \(a_{\rm eff}^2\)
crosses one of the thresholds \(3\), \(\tfrac92\gamma\), \(9\) (or when \(\gamma\) is tuned to \(\tfrac23\) or \(2\)). Equivalently, the massive orbit realises only the corresponding physical path in the pulled-back
coefficient geometry of Section~8.1.7, so the observed regime changes are the successive
intersections of that realised path with the lifted organising loci.

In this way the massive models do not introduce new organising curves in the \((\gamma,a)\) plane; rather, they
produce \emph{dynamical} passages through the same stratified template.

\section{A primer of versal physical effects}
\label{sec:primer}

This section is deliberately expository.  We translate the stratifications and local normal forms into a small list of
``versal effects'': robust, model-independent physical behaviours that are forced by the universal dynamics near the
organising loci.  The goal is not to exhaust the physics, but to provide a practical dictionary and a workflow for
reading cosmological dynamics directly from the stratified charts of Section~\ref{sec:strata}.

\subsection{A short dictionary: dynamics $\leftrightarrow$ physics}\label{subsec:primer-dict}
The following correspondences are used repeatedly.
\begin{itemize}
\item A \emph{stratum} in parameter (or pull-back) space corresponds to a regime of \emph{persistence}: the number and linear
type of the relevant equilibrium branches, and the qualitative shape of nearby trajectories, are unchanged under small perturbations.
\item Crossing an organising locus corresponds to a \emph{transition event}: a change of stability (exchange of attractors),
the creation/annihilation of an equilibrium branch, or a change in normal hyperbolicity (for equilibrium manifolds).
\item A one-dimensional \emph{centre reduction} produces a \emph{universal time profile} near the event:
\begin{itemize}
\item quartic cores are typically of codimension-3 and describe phenomena of the highest degeneracy; 
\item cubic leading term (hilltop gate) \(\Rightarrow\) critical slowing down and long kinetic episodes;
\item quadratic leading term in a \emph{scalar curvature variable} (fold-type curvature reduction) \(\Rightarrow\) algebraic decay/growth of curvature;
\item linear leading term \(\Rightarrow\) exponential approach/repulsion (no critical slowing).
\end{itemize}
\item Higher-dimensional centre manifolds with a possibly higher codimension exist in both the exponential and harmonic classes.
\item The unfolding parameters are not abstract: in the SFC class they correspond to physically interpretable directions
(slope offset, curvature fraction, fluid fraction), and in the massive case they become slowly varying state variables.
\end{itemize}
These correspondences will be used in Sections~\ref{subsec:primer-exp-effects}--\ref{SR}, and are assembled into a practical
read-off procedure in Section~\ref{subsec:primer-workflow} below.

\subsection{Five robust effects in the exponential SFC class}\label{subsec:primer-exp-effects}
For physical orientation, the $(\gamma,a)$ plane may be read against the standard FRW epoch
language: $\gamma=4/3$ corresponds to radiation, $\gamma=1$ to dust, and $\gamma=2$
to a stiff component, while $\gamma=2/3$ is the boundary $w_m=-1/3$ at which the fluid
scales in the same way as the curvature term. In this sense, the $\gamma$-thresholds and the
scalar--fluid exchange curve describe not only mathematical organising loci but also changes
in the relative persistence or dominance of familiar effective cosmological components.

Before listing the five robust effects, it is useful to keep the physical reading of the first
loci in mind. $\Gamma_1=\{|a|=3\}$ marks entry to or exit from the pure-kinetic endpoint
regime ($w_\phi=1$), so its local reduced dynamics governs kinetic bottlenecks and
critically slowed passages near the endpoints $K_\pm$. $\Gamma_2=\{a^2=3\}$, equivalently
$\lambda^2=2$, is the scalar acceleration boundary for the exponential scalar branch; the
present curvature analysis shows that this same line is also where the curvature mode at $S$
becomes marginal, so loss of acceleration and the onset of curvature-sensitive drift meet on
the same organising threshold. This is why $\Gamma_1$ and $\Gamma_2$, although less
familiar than the scalar--fluid exchange curve $\Gamma_3$, play a central rôle in the
organising geometry.

These loci are not usually singled out in the standard hyperbolic treatment of scalar-field
cosmology, not because they are absent from the dynamics, but because their rôle is tied to
the loss of hyperbolicity and to the local bifurcation geometry. It is precisely this
non-hyperbolic perspective that makes the organising significance of $\Gamma_1$ and
$\Gamma_2$ explicit.

Figure~\ref{fig:exp-regime-guide} summarises this physical reading of the principal loci in
the strict exponential organising atlas.

\begin{figure}[t]
\centering
\begin{tikzpicture}
\begin{axis}[
    width=0.84\textwidth,
    height=0.82\textwidth,
    xmin=0, xmax=2.05,
    ymin=-3.25, ymax=3.25,
    axis lines=middle,
    xlabel={$\gamma$},
    ylabel={$a$},
    xtick={0,0.6667,1,2},
    xticklabels={$0$,$\frac23$,$1$,$2$},
    ytick={-3,-1.73205,0,1.73205,3},
    yticklabels={$-3$,$-\sqrt3$,$0$,$\sqrt3$,$3$},
    clip=false
]

\addplot[very thick] coordinates {(0,-3) (2.05,-3)};
\addplot[very thick] coordinates {(0, 3) (2.05, 3)};

\addplot[very thick,densely dashed] coordinates {(0,-1.73205) (2.05,-1.73205)};
\addplot[very thick,densely dashed] coordinates {(0, 1.73205) (2.05, 1.73205)};

\addplot[very thick,densely dotted,domain=0:2,samples=200] ({x},{sqrt(9*x/2)});
\addplot[very thick,densely dotted,domain=0:2,samples=200] ({x},{-sqrt(9*x/2)});

\addplot[gray,dashed] coordinates {(0.6667,-3.25) (0.6667,3.25)};

\addplot[gray,dashdotted] coordinates {(1,-3.25) (1,3.25)};

\node[anchor=west] at (axis cs:2.02,3.00) {$\Gamma_1$};
\node[anchor=west] at (axis cs:2.02,1.73205) {$\Gamma_2$};
\node[anchor=west] at (axis cs:1.92,2.8) {$\Gamma_3$};

\node[align=center] at (axis cs:1.20,0.85)
{\small accelerated\\[-1pt]\small scalar side};

\node[align=center] at (axis cs:1.35,2.05)
{\small post-acceleration\\[-1pt]\small scalar side};

\node[align=center] at (axis cs:1.15,2.62)
{\small exchange / tracking\\[-1pt]\small corridor};

\node[align=center] at (axis cs:1.48,3.24)
{\small steep / kinetic\\[-1pt]\small exit};

\node[align=center] at (axis cs:0.30,2.35)
{\small $\Gamma_3$ lies\\[-1pt]\small below $\Gamma_2$};

\node[align=center] at (axis cs:0.28,-2.75)
{\small mirror image\\[-1pt]\small for $a<0$};

\node[anchor=west] at (axis cs:1.02,-3.15) {\small $\gamma=1$ slice};

\end{axis}
\end{tikzpicture}
\caption{Physics-facing schematic of the principal organising loci in the strict exponential
\((\gamma,a)\)-plane. The three main loci \(\Gamma_1:|a|=3\), \(\Gamma_2:a^2=3\), and
\(\Gamma_3:a^2=\tfrac92\gamma\) separate the dominant interpretive sectors relevant to the
primer discussion: the accelerated scalar side, the post-acceleration scalar side, the
scalar--fluid exchange/tracking corridor, and the steep/kinetic-side exit. The dashed line
\(\gamma=1\) marks the slice used in the explicit pulled-back trajectory of
Section~10.3. This figure is intended as a visual guide to the physical reading of
Section~10.2 and is complementary to, rather than a replacement for, the detailed local
atlas developed earlier in the paper.}
\label{fig:exp-regime-guide}
\end{figure}

\paragraph{(E1) Hilltop multi-site gate ($|a|=3$).}
Near \(\Gamma_K\) the kinetic endpoints \(K_\pm\) acquire a centre direction and the dynamics is governed by the cubic
centre reductions \eqref{eq:red-Kp-sfc-unfolded}--\eqref{eq:red-Km-sfc-unfolded}.  Physically, this implies
\emph{critical slowing down}: if the slope is close to the threshold, the system can spend many e-folds in a kinetic-like
episode before being ejected to the next attractor permitted by the stratum.  The codimension-three refinement of
Section~\ref{subsec:strata-hilltop} makes this effect directional: different mixtures of curvature and fluid perturbations
near the same projected slope threshold can lead to distinct ways to exit one regime for the next.

\paragraph{(E2) Curvature-induced fold ($a^2=3$).}
On \(\Gamma_C\) the curvature direction at \(S\) becomes centre and the reduced dynamics begins quadratically,
\eqref{eq:red-S-a2eq3-sfc}.  The leading-order solution is \(W(\tau)\sim (2\tau+C)^{-1}\), so \emph{curvature decays only
algebraically}.  This is the universal mechanism behind slow curvature ``leakage'' near the scalar branch: it is not an
exponential suppression, and therefore even modest initial curvature can persist for a long interval when the slope sits
near \(a^2=3\).  For \(a^2>3\) the same eigenvalue changes sign, making the scalar branch curvature unstable and opening the
curvature-scaling ``channel''.

\paragraph{(E3) Scalar--fluid exchange ($a^2=\tfrac92\gamma$).}
On \(\Gamma_{SF}\) one eigenvalue at \(S\) vanishes (cf.\ \eqref{eq:eigs-S-sfc}), and one eigenvalue at \(F\) vanishes
(cf.\ \eqref{eq:trdet-F2-sfc}),  the transcritical exchange locus.  Physically this is the boundary between
scalar-dominated late-time behaviour and fluid-tracking (scaling) behaviour.  In particular, in the strata where \(F\) is
the sink, the scalar energy density follows the fluid in a rigid way determined by \((\gamma,a)\). Conversely, when \(S\)
is the sink the fluid becomes dynamically irrelevant at late times.  This ``tracker versus domination'' dichotomy is the
clearest example of a versal effect: it is forced by the normal form, and is insensitive to model details away from the
organising set.

\paragraph{(E4) Fluid--curvature degeneracy ($\gamma=\tfrac23$).}
On \(\Gamma_{MC}\) the equilibria \(M\) and \(C\) lie on an equilibrium line \(\mathcal L_{MC}\), and the organising issue is
transverse stability (normal hyperbolicity) rather than an isolated equilibrium bifurcation.  Physically this is a
marginal equation of state threshold (\(w=\gamma-1=-\tfrac13\)) at which curvature and the one-fluid mode have identical
scale factor dependence.  As a result, the system admits a continuous family of ``mixed'' curvature--fluid balances,
and small perturbations can drift along the line without selecting a unique asymptotic state.

\paragraph{(E5) Stiff-fluid coincidences ($\gamma=2$).}
On \(\Gamma_{\rm stiff}\) additional neutral directions occur at several sites, generating higher-dimensional centre
dynamics and therefore additional (non-generic) universal behaviours.  We treat \(\gamma=2\) as a boundary case, the main
message being that any physical model approaching stiff behaviour must be analysed with the full higher-codimension centre
manifold, rather than by perturbing the generic \(\gamma\neq 2\) picture.

\subsection{An explicit pulled-back trajectory and its selection rules}
\label{subsec:explicit-path}

To make the pull-back viewpoint completely explicit, it is useful to display one
simple massive-field path and the corresponding pulled-back crossings of the
organising loci. The purpose of this example is not to provide a full
observational fit, or a controlled approximation through every later threshold,
but to show how the organising atlas, once combined with a realised path in
coefficient space, produces an ordered cosmological regime sequence and associated
near-threshold dwell laws.

Fix, for simplicity, the fluid parameter as
\begin{equation}
\gamma=1.
\label{eq:traj-gamma}
\end{equation}
Consider the quadratic potential
\begin{equation}
V(\phi)=\frac12 m^2\phi^2,
\qquad
\lambda_{\mathrm{eff}}(\phi)=-\frac{2}{\phi}.
\label{eq:traj-quad-pot}
\end{equation}
As a simple analytic representative of a massive-field history, take the leading
large-field slow-roll expression

\begin{equation}
\frac{d\phi}{dN}\simeq-\frac{2}{\phi},
\qquad
\phi^2(N)=\phi_0^2-4N,
\qquad
\phi_0>\sqrt{2},
\label{eq:traj-sr-history}
\end{equation}
where \(N\) denotes the number of e-folds. We use \eqref{eq:traj-sr-history} here only as a transparent path in coefficient
space; it is not meant as a controlled slow-roll approximation through every
later threshold, where \(|\lambda_{\mathrm{eff}}|\) is no longer small. This yields the explicit pulled-back
path
\begin{equation}
\Sigma(N)=\bigl(\gamma,\lambda_{\mathrm{eff}}(N)\bigr)
=
\left(1,-\frac{2}{\sqrt{\phi_0^2-4N}}\right).
\label{eq:traj-path}
\end{equation}
Equivalently, in the bounded slope variable \(\zeta=\arctan\lambda_{\mathrm{eff}}\),
\begin{equation}
\widetilde\Sigma(N)=\bigl(1,\zeta(N)\bigr),
\qquad
\zeta(N)=\arctan\!\left(-\frac{2}{\sqrt{\phi_0^2-4N}}\right).
\label{eq:traj-zeta-path}
\end{equation}
Other monotone histories, for example exponential ans\"atze of the form
\(\phi(N)=\phi_0e^{-\alpha N}\), lead to the same qualitative ordering
mechanism; the present choice is simply a convenient analytic representative.

For \(\gamma=1\), the three principal loci in the exponential organising atlas are
\begin{equation}
\Gamma_2:\lambda^2=2,
\qquad
\Gamma_3:\lambda^2=3,
\qquad
\Gamma_1:\lambda^2=6.
\label{eq:traj-main-loci}
\end{equation}
The corresponding pulled-back path maps of Section~\ref{sec:paths} are therefore
\begin{equation}
h_{\Gamma_2}(N)=\lambda_{\mathrm{eff}}^2(N)-2,\qquad
h_{\Gamma_3}(N)=\lambda_{\mathrm{eff}}^2(N)-3,\qquad
h_{\Gamma_1}(N)=\lambda_{\mathrm{eff}}^2(N)-6.
\label{eq:traj-path-maps}
\end{equation}
In the present example,
\begin{equation}
\lambda_{\mathrm{eff}}^2(N)=\frac{4}{\phi_0^2-4N},
\label{eq:traj-lambda-sq}
\end{equation}
so the zeros of the three path maps occur at
\begin{equation}
N_2=\frac{\phi_0^2-2}{4},
\qquad
N_3=\frac{\phi_0^2-\frac43}{4},
\qquad
N_1=\frac{\phi_0^2-\frac23}{4}.
\label{eq:traj-crossing-times}
\end{equation}
Hence
\begin{equation}
N_2<N_3<N_1.
\label{eq:traj-crossing-order}
\end{equation}
For the chosen positive-\(\phi\) branch, \(\lambda_{\mathrm{eff}}<0\), so the relevant kinetic endpoint is the negative-slope branch of the \(\Gamma_1\) gate.
Thus the path crosses the pulled-back loci in the definite order
\begin{equation}
\widetilde\Gamma_2\longrightarrow \widetilde\Gamma_3\longrightarrow \widetilde\Gamma_1.
\label{eq:traj-crossing-sequence}
\end{equation}

This explicit ordering has a direct regime interpretation. For \(N<N_2\), one has
\(\lambda_{\mathrm{eff}}^2<2\), so the history lies on the accelerated scalar side of the atlas.
For \(N_2<N<N_3\), one has \(2<\lambda_{\mathrm{eff}}^2<3\), so the path has crossed the
acceleration threshold but not yet the scalar--fluid exchange threshold. For
\(N_3<N<N_1\), one has \(3<\lambda_{\mathrm{eff}}^2<6\), so the path lies on the tracking-side
portion of the atlas while still remaining below the kinetic endpoint threshold. Finally, for
\(N>N_1\), one has \(\lambda_{\mathrm{eff}}^2>6\), and the path has entered the steep/kinetic-side
exit corridor. In this sense, the pull-back maps \(h_{\Gamma_i}\) turn the organising loci into
explicit ordered episode thresholds along the realised history.

A nearby but qualitatively distinct situation is obtained from a near-grazing path. For example,
near \(\widetilde\Gamma_2\), one may consider
\begin{equation}
\lambda_{\mathrm{eff}}^2(N)=2+\varepsilon+a(N-N_\ast)^2,
\qquad
0<\varepsilon\ll1,
\qquad
a>0.
\label{eq:traj-near-grazing}
\end{equation}
This path does not cross \(\widetilde\Gamma_2\), since the minimum value of
\(\lambda_{\mathrm{eff}}^2\) is \(2+\varepsilon\), but it comes within distance \(\varepsilon\) of the
threshold. The associated reduced normal form then gives the familiar bottleneck scaling
\begin{equation}
T_{\mathrm{bottle}}\sim \varepsilon^{-1/2}.
\label{eq:traj-bottleneck}
\end{equation}
Thus the path formulation distinguishes clean crossings from prolonged near-threshold
lingering, and does so directly at the level of the pulled-back detuning functions
\(h_{\Gamma_i}\).

These two situations are summarised in Figure~\ref{fig:explicit-path}: panel~(a) shows the
realised path and its ordered crossings of the pulled-back loci, while panel~(b) shows the
corresponding threshold readout \(\lambda_{\mathrm{eff}}^2(N)\) together with a dashed
near-grazing alternative.

\begin{figure}[t]
\centering

\begin{subfigure}[t]{0.48\textwidth}
\centering
\begin{tikzpicture}
\begin{axis}[
    width=\textwidth,
    height=0.80\textwidth,
    xmin=0.92, xmax=1.08,
    ymin=0, ymax=6.6,
    axis lines=left,
    xlabel={$\gamma$},
    ylabel={$\lambda_{\mathrm{eff}}^2$},
    xtick={1},
    xticklabels={$1$},
    ytick={2,3,6},
    clip=false
]

\addplot[thin,densely dashed] coordinates {(0.92,2) (1.08,2)};
\addplot[thin,densely dashed] coordinates {(0.92,3) (1.08,3)};
\addplot[thin,densely dashed] coordinates {(0.92,6) (1.08,6)};

\addplot[very thick,-{Stealth[length=3mm]}] coordinates {(1,0.55) (1,6.55)};

\addplot[only marks,mark=*,mark size=1.8pt] coordinates {(1,2) (1,3) (1,6)};

\node[anchor=west] at (axis cs:1.015,1.6) {$\widetilde\Gamma_2$};
\node[anchor=west] at (axis cs:1.015,3.35) {$\widetilde\Gamma_3$};
\node[anchor=west] at (axis cs:1.015,5.6) {$\widetilde\Gamma_1$};
\node[anchor=east] at (axis cs:0.995,0.90) {\small increasing $N$};

\end{axis}
\end{tikzpicture}
\caption{Realised path in the pulled-back chart.}
\end{subfigure}
\hfill
\begin{subfigure}[t]{0.48\textwidth}
\centering
\begin{tikzpicture}
\begin{axis}[
    width=\textwidth,
    height=0.80\textwidth,
    xmin=0, xmax=2.10,
    ymin=0, ymax=6.6,
    axis lines=left,
    xlabel={$N$},
    ylabel={$\lambda_{\mathrm{eff}}^2(N)$},
    xtick={1.75,1.9167,2.0833},
    xticklabels={$N_2$,$N_3$,$N_1$},
    ytick={2,3,6},
    clip=false
]

\addplot[thin,densely dashed] coordinates {(0,2) (2.2,2)};
\addplot[thin,densely dashed] coordinates {(0,3) (2.2,3)};
\addplot[thin,densely dashed] coordinates {(0,6) (2.2,6)};

\addplot[very thick,domain=0:2.09,samples=200] {4/(9-4*x)};

\addplot[only marks,mark=*,mark size=1.8pt]
coordinates {(1.75,2) (1.9167,3) (2.0833,6)};

\addplot[thick,dashed,domain=1.15:2.10,samples=100]
{2.18 + 1.40*(x-1.55)^2};

\node[anchor=west] at (axis cs:0.05,6.30) {\small $\widetilde\Gamma_1$};
\node[anchor=west] at (axis cs:0.05,3.30) {\small $\widetilde\Gamma_3$};
\node[anchor=west] at (axis cs:0.05,1.6) {\small $\widetilde\Gamma_2$};
\node[align=center] at (axis cs:0.9,2.95) {near-grazing\\variant};

\end{axis}
\end{tikzpicture}
\caption{Threshold readout and near-grazing alternative.}
\end{subfigure}

\caption{Explicit pulled-back trajectory for the quadratic potential with \(\gamma=1\).
Panel~(a) shows the realised path in the pulled-back organising chart, together with its
ordered crossings of the loci \(\widetilde\Gamma_2\), \(\widetilde\Gamma_3\), and
\(\widetilde\Gamma_1\). Panel~(b) shows the corresponding threshold readout
\(\lambda_{\mathrm{eff}}^2(N)\), whose intersections with the levels \(2,3,6\) occur at the
ordered e-fold times \(N_2<N_3<N_1\). The dashed curve illustrates a near-grazing variant,
which approaches \(\widetilde\Gamma_2\) without crossing it. The figure makes explicit how
the pull-back maps \(h_{\Gamma_i}\) convert the organising loci into ordered episode
thresholds along a realised massive history.}
\label{fig:explicit-path}
\end{figure}

This example also makes clear that the path formulation carries several layers of selection. Here ``realised'' means realised by the chosen model and approximation scheme: the curve is induced by the quadratic-potential relation \(\lambda_{\mathrm{eff}}=-2/\phi\) together with the representative history \eqref{eq:traj-sr-history}, rather than chosen arbitrarily in the ambient unfolding or parameter space.
First, there are \emph{geometric selection rules}: only paths compatible with the physical
region, the invariant gates, and the admissible pull-back geometry can occur. Second, there
are \emph{dynamical selection rules}: once a realised history is monotone, the detuning
functions \(h_{\Gamma_i}\) inherit a definite ordering, so not every diagrammatically available
transition is dynamically accessible, nor can the thresholds be crossed in an arbitrary order.
Third, there are \emph{phenomenological selection rules}: among dynamically admissible
histories, only some have durations and regime sequences compatible with a plausible
cosmological interpretation. 

The organising atlas therefore determines which regime changes
are locally available; the realised path determines which of them are dynamically accessible
and in what order; and phenomenology imposes a further selection by excluding histories
whose sequencing or durations are cosmologically implausible.

The present subsection is not intended to identify the unique path selected by observation.
Rather, it shows that the framework already predicts part of the observational question:
not a single fitted trajectory, but the admissible orderings of regimes, the loci at which
transitions may occur, and the local dwell-time laws governing near-threshold passages.
Confronting these admissible histories with phenomenology is a natural next step, but the
underlying selection structure is already visible at the level of the pulled-back organising
geometry.

\subsection{Massive extensions: dynamical passage through the template}\label{subsec:primer-massive}
In the massive extensions the projection \(\pi\) in \eqref{eq:pi-frozenslope} turns an orbit into a curve in the
\((\gamma,a)\) plane.  A single cosmological solution can therefore \emph{experience} several of the effects (E1)--(E5)
sequentially, without any external parameter change: the ``parameter drift'' is internal, carried by \(a_{\rm eff}\).

In particular, whenever \(a_{\rm eff}\) crosses a threshold, the orbit crosses a pulled-back stratum boundary and the
``dominant balance'' (scalar-dominated, tracking, curvature-scaling, kinetic episode) can change abruptly.  This is the
mechanism by which massive models naturally generate multi-stage histories: a sequence of persistent regimes separated by
versal transition events. In addition, the massive closure introduces the slope gates \(Y=0\) and \(\zeta=0\) (Section~7), which control when the internal drift in \(a_{\rm eff}\) stalls or reverses direction.

\subsection{Slow-roll, ultra slow-roll, and oscillatory regimes}\label{SR}

It is useful to supplement the dictionary of Section~\ref{subsec:primer-dict} with a regime-level reading of the
stratified picture. In \cite{MI-I}, for the quadratic scalar-field problem near the Hopf--steady--state
organising centre, the asymptotically stable invariant set on the reduced centre manifold determines
the physical regime: a hyperbolic attracting equilibrium corresponds to slow roll (SR), a transient
passage near the non-attractor organising centre and its unstable continuation corresponds to
ultra slow roll (USR), and a stable periodic orbit in the reduced system lifts to an invariant torus
in the full flow, giving a robust oscillatory regime. The present paper keeps this interpretive
principle, but applies it to a wider organising geometry.

In the strict exponential scalar--fluid--curvature class, the loci \(\Gamma_{1},\ldots,\Gamma_{5}\)
do not generate a Hopf-type oscillatory attractor. Instead, they organise changes between persistent
equilibrium balances and nonhyperbolic transition episodes. Accordingly, the natural regime
reading is as follows.

\paragraph{Slow roll.}
A slow-roll regime corresponds here to a stratum in which the attracting local balance is a
hyperbolic equilibrium with small kinetic contribution and accelerated expansion. In practice,
this means a scalar-dominated or scalar-tracking branch whose sink character persists throughout
the stratum. In the language of Section~\ref{subsec:primer-exp-effects}, such regimes lie away from the organising loci and
are read from the persistent attracting site selected by the sign conditions in the corresponding
stratum.

\paragraph{Ultra slow roll.}
Ultra slow roll is not represented by a separate permanent attractor, but by prolonged passage
near a nonhyperbolic organising threshold or along its weakly unstable continuation. In the
present exponential setting, the clearest examples are the hilltop gate \(\Gamma_{1}\), where the
cubic centre reduction produces critical slowing and long kinetic-like dwell times, and marginal
situations near the loss of normal hyperbolicity loci \(\Gamma_{4}\) and \(\Gamma_{5}\), where the flow
can remain trapped near a non-isolated set before peeling away. Thus, in the same spirit as in
\cite{MI-I}, USR should be read as a bottleneck regime organised by nonhyperbolicity rather than
as a distinct asymptotic phase.

\paragraph{Oscillatory regimes.}
For the strict exponential class, the local organising templates of Sections~\ref{sec:exp-Z2-gate}--\ref{sec:curvature} do not by
themselves produce a robust oscillatory attractor analogous to the \(\alpha\)-stratum of \cite{MI-I}.
The robust local effects are instead critical slowing, curvature leakage, tracker exchange, and drift
along equilibrium continua. Oscillatory behaviour enters the present work primarily through the
massive quadratic side. Indeed, the bounded-slope formulation of Section~\ref{sec:massive-extensions} recovers the lower-dimensional
quadratic invariant subcases, so the oscillatory sector identified in \cite{MI-I}, including the
periodic-orbit / invariant-torus mechanism, remains available on the corresponding invariant slices.
From the present viewpoint, these oscillatory states should therefore be regarded as part of the
massive pull-back picture rather than as new local attractors created by the exponential loci.

Putting these points together, the regime dictionary in the present paper is:
\[
\text{SR} \Longleftrightarrow \text{persistent hyperbolic attracting balance},
\]
\[
\text{USR} \Longleftrightarrow \text{passage near a nonhyperbolic organising set},
\]
while
\[
\text{oscillations} \Longleftrightarrow \text{periodic or quasi-periodic invariant set on the massive side}.
\]
This extends the interpretation developed in \cite{MI-I}: there the emphasis falls on the local
Hopf--steady--state organising centre and its oscillatory sector, whereas here the same philosophy
is embedded into a larger stratified geometry with fluid, curvature, gates, and pulled-back
organising loci.

Finally, the pull-back formulation of Section~\ref{sec:strata} gives a dynamical version of this dictionary.
A massive orbit need not remain in a single regime: as \(a_{\mathrm{eff}}\) drifts, it can cross
preimages of the exponential organising loci and hence move through successive SR-like,
USR-like, tracking, curvature-drift, or kinetic-like episodes. When the orbit enters an invariant
quadratic slice carrying the MI-I oscillatory sector, this same trajectory may also pass into or
out of an oscillatory phase. Thus the present framework organises not only isolated regimes but
also their admissible sequences.

\subsection{A practical workflow}\label{subsec:primer-workflow}
For later use (including the applications envisaged in the Introduction), we record a minimal ``how-to''.
\begin{enumerate}
\item \textbf{Locate the regime.}  Compute \(\gamma\) for the relevant fluid sector and the effective slope \(a\) (or
\(a_{\rm eff}\)) for the scalar sector.  Place the point in Figure~\ref{fig:bifn-map-gamma-lambda} and determine the stratum.
\item \textbf{Read off the dominant balance.}  In the interior of a stratum, the attracting site(s) are fixed; use the sign
conditions encoded in \eqref{eq:eigs-S-sfc}--\eqref{eq:trdet-F2-sfc} (and the existence conditions of the curvature branch)
to decide which balance is selected.
\item \textbf{If near an organising locus, replace by the normal form.}  Use the appropriate centre reduction:
cubic at \(\Gamma_K\), fold-type curvature reduction \eqref{eq:red-S-a2eq3-sfc} at \(\Gamma_C\), and transcritical exchange
near \(\Gamma_{SF}\).  The normal form immediately yields scaling estimates for dwell times and transient amplitudes
(critical slowing near \(\Gamma_K\), algebraic curvature decay near \(\Gamma_C\), etc.).
\end{enumerate}

\section{Discussion}
\label{sec:discussion}
A central point of the present analysis is that the additional fluid and curvature modes do
not merely enlarge the state space: after reduction, they provide the extra deformation
directions needed to realise the relevant versal effects and unfolding parameters. In this way,
the full organising geometry needed to describe the dynamics of nearby systems is already
visible within the FRW class itself, without passing to more general cosmological models
outside the homogeneous and isotropic setting. An analogous analysis for more general
homogeneous cosmologies, notably the Bianchi classes, appears entirely feasible, but lies
beyond the scope of the present work and is left for future study.

The framework developed here is therefore local-to-global in a controlled sense: the versal
and normal-form language applies directly to the reduced local germs, while the wider
``organising'' and stratification language refers to the way these local models assemble into
a coherent geometric description of the cosmological dynamics.

Instead of listing the many equilibria and many subcases, here that summary object is the organising geometry: the small transition set in the strict exponential \((\gamma,a)\) plane, together with the massive equilibrium continua, their gates, and the vertical \(\gamma\)-thresholds controlling normal hyperbolicity.
Once these loci are known, the local diagram types and their persistence regions follow systematically from centre reductions
and standard normal-form logic.

The organising geometry displayed in Section~10, and made explicit in the path example of
Section~10.3, also carries a natural hierarchy of selection rules. First, only paths compatible with the physical region and
the invariant gates are geometrically admissible. Second, once a realised history is monotone,
the pull-back maps \(h_{\Gamma_i}\) impose a definite ordering on the accessible threshold
crossings. Third, phenomenology imposes a further selection: among geometrically and
dynamically admissible histories, only some have durations and regime sequences compatible
with a plausible cosmological interpretation. In this sense, the present framework already
predicts part of the observational problem---namely the admissible orderings of episodes, the
locations of transition thresholds, and the local dwell-time laws near them---even though a
full confrontation with data lies beyond the scope of the present paper.

A natural next step is to ask how much of this organising picture survives for non-canonical
scalar fields, such as DBI~\cite{SilversteinTong} or \(k\)-essence~\cite{ArmendarizPiconMukhanovSteinhardt} models. The present paper is restricted to canonical
scalar dynamics, where the effective slope variable and the associated organising loci admit a
particularly transparent formulation. For non-canonical kinetic terms one expects the local
reduced normal forms, detuning functions, and invariant gates to be modified, sometimes
substantially, by the altered kinetic structure. Nevertheless, the underlying strategy developed
here---local reduction near non-hyperbolic organisers, assembly into an organising atlas, and
path selection inside the corresponding unfolding geometry---should still provide a useful
template for such extensions. In particular, it would be interesting to understand to what extent the same organising/path
viewpoint can accommodate both inflationary episode sequences and late-time dark-energy
transitions within a common geometric language, thereby linking early- and late-universe
scalar-field phenomenology at the level of admissible regime histories.

In the strict exponential class, the slope is an external parameter and transitions are read off as crossings of loci in
parameter space. In the quadratic massive class the slope becomes a bounded state coordinate, and transitions occur dynamically
along trajectories through gates and through the pull-back of the exponential stratification. This distinction clarifies which
phenomena are genuinely ``bifurcations in parameter space'' and which are ``versal transitions under internal drift''. In the language of Section~\ref{sec:paths}, this means that the cosmological model does not explore the
full unfolding space arbitrarily, but only the physically realised path or pulled-back slice
selected by the available detunings and invariant directions.

A recurring theme is that the fluid and curvature directions are not treated as external knobs; rather, near a chosen organising
site they enter the translated jet as additional coordinates and contribute to the coefficients of the reduced germ after elimination
of hyperbolic variables. This is the precise sense in which full modes supply physical unfolding directions for the scalar-only
degeneracies. The resulting stratified hilltop picture near \(|a|=3\) is therefore not a single bifurcation curve but a family
of persistent diagram types indexed by approach direction in unfolding space.

The bounded slope massive closure adopted here is designed to remain comparable with earlier scalar-only massive analyses and to
make the additional full modes explicit. In particular, the robust equilibrium lines and gates identified in Section~7 persist
under restriction to invariant subcases (scalar-only, SF, SC), and provide a clean ambient framework for comparison with the
``undressed'' massive dynamics (no fluid, no curvature). Likewise, the distinction between (i) spectral mode interactions at a
single equilibrium and (ii) full-mode extensions of the state space prevents confusion in the interpretation of ``mode'' in
cosmology: the organising episodes described here are dominated by loss of hyperbolicity and normal hyperbolicity thresholds,
with spectral collisions entering only through the usual centre-manifold reductions.

\paragraph{Limitations and extensions.}
We have focused on two canonical potentials (strict exponential and quadratic) as universality classes for parameter-driven and
state-variable-driven organisation. A natural extension is to treat broader potential families by introducing additional bounded
slope variables and analysing the resulting higher-dimensional organising skeleton. Another extension is to include multiple fluids,
anisotropies, or additional geometric degrees of freedom. From the present viewpoint, these add further full-mode directions and
are expected to enlarge the unfolding space and enrich the stratification. Finally, while our ``primer'' section translates normal
forms into robust physical effects, quantitative confrontation with data requires connecting these versal mechanisms to specific
parameterisations of the potential and initial conditions. The stratified charts provide the qualitative background for such work.

\paragraph{On extensions in \(\gamma\).}
Although we restrict attention to \(\gamma\in(0,2]\) (standard barotropic matter), the organising locus and reduction calculations
are algebraic in \(\gamma\) and can be extended formally beyond this interval.  Allowing \(\gamma\le 0\) corresponds to effective
components with \(w=\gamma-1\le -1\) and changes the physical interpretation (and in some cases the ordering of stability regions),
but the same organising geometry framework applies.  We defer a systematic discussion of such effective fluid extensions to future work.

\bigskip

The main message is that FRW scalar field mixtures admit a small number of organising thresholds whose local normal forms control
persistence and transitions. Treating these thresholds as a unified bifurcation geometry rather than as a disconnected list of
special cases yields a compact and structurally stable dynamics for both strict exponential and harmonic extensions.

\paragraph{On the fate of the Ref.~\cite{MI-I} invariant torus}
The bounded-slope quadratic SF--fluid--curvature system contains the lower-dimensional
quadratic subcases as invariant slices, notably \(\Omega_k=0\), \(\Omega_m=0\), and
\(\Omega_k=\Omega_m=0\); hence the oscillatory periodic-orbit / invariant-torus mechanism
identified in \cite{MI-I} remains available on the corresponding invariant quadratic sectors.
What the present paper establishes is therefore the persistence of the \emph{torus-producing
mechanism} inside the enlarged ambient state space, not a Cantor family of invariant tori in
the sense of Wiggins (cf. \cite{wig88}, p. 322).

A Wiggins-type theorem would require substantially more: one must pass from the flow to an
appropriate return map, identify a compact invariant torus \(V\), prove its normal hyperbolicity,
construct its stable and unstable manifolds, and then verify the existence of a transverse
homoclinic torus. These ingredients are not established here. Accordingly, the Cantor-set
question should be regarded as a natural continuation problem rather than as a theorem proved
in the present work.

\appendix
\section{Glossary and correspondences used in this paper}
\label{app:glossary}
In this Appendix, some notions of bifurcation and singularity theory will be reviewed. Everything below refers to an autonomous smooth dynamical system in finite dimensions,
\be\label{f-sys}
\dot{x}=f(x,\lambda),
\ee
where $x(t)$ is a vector function of the time and the map (germ) $f$  also depends on some parameters collectively denoted by $\lambda$, as is the case with  the gravity problems analysed in this paper. We shall consider the graph of all equilibrium solutions\footnote{It is perhaps interesting to note in this connection the abuse of language when the words "critical point" or "singular point" are sometimes used in the literature of hyperbolic theory (and also its applications to gravity/cosmology problems) to denote an equilibrium $(x_0,\lambda_0)$ of the system \eqref{f-sys}, that is when $\dot{x}=0$, i.e.,  the point belongs to the landscape \eqref{eq:land}. This is especially misleading for hyperbolic analysis because in this way, a "critical" (or singular) point of the system \eqref{f-sys} is \emph{not} a critical (singular) point of the right-hand-side, i.e., it does not satisfy the condition that the rank of the derivative $D_xf(x,\lambda)$ at that point be less than the maximum possible value (i.e., $n$). This terminology is however, particularly useful in the context of bifurcation analyses because there the said rank is never maximal.} 
\be\label{eq:land}
\mathcal{Z}=\{\,(x,\lambda)\mid f(x,\lambda)=0\},
\ee
in the combined $(x,\lambda)$-space; in  the `landscape' point of view a non-hyperbolic equilibrium point $(x_0,\lambda_0)$ (simplest form of non-hyperbolic set) corresponds to a valley floor that has been `flattened' in the sense that the Jacobian matrix $D_x f(x_0,\lambda_0)$ is singular.

\subsection{Some standard definitions from bifurcation theory}
\begin{description}
\item[\texttt{non-hyperbolic set/object}.]
Typically an equilibrium point, line, or other continuum (also called a `site' - see below),  where the Jacobian is non-invertible; this implies a \emph{necessary} condition for the existence of multiple solutions $x(\lambda)$ of the landscape equation \eqref{eq:land}.

\item[\texttt{centre reduction}.]
Reduction of the flow near a non-hyperbolic point to its centre manifold, yielding a lower-dimensional reduced system whose leading
terms determine the local phase portrait.

\item[\texttt{normal form}.]
A simplified representative (obtained by near-identity changes of variables on the centre manifold) in which the reduced dynamics
takes a canonical polynomial form (e.g.\ cubic hilltop, fold-type quadratic, transcritical exchange).

\item[\texttt{translated jet}.]
The Taylor expansion of the vector field about a chosen site after translation to local coordinates, typically kept to quadratic
(or cubic) order to determine the leading reduced dynamics and the coefficients entering the normal form.

\item[\texttt{germ} / \texttt{strong equivalence}.]
The local dynamical behaviour near a site considered up to smooth changes of coordinates (and admissible reparametrisations) that
preserve qualitative structure; used to classify reduced dynamics into canonical types.

\item[\texttt{Distinguished vs.~unfolding parameters}.]
We shall call the original parameters of a model `distinguished', or external, or control parameters. By contrast the unfolding (or, `auxiliary') parameters are the coefficients \(\mu\) appearing in the topological normal form. In the strict exponential case some correspond to genuine external
parameters (e.g.\ \(a-3\)), while others arise as induced coefficients coming from motion in extra state-space directions
(fluid/curvature fractions) prior to reduction. 

\item[\texttt{(versal) unfolding}.]
A family of reduced vector fields depending on parameters \(\mu\) that realises, up to equivalence, all small perturbations of the
reduced germ. ``Versal'' means no essential perturbation direction is missed.

\item[\texttt{unfolding parameters and codimension}.]
The number of unfolding parameters  is the `codimension' of a versal unfolding and typically measures the degree of degeneracy of the dynamical system;  the larger the codimension  the higher the degeneracy or complexity of the dynamics. Zero codimension corresponds to hyperbolic dynamics.

\item[\texttt{transition set}.]
The subset of unfolding-parameter space (or its projection to physical parameter space) where the reduced diagram type fails to be
structurally stable (e.g.\ multiplicity changes, stability exchange, higher degeneracy). The complement is partitioned into strata.

\item[\texttt{stratum} / \texttt{stratification}.]
A stratum is a connected component of parameter space with the organising set removed; within a stratum the qualitative local phase
portrait persists. A stratification is the decomposition into all strata plus their organising boundaries.

\item[\texttt{transition variety}.]
A higher-codimension subset in unfolding-parameter space (e.g.\ fold, cusp, or swallowtail surfaces) that refines the macroscopic
stratification near a locus such as \(|a|=3\), distinguishing different persistent diagram types by approach direction in full
unfolding space.

\item[\texttt{bifurcation diagram}.]
A pictorial depiction of  bifurcation sets separating the different strata in unfolding-parameter space together with the corresponding phase diagrams. The bifurcation diagram describes the full dynamics of the versal unfolding.  In older terminology,  `bifurcation diagram' was used in a similar sense to `landscape' as in Eq. \eqref{eq:land}.

\item[\texttt{persistence domain}.]
A parameter-space region (equivalently, a stratum) on which the qualitative phase portrait (types and stability of relevant sites
and their local connections) is unchanged under small perturbations.

\item[\texttt{frozen-slope projection}.]
In the massive (quadratic) class, the map that associates to a state its instantaneous effective slope, producing a curve
\((\gamma,a_{\rm eff}(N))\) in the distinguished parameter plane \((\gamma,a)\) along an orbit.

\item[\texttt{pull-back stratification}.]
Given the frozen-slope map \(\pi:(X,Y,\Omega_k,\zeta;\gamma)\mapsto (\gamma,a_{\rm eff}(\zeta))\), the pull-back stratification is
the decomposition of the massive state space by the preimages \(\pi^{-1}(\mathcal S)\) of strata \(\mathcal S\) in the exponential
\((\gamma,a)\) stratification. Crossing a pulled-back organising hypersurface \(\pi^{-1}(\Gamma)\) produces the same \emph{local}
diagram change as crossing \(\Gamma\) in the parameter-driven exponential setting.

\end{description}

\subsection{Further  organising notions}
\begin{description}

\item[\texttt{Episode}.]
A local organising `event' in which one or more equilibrium objects  and/or invariant boundaries interact as a control
quantity is varied.  In the strict exponential class, this variation is parameter-driven (e.g.\ moving in \((\gamma,a)\));
in the massive class it can be state-variable-driven (e.g.\ drift of \(a_{\rm eff}(N)\)).
Typical episodes include: collision/creation/annihilation of equilibria, loss/recovery of (normal) hyperbolicity,
or a coupled event involving several sites and an existence boundary.

\item[\texttt{organising locus}.]
A subset of parameter space (e.g.\ a curve or axis in \((\gamma,a)\)) and/or an invariant subset in state space on which a
distinguished equilibrium object (a \texttt{site}) loses (or recovers) hyperbolicity, or an equilibrium manifold loses
(or recovers) normal hyperbolicity.  Crossing an organising locus forces a qualitative change in the local phase portrait.

\item[\texttt{organising set}.]
The union of the organising loci relevant to a given class (for the strict exponential SFC class: the five loci listed in
Theorem~\ref{thm:Ri-summary}\textbf{(R1)}).

\item[\texttt{organising geometry}.]
The organising set together with the associated local reduced normal forms and their transition/stratification structure.
Equivalently: the collection of (i) the distinguished loci where qualitative change is possible, and (ii) the universal
local models that determine what changes when those loci are crossed.

\item[\texttt{site}.]
A distinguished equilibrium object (point, line, segment, etc.) viewed as part of the organising geometry: typically an
equilibrium that is non-hyperbolic on an organising locus (or becomes non-hyperbolic at a threshold), so that a
centre-manifold or normal-hyperbolicity analysis is required.  Away from organising loci the same object may be hyperbolic;
we still refer to it as a site when it is being tracked as part of an organising episode.

\item[\texttt{gate}.]
An invariant subset (often a boundary component or invariant hypersurface) through which trajectories can enter or exit a
persistent regime.  Gates are typically attached to, or emanate from, sites where hyperbolicity/normal hyperbolicity is lost,
and they control admissible transition routes between strata in unfolding parameter space.

\item[\texttt{organising centre}.]
A coupled local organising episode involving more than one site (or a site together with an existence boundary) whose
interaction controls the local reduced dynamics.  In this paper the canonical example is the multi-site episode at \(|a|=3\).

\item[\texttt{organising degeneracy}.]
The nature of degeneracy present in the normal form as measured by codimension. A codim-$k$ organising degeneracy  implies the present of exactly $k$ unfolding parameters in the versal unfolding, and a $k$-dimensional bifurcation diagram. For example, a saddle-node bifurcation shows a codim-1 organising degeneracy,  a $\mathbb{Z}_2$-equivariant normal form has a codim-2 degeneracy, whereas a hilltop degeneracy is of codim-3.

\item[\texttt{equilibrium manifold / continuum}.]
A set of equilibria forming a smooth submanifold (line/segment) in state space. The tangential directions are neutral for geometric
reasons; qualitative change is controlled by transverse (normal) directions.

\item[\texttt{normal hyperbolicity}.]
An equilibrium manifold is normally hyperbolic if the linearisation has no centre spectrum transverse to the manifold.
Loss/recovery of normal hyperbolicity occurs when a \emph{normal} eigenvalue crosses \(0\); this is the organising mechanism on
axes such as \(\gamma=\tfrac23\) for the \(M\)–\(C\) equilibrium manifold.
\end{description}

\section{Toy examples for the organising notions used in the text}
\label{app:toy}

This appendix briefly illustrates, on simple low-dimensional systems, a few of the organising
notions used throughout the paper. Its purpose is not to provide a general introduction to
bifurcation or centre-manifold theory, but to give a compact entry guide to the specific
vocabulary employed in the main text. In particular, the examples below are chosen to mirror
three recurrent mechanisms in the paper: transition sets separating persistent strata, exchange
thresholds, and loss of normal hyperbolicity along an equilibrium continuum. In the main text,
these appear respectively in the local transition geometry near loci such as $\Gamma_1$, in the
scalar--fluid exchange curve $\Gamma_3$, and in the equilibrium-continuum threshold
$\Gamma_4$.

\subsection{Transition sets, strata, and persistence: a fold example}

Consider the one-parameter scalar family
\be 
\dot x=\mu-x^2.
\label{eq:s-n}
\ee
The equilibria satisfy \(x^2=\mu\). Hence:
for \(\mu<0\) there are no real equilibria, for \(\mu=0\) there is one double equilibrium at
\(x=0\), and for \(\mu>0\) there are two equilibria \(x=\pm\sqrt{\mu}\). Thus the distinguished
parameter value \(\mu=0\) is the transition set, while the open regions \(\mu<0\) and \(\mu>0\)
are strata on which the qualitative phase portrait persists. This is the simplest toy model for
the rôle played in the main text by transition loci separating persistent parameter regions.

Equivalently, the family \eqref{eq:s-n}
may be viewed as the standard codimension-one versal unfolding of the degenerate
vector field \(\dot x=-x^2+O(x^3)\): the single parameter \(\mu\) is enough to realise the
generic local diagram change. In this sense the fold is structurally stable as a local
one-parameter bifurcation pattern: higher-order perturbations do not change its qualitative
nearby fixed-point geometry.

\subsection{Exchange thresholds: a transcritical example}

Consider next
\be 
\dot x=\mu x-x^2.
\label{eq:trans}
\ee
The equilibria are \(x=0\) and \(x=\mu\), and they exchange stability at \(\mu=0\). This is the
simplest toy model of an exchange threshold: the set \(\mu=0\) marks the parameter value at
which the diagram changes through a balance-exchange mechanism. In the main text, the
curve \(\Gamma_3\) plays this rôle for the scalar--fluid exchange.

Unlike the fold, however, the transcritical pattern is not structurally stable under generic
perturbation. A perturbation by a constant term,
\be 
\dot x=\varepsilon+\mu x-x^2,
\label{eq:trans-pert}
\ee
breaks the exact crossing of the two equilibrium branches: depending on the sign of
\(\varepsilon\), one obtains either separated branches with no exchange at \(\mu=0\) or a pair of
saddle-node folds. Thus the transcritical should be viewed as a more degenerate exchange
threshold than the fold. This is why, in the main text, exchange loci such as \(\Gamma_3\)
must be read through their specific reduced normal forms rather than as generic codimension-one
bifurcations of fold type.

\subsection{Loss of normal hyperbolicity: an equilibrium-line example}

Finally, consider the planar system
\begin{equation}
\dot x=\mu x,\qquad \dot y=0.
\label{eq:loss-nor-hyp}
\end{equation}
For every value of \(\mu\), the line \(x=0\) is a continuum of equilibria. The \(y\)-direction is
tangential and neutral, while the normal eigenvalue is \(\mu\). Thus for \(\mu\neq0\) the line
\(x=0\) is normally hyperbolic, whereas at \(\mu=0\) this normal hyperbolicity is lost. In this
sense, \(\mu=0\) is not an isolated-equilibrium bifurcation value, but the threshold at which an
equilibrium continuum ceases to be normally hyperbolic.

This is the toy mechanism relevant to the \(\Gamma_4=\{\gamma=2/3\}\) threshold in the main
text, where the distinguished object is likewise an equilibrium continuum rather than an isolated
site. It is also the simplest local prototype behind Fenichel-type persistence results for
overflowing or inflowing invariant manifolds: compare, for example, Wiggins' discussion of the
segment \(y=0\) in the planar linear system \(\dot x=ax,\ \dot y=-by\), where the invariant
segment persists under sufficiently small perturbation as long as the appropriate normal
hyperbolicity inequalities remain valid~\cite{wig88}, Example 1.3.5.

\subsection{Paths in unfolding space: a toy illustration}

In this work, bifurcation problems often arise in families of equations of the form
\(\dot x=g(x,\lambda)\), at distinguished points \((x_0,\lambda_0)\) where the reduced
equation \(g(x,\lambda)=0\) is most degenerate in an appropriate local sense. At such
organising centres one applies bifurcation-theoretic methods to describe the nearby diagram.

Consider, for example, the pitchfork bifurcation
\begin{equation}
g(x,\lambda)=x^3+\lambda x=0.
\label{eq:pitch}
\end{equation}
Its core is the cubic germ \(x^3\), whose standard two-parameter versal deformation may be
written as
\begin{equation}
\dot x=\mu_1+\mu_2 x+x^3.
\label{eq:two-par}
\end{equation}

As a family in the unfolding variables \((\mu_1,\mu_2)\), this system has a transition set in
the \((\mu_1,\mu_2)\)-plane separating regions with different equilibrium diagrams. However, a
given physical model need not explore this plane freely: it may only realise a one-parameter
path
\begin{equation}
(\mu_1,\mu_2)=h(\lambda)=(0,\lambda).
\label{eq:path}
\end{equation}
In the usual \((u,v)\) notation for the versal family \(x^3+ux+v\), the pitchfork corresponds to
the path \((u,v)=(\lambda,0)\) (cf.~\cite{montaldi21}, Chapter~18); equation~\eqref{eq:path}
is the same statement written in the present \((\mu_1,\mu_2)\) ordering.

The observable transitions are then those encountered along the image of this path, not all
those available in the ambient unfolding diagram. This is the basic idea behind the path
formulation used in Section~\ref{sec:paths}: the full unfolding geometry serves as the
organising chart, while the cosmological model itself realises only distinguished slices or
paths inside it.

\addcontentsline{toc}{section}{Acknowledgments}
\section*{Acknowledgments}
The author is grateful to Tony Roberts for useful discussions and for spotting an error in the coefficient of the cubic term in the centre manifold reduction equation \eqref{eq:red-Kp-sfc-worked}.


\addcontentsline{toc}{section}{References}

\begin{thebibliography}{99}
\bibitem{ycb09}Y. Choquet-Bruhat, \emph{General relativity and the Einstein equations} (Oxford University Press, 2009).
\bibitem{thom72}R. Thom, \emph{Structural Stability and Morphogenesis} (CRC Press, 2018, translation of the original 1972 French edition).
\bibitem{arnold85}
V.~I. Arnol'd, S.~M. Guse\u{\i}n-Zade and A.~N. Varchenko,
\emph{Singularities of Differentiable Maps, Vol.~I}
(Birkh\"auser, 1985).
\bibitem{kuz23}Yu. A. Kuznetsov, \emph{Elements of Applied Bifurcation Theory,} Fourth Ed. (Springer, AMS 112, 2023)
\bibitem{montaldi21}
J. Montaldi,
\emph{Singularities, Bifurcations and Catastrophes}
(Cambridge University Press, 2021).

\bibitem{coley1}A. A. Coley, \emph{Dynamical Systems and Cosmology} (Springer, 2003).
\bibitem{SHS2014}M. Szydlowski, O. Hrycyna and A. Stachowski, \emph{Scalar field cosmology -- geometry of dynamics}, Int. J. Geom. Methods Mod. Phys. 11 (2014) 1460012; arXiv:1308.4069 [gr-qc].
\bibitem{CopelandReview2018}S. Bahamonde, C. G. Boehmer, S. Carloni, E. J. Copeland, W. Fang, N. Tamanini, \emph{Dynamical systems applied to cosmology: dark energy and modified gravity,} Phys. Rep.  775-777 (2018) 1-122.
\bibitem{ChervonBook}S. Chervon, I. Fomin, V. Yurov and A. Yurov, \emph{Scalar Field Cosmology}  (World Scientific, 2019).

\bibitem{6} V. A. Belinski, L. P. Grishchuk, I. M. Khalatnikov,  and Ya. B. Zel'dovich, \emph{Inflationary stages in cosmological models with a scalar field}, Phys. Lett. B \textbf{155B} (1985) 232.
\bibitem{hal87}Halliwell, J, \emph{Scalar Fields in Cosmology with an Exponential Potential,} Phys. Lett. B185 (1987)
341.
\bibitem{8} G. W. Gibbons, S. W. Hawking, and J. M. Stewart, \emph{A Natural Measure on the Set of All Universes,} Nucl. Phys. B \textbf{281} (1987) 736.    
\bibitem{ahu15} A. Alho, J. Hell, and C. Uggla, \emph{Global dynamics and asymptotics for monomial
scalar field potentials and perfect fluids}, Class. Quant. Grav., 32(14):145005, 2015; 	arXiv:1503.06994 [gr-qc]
\bibitem{cop1}Copeland, E. J., Liddle, A. R., and Wands, D.,  \emph{Exponential potentials and cosmological
scaling solutions,} Phys. Rev. D57 (1998) 4686; arXiv:gr-qc/9711068.
\bibitem{van1}van den Hoogen, R., Coley, A. A., and Wands, D.,  \emph{Scaling solutions in Robertson-Walker
space-times,} Class. Quant. Grav., 16 (1999) 1843; arXiv:gr-qc/9901014.


\bibitem{cop2}Copeland, E. J., Mizuno, S., and Shaeri, M., \emph{Dynamics of a scalar field in Robertson-Walker
spacetimes} Phys. Rev. D79 (2009) 103515; arXiv:0904.0877.

\bibitem{u1} A. Alho, C. Uggla, and J. Wainwright, \emph{Quintessence from a state space perspective}, Phys. Dark Universe 39 101146; arXiv:2209.09684
\bibitem{u2} A. Alho, E. Bernholm, C. Uggla, \emph{Quintessence}, arXiv:2511.02727
\bibitem{u3} A. Alho and C. Uggla, \emph{Quintessence: Quadratic potentials}, arXiv:2511.12244

\bibitem{MI-I}
S. Cotsakis and I. Antoniadis, \emph{Mode interactions in scalar field cosmology}, Phil. Trans. R. Soc. (2026) (accepted); arXiv:2512.04607v2


\bibitem{ar93}V. I. Arnold, V. V. Goryunov, O. V. Lyashko, V. A. Vasiliev, \emph{Singularity Theory II: Classification and Applications}, In: Dynamical Systems VIII, V. I. Arnold (ed.) (Springer, 1993)
\bibitem{ar94}V. I. Arnold, \emph{Dynamical Systems V: Bifurcation Theory and Catastrophe Theory }(Springer, 1994)
\bibitem{GolubitskySchaefferI}M. Golubitsky,  D. G. Schaeffer, \emph{Singularities and Groups in Bifurcation Theory,} Volume I (Springer, 1988)
\bibitem{gh83}J. Guckenheimer and P. Holmes, \emph{Nonlinear oscillations, dynamical systems, and bifurcations of vector fields} (Springer, 1983)
\bibitem{wig03}S. Wiggins, \emph{Introduction to applied nonlinear dynamical systems and chaos,} 2nd. Ed. (Springer, 2003)
\bibitem{ar83}V. I. Arnold, \emph{Geometrical Methods in the Theory of Ordinary Differential Equations} (Springer, 1983)
\bibitem{roberts} A. J.	Roberts,  \emph{Model emergent dynamics in complex systems} 1st. Ed. (SIAM, 2015)
\bibitem{wig88}S. Wiggins, \emph{Global Bifurcations and chaos,} 2nd. Ed. (Springer, 1988)
\bibitem{Zholondek1983}K. Zholondek, \emph{On the versality of a family of symmetric vector fields in the plane}, Math. USSR Sbornik 48 (1984) 463
\bibitem{CotsakisFL}S. Cotsakis, \emph{Friedmann-Lema\^itre universes and their metamorphoses}, Eur. Phys. J. C (2025) 85:579; arXiv:2411.17286v2
\bibitem{an66}A. A. Andronov, A. A. Vitt, and S. E. Khaikin, \emph{Theory of Oscillators} (New York: Dover, 1987)
\bibitem{SilversteinTong}
E.~Silverstein and D.~Tong,
\emph{Scalar speed limits and cosmology: Acceleration from D-cceleration},
Phys.\ Rev.\ D {\bf 70} (2004) 103505.

\bibitem{ArmendarizPiconMukhanovSteinhardt}
C.~Armendariz-Picon, V.~Mukhanov and P.~J.~Steinhardt,
\emph{Essentials of k-essence,}
Phys.\ Rev.\ D {\bf 63} (2001) 103510.
\end{thebibliography}
\end{document}